\documentclass[AMA,Times1COL]{WileyNJDv5} 

\usepackage{cleveref}
\usepackage{comment}
\usepackage{multirow}
\usepackage{array, blkarray, makecell}
\usepackage{rotating}
\usepackage{adjustbox}
\usepackage{graphicx}
\usepackage{tabularx}
\usepackage{caption}
\usepackage{tikz}
\usetikzlibrary{fit, arrows.meta}
\usepackage[section]{placeins}   

\def\bSig\mathbf{\Sigma}

\newcommand{\indep}{\perp \!\!\! \perp}
\newcommand{\E}{\mathbb{E}}
\newtheorem{objective}{Objective}

\DeclareMathOperator*{\argmin}{arg\,min}
\DeclareMathOperator*{\logit}{\text{logit}}

\articletype{Original Article}%

\received{Date Month Year}
\revised{Date Month Year}
\accepted{Date Month Year}

\begin{document}

\title{Adaptive sparsening and smoothing of the treatment model for longitudinal causal inference using outcome-adaptive LASSO and marginal fused LASSO}

\author[1,2]{Mireille E Schnitzer}
\author[3,4]{Denis Talbot}
\author[1]{Yan Liu}
\author[5]{David Berger}
\author[6,7]{Guanbo Wang}
\author[2,8]{Jennifer O'Loughlin}
\author[2,8]{Marie-Pierre Sylvestre}
\author[9]{Ashkan Ertefaie}

\authormark{Schnitzer \textsc{et al.}}
\titlemark{Adaptive sparsening and smoothing of the treatment model for longitudinal causal inference using outcome-adaptive LASSO and marginal fused LASSO}

\address[1]{\orgdiv{Faculté de pharmacie}, \orgname{Université de Montréal}, \orgaddress{\state{QC}, \country{Canada}}}

\address[2]{\orgdiv{Département de médecine sociale et préventive}, \orgname{Université de Montréal}, \orgaddress{\state{QC}, \country{Canada}}}

\address[3]{\orgdiv{Département de médecine sociale et préventive}, \orgname{Université Laval}, \orgaddress{\state{QC}, \country{Canada}}}

\address[4]{\orgdiv{Centre de recherche du CHU de Québec}, \orgname{Université Laval}, \orgaddress{\state{QC}, \country{Canada}}}

\address[5]{\orgdiv{Department of Computer Science}, \orgname{Université de Montréal}, \orgaddress{\state{QC}, \country{Canada}}}

\address[6]{\orgdiv{The Dartmouth Institute for Health Policy and Clinical Practice}, \orgname{Geisel School of Medicine at Dartmouth}, \orgaddress{\state{NH}, \country{United States}}}

\address[7]{\orgdiv{Department of Biomedical Data Science}, \orgname{Geisel School of Medicine at Dartmouth}, \orgaddress{\state{NH}, \country{United States}}}

\address[8]{\orgdiv{Centre de recherche du centre hospitalier}, \orgname{Université de Montréal}, \orgaddress{\state{QC}, \country{Canada}}}

\address[9]{\orgdiv{Department of Biostatistics, Epidemiology \& Informatics, Division of Biostatistics}, \orgname{University of Pennsylvania}, \orgaddress{\state{PA}, \country{United States}}}

\corres{Mireille Schnitzer, Faculté de pharmacie, Université de Montréal, Montréal, QC, H3C 3J7, Canada\email{mireille.schnitzer@umontreal.ca}}


\abstract[Abstract]{Causal variable selection in time-varying treatment settings is challenging due to evolving confounding effects. Existing methods mainly focus on time-fixed exposures and are not directly applicable to time-varying scenarios. We propose a novel two-step procedure for variable selection when modeling the treatment probability at each time point. We first introduce a novel approach to longitudinal confounder selection using a Longitudinal Outcome Adaptive LASSO (LOAL) that will data-adaptively select covariates with theoretical justification of variance reduction of the estimator of the causal effect. We then propose an Adaptive Fused LASSO that can collapse treatment model parameters over time points with the goal of simplifying the models in order to improve the efficiency of the estimator while minimizing model misspecification bias compared with naive pooled logistic regression models. Our simulation studies highlight the need for and usefulness of the proposed approach in practice. We implemented our method on  data from the Nicotine Dependence in Teens study to estimate the effect of the timing of alcohol initiation during adolescence on depressive symptoms in early adulthood.}



\maketitle



\section{Introduction}
\label{s:intro}

Near practical positivity violations are a common problem when conducting causal inference with time-varying binary treatments, especially when treatment can change over many time-points. For marginal structural models (MSMs), which are models for the counterfactual outcome under a longitudinal treatment intervention, the typical sequential positivity assumption requires that, at each time-point, the probability of accessing either level of treatment (propensity score) is non-zero for any possible covariate values and prior treatments. The related practical condition is that one must have observed outcome information under all relevant treatment patterns and baseline and time-varying covariate values. In typical finite sample applications, this condition is difficult to satisfy.~\citep{Rudolphnonpositivity, SchomakerLTMLEchalleges} Under such data sparsity, outcome regression-based methods can smooth over time points where treatments are not observed for certain covariate strata. However, unless all outcome models are correctly specified -- which is difficult to achieve when extrapolation is necessary -- this approach leads to biased estimation. Alternatively, many methods involve weighting by the inverse of the probability of treatment.~\citep{Bang, RobinsMSM2000, vdLandGruber2012LTMLE} This involves directly modeling the probability of treatment for each time-point. When smoothing is desirable, it is possible to pool these treatment models over time~\citep{ColeHernanIPW, HernanMSM2000, Rudolphnonpositivity} or covariate information, \citep{ColeHernanIPW} but this can also lead to bias if the resulting models are not correctly specified for the probability of treatment at each time-point. \textcolor{black}{Doubly robust causal inference approaches  such as Targeted Maximum Likelihood Estimation~\cite{van2011targeted,vanderLaanLTMLE} (TMLE), Augmented-Inverse Probability of Treatment Weighting,~\cite{Bang} and Double Machine Learning,~\cite{ChernozhukovDML} which can leverage machine learning methods for the estimation of nuisance quantities, perform well in the absence of data sparsity. However, while they avoid parametric assumptions, machine learning applied directly to estimating the probability of treatment may exacerbate issues related to sparsity.~\cite{SchnitzerCTMLE, diop2022confounding}} It is, therefore, of interest to develop data-adaptive approaches to model selection that can trade off bias and variance under sparse conditions.

A related challenge involves defining and selecting covariates to satisfy the sequential conditional exchangeability (``no unmeasured confounders'') assumption.~\citep{ColeHernanIPW}  Beyond adjustment for confounders, excluding covariates that only affect the treatment (i.e. instruments) and adjusting for pure causes of the outcome can reduce estimation variance with inverse probability of treatment weighting.~\citep{Brookhartvarselection, Rotnitzkyoveradjustment, Schistermanoveradjustment} Previous work related to fitting MSMs with time-varying treatments demonstrated quantitatively \citep{Lefebvre} and theoretically \citep{RotnitzkySmucler, LongEffAdenyo} that estimation variance can be inflated when adjusting for variables that only cause treatment. In particular, Rotnitzky and Smucler (2020)\cite{RotnitzkySmucler} established that, under a nonparametric model with a given directed acyclic graph (DAG), and a time-dependent adjustment set, the removal of certain types of non-confounding covariates will reduce the asymptotic variance of nonparametric efficient estimators.

Many approaches for covariate selection or reduction \textcolor{black}{that target reduced variance of the causal estimator} have been developed in a single time-point treatment setting,~\citep{LohandVansteelandt, PERSSON2017280, SchneeweissHDPS, TangCausalBall} including Bayesian approaches.~\citep{TalbotBCEE, WangBAC,WilsonBayesVarSel} In particular, Shortreed and Ertefaie's outcome-adaptive LASSO~\citep{SE} uses the inverse magnitude of the outcome model regression coefficients to penalize the corresponding covariates in an adaptive LASSO for the treatment model. This aims to exclude any variable \textcolor{black}{from the treatment model} that does not have a conditional association with the outcome. 
One version of Collaborative Targeted Maximum Likelihood Estimation (C-TMLE)~\citep{GruberCTMLE} greedily selects covariates into the treatment model when the inclusion improves the fit of the propensity score-updated outcome model, and uses cross-validation to select an optimal number of selection steps.
\textcolor{black}{In the longitudinal setting, there are few options for dealing with data sparsity.} Schnitzer et al. (2020) \cite{SchnitzerCLTMLE} extended C-TMLE to the longitudinal treatment setting but noted its computational complexity. \textcolor{black}{Other approaches \cite{Diaz03042023,KennedyIncrementalInterventions} have been developed that modify the target causal parameter, leading to a different scientific interpretation, but allowing for greater robustness to sparsity.}

While not yet applied to any causal setting, the fused LASSO~\citep{Hoefling, FusedLASSOTib,Viallon_StatsComp} was proposed to smooth over spacial or temporal structures by penalizing both coefficient magnitudes and the distance between coefficients of neighboring or grouped covariates in a linear regression model. Viallon et al. \cite{Viallon2013,Viallon_StatsComp} proposed an extension for generalized linear models with adaptive weights~\citep{ZouAdaptive} resulting in oracle properties. The optimization problem is solved with a coordinate-wise optimization algorithm.~\citep{Hoefling}

In this study, we explore a two-phase approach to treatment model dimension reduction in the longitudinal context with a causal objective of estimating the parameters of an MSM. We first define a ``saturated'' model for the probability of treatment at all time-points that adjusts for the complete history of covariates and completely stratifies the model by time. Then, we propose to remove covariates from this model using a longitudinal extension of the outcome-adaptive LASSO applied to the saturated model. Our selection criteria use parametric working outcome models to operationalize the variance reduction results. \cite{RotnitzkySmucler,LongEffAdenyo} The second step, carried out after the initial covariate reduction, involves model smoothing over time such that covariate-treatment associations can have shared parameters over time. This step utilizes an implementation of the adaptive fused LASSO for a logistic regression that penalizes the distance between coefficients of a given covariate and the treatment at different time points. We then perform extensive simulation studies to evaluate the performance and robustness of our approach compared to non-adaptive and oracle estimators and longitudinal C-TMLE. 
Finally, we apply our approach in a complex longitudinal setting to estimate the effect of drinking initiation in adolescents on scores of depression symptoms in early adulthood.

\section{Data, target parameter, and estimation}\label{sec 2}

In this section, we present the target of estimation and give preliminaries that will allow us to describe our proposed model selection procedure.
\subsection{Data and target parameter}
Suppose that we have a data structure $\boldsymbol{O}=(\boldsymbol{L}_0,A_0,\boldsymbol{L}_1,A_1,...,A_T,Y)$ where the outcome $Y$ is continuous, the treatments $A_t; t=0,...,T$ are binary, and the covariates $\boldsymbol{L}_t; t=0, ..., T$ are multivariate. We use $\overline{\boldsymbol{L}}_t=(\boldsymbol{L}_0,...,\boldsymbol{L}_t)$ to indicate the history of covariates up to and including time $t$. The covariates potentially include both binary and continuous components. We take $n$ independent identically distributed samples, with realizations denoted by lowercase letters, e.g. $\boldsymbol{o}_i$ and $a_{0,i}$ are realizations of $\boldsymbol{O}$ and $A_0$, respectively, for $i=1,...,n$. 

We consider an intervention that sets each treatment node $A_t$ to some fixed value, either zero or one, i.e. setting $\overline{A}=(A_0,\dots,A_T)$ to the treatment pattern $\overline{a}=(a_0,...,a_T)$ where each $a_t$ is either zero or one. 
Define the counterfactual variable $Y^{\overline{a}}$ to be the potential outcome under treatment pattern $\overline{a}$. 
Our interests lie in modeling the marginal expectation of the counterfactual outcome potentially conditional on some subset of the baseline covariates $\boldsymbol{L}_0$, i.e. modeling $\E(Y^{\overline{a}}\mid \boldsymbol{L}_0)$. 
As an example, define the MSM of interest as 
\begin{align}
   \E(Y^{\overline{a}}\mid {L}^{1}_0 )=\mu_0 + \mu_1 {L}^{1}_0 + {\mu}_2 cum(\overline{a}), 
    \label{MSM}
\end{align}
where ${L}^{1}_0\subseteq \boldsymbol{L}_0$ is a single (one-dimensional) baseline covariate, and $cum(\overline{a})$ represents the cumulative function that counts the number of ones (i.e. number of treated time-points) in the treatment pattern $\overline{a}$. Thus, $\mu_2$ represents the change in the expected outcome from one additional treated time point under the linear model and $\{\mu_0, \mu_1, \mu_2\}$ is our target parameter. 
We can more appropriately define the MSM as a working model, and the true parameter values as projections of the true counterfactual regression curve $\E(Y^{\overline{a}}\mid L_0^1)$ onto the working model \textcolor{black}{under a given loss function}.~\citep{NEUGEBAUER2007419} See Petersen et al.\cite{Petersen2014} for complete details in the time-varying treatment context.

\subsection{Identifiability}

The MSM parameters are identifiable under typical causal assumptions including consistency, sequential positivity, and sequential conditional exchangeability.~\cite{MSMreference_Synthese} Sequential conditional exchangeability  is given as
 \begin{align*}
     Y^{\overline{a}} \indep A_t \mid (\overline{A}_{t-1}=\overline{a}_{t-1}, \overline{\boldsymbol{L}}_t), \quad \quad t=0,...,T, 
 \end{align*}
 where variables with negative subscripts should be discarded here and subsequently.
 Positivity means that $P(A_t\mid \overline{\boldsymbol{L}}_t, \overline{A}_{t-1})>0$ for all supported values of $(\overline{\boldsymbol{L}}_t, \overline{A}_{t-1})$. Consistency means that we can equate the potential outcome under the observed treatment to the observed outcome, i.e. $Y^{\overline{a}}=Y$ if $\overline{A}=\overline{a}$.

In order to review the identifiability of the MSM parameters, we define nested expectations of the outcome conditional on the treatment pattern of interest.~\citep{Bang} We initialize $q_{T+1}(\overline{a}_{T+1},\overline{\boldsymbol{L}}_{T+1})=Y$. We recursively define 
\begin{align*}
  q_t(\overline{a}_t,\overline{\boldsymbol{L}}_t) = \E\{ q_{t+1}(\overline{a}_{t+1},\overline{\boldsymbol{L}}_{t+1}) \mid \overline{A}_t=\overline{a}_t, \overline{\boldsymbol{L}}_t\}, \quad \quad t=T,...,0.
\end{align*}
We will use the notation $q^{\overline{a}}_t=q_t(\overline{a},\overline{\boldsymbol{L}}_t)$ for brevity. Under the above causal assumptions, the g-formula $\E(q_0^{\overline{a}}\mid L_0^1)=\E(Y^{\overline{a}}\mid L_0^1)$ identifies the true regression curve for $Y^{\overline{a}}$ with respect to $\overline{a}$ and $L_0^1$.~\cite{Bang} Under the MSM in \cref{MSM}, this also identifies the true values of the parameters $\boldsymbol{\mu}=(\mu_0,\mu_1,\mu_2)$.

If we do not want to make any causal assumptions, we can alternatively define our parameter of inference statistically through the function $q_0^{\overline{a}}$. 
The parameters $\boldsymbol{\mu}$ can be directly defined as minimizing the least-squares risk function of the model with $q_0^{\overline{a}}$ as the outcome with a regression specification according to the right-hand side of \cref{MSM}. 

\subsection{Estimators}
There are many available estimators of $\boldsymbol{\mu}$. One such estimator is G-computation, which uses regressions to sequentially estimate each $q^{\overline{a}}_t$, starting at time $T$.~\citep{Bang,SchnitzerLTMLE} Then, define ${q}^s_t$ to be the vector composed of the stacked quantities $q^{\overline{a}}_t$ of each possible pattern $\overline{a}$. The final step involves regressing the estimate  of the stacked vector ${q}^s_0$ according to the MSM in \cref{MSM}. 
 
Inverse probability of treatment weighting (IPTW) is an alternative approach that involves estimating the functions $g_t(\overline{a}_t,\overline{\boldsymbol{L}}_{t})=P(A_t=a_t\mid \overline{\boldsymbol{L}}_t,\overline{A}_{t-1}=\overline{a}_{t-1})$ for each time-point $t=0,...,T$. One implementation then involves regressing the observed $Y$ on the covariates in Model (\ref{MSM}) using weights equal to estimates of $w_t(\overline{a}_t,\overline{\boldsymbol{L}}_t)= \prod_{k=0}^t  \frac{g_k(a_k,{L}^1_0)}{g_k(\overline{a}_k,\overline{\boldsymbol{L}}_k)}$
 for $t=T$, where $g_t(a_t,L_{0}^1)$ is defined as the stabilizing probability $P(A_t=a_t\mid L_0^1)$. Longitudinal targeted maximum likelihood estimation (LTMLE)~\citep{Petersen2014,SchnitzerLTMLE, vanderLaanLTMLE} is an approach that uses estimates of the functions $q_t^{\overline{a}}$ in addition to the weights $w_t(\overline{a}_t,\overline{\boldsymbol{L}}_t)$ in order to estimate the parameters of the MSM.
 
 A primary question across all methods that use inverse probability of treatment weights (such as IPTW and LTMLE) is how to approach modeling the treatment process to estimate the functions $g_t(\overline{a}_t,\overline{\boldsymbol{L}}_{t})$. The two primary approaches are to model the treatment separately at each time-point or to pool the model over the $T+1$ time-points. The latter is interesting because it allows for model simplification under sparsity while still allowing for greater model complexity with sufficient data support. But without a priori restrictions on the conditioning of the pooled treatment model, it is clear that the number of covariates can become large as the number of time-points increases.  And indeed, incorrect pooling and modeling decisions can lead to bias in the estimation of the MSM parameters.

\section{Variable Selection}\label{Section: LOAL}
In this section, we describe covariate reduction of the treatment model using an implementation of the outcome-adaptive LASSO. 

\subsection{Selection goal}\label{Selection_goal}
We consider parametric logistic regressions for the treatment models. For simplified illustration, we consider two time-points ($T=1$), with the associated data structure $O=(\boldsymbol{L}_0,A_0,\boldsymbol{L}_1,A_1,Y)$. We consider a model for the probability of treatment, stratified on time, written as
 \begin{align}
   \text{logit}\left\{{P}( A_0 = 1 \mid {\boldsymbol{L}}_{0})\right\} &=  \alpha_{0,-1}+ \boldsymbol{\alpha}_{0,0}\boldsymbol{L}_0, \label{strattxmod0}\\
 \text{logit} \left\{{P}( A_1 = 1 \mid \overline{\boldsymbol{L}}_{1}, {A}_{0})\right\} &=  \alpha_{1,-1} +  \boldsymbol{\alpha}_{1,0} \boldsymbol{L}_0 + 
     \boldsymbol{\alpha}_{1,1} \boldsymbol{L}_1 +
   \alpha_{1,2} A_0, \label{strattxmod1}
 \end{align}
where the coefficients may be vectors when the corresponding $\boldsymbol{L}_t$ is multivariate. We can represent the same restriction on the mean as the pooled model
 \begin{align}
    m_t(\overline{\boldsymbol{L}}_t,\overline{A}_{t-1};\boldsymbol\alpha)=& {P}(A_t = 1\mid \overline{A}_{t-1},\overline{\boldsymbol{L}}_t) = \text{logit}^{-1}\left\{\mathbb{I}(t=0) \left(\alpha_{0,-1}+ \boldsymbol{\alpha}_{0,0}\boldsymbol{L}_0 \right) 
    + \mathbb{I}(t=1)  \left(\alpha_{1,-1} +\boldsymbol{\alpha}_{1,0} \boldsymbol{L}_0+\boldsymbol{\alpha}_{1,1} \boldsymbol{L}_1  +  \alpha_{1,2} A_0\right)\right\}, \label{PM}
 \end{align}
 where $t \in \{0, 1\}$ and $\mathbb{I}(\cdot)$ is the indicator function. 

Define $\boldsymbol{\alpha}=(\alpha_{0,-1},...,\alpha_{1,2})$ which are the coefficients of the covariates in the pooled propensity score model in \eqref{PM}. 

Now consider the working regression models
\begin{align}
    &{E}(q_0^{\overline{a}}\mid \boldsymbol{L}_0)=\beta_{0,-1} +  \boldsymbol{\beta}_{0,0} \boldsymbol{L}_0, \label{SM1}\\
    &{E}(q_1^{\overline{a}}\mid \overline{\boldsymbol{L}}_1)=\beta_{1,-1}+ \boldsymbol{\beta}_{1,0} \boldsymbol{L}_0 +  \boldsymbol{\beta}_{1,1} \boldsymbol{L}_1 + \beta_{1,2} a_0, \label{SM2}
\end{align}
with true parameter values minimizing the risk under a squared-error loss function. Note that under the causal assumptions, these correspond to working structural models for $Y^{\overline{a}}$, i.e. Model~(\ref{SM1}) for ${E}(Y^{\overline{a}}\mid \boldsymbol{L}_0)$ and Model~(\ref{SM2}) for ${E}(Y^{\overline{a}}\mid \overline{\boldsymbol{L}}_1)$.
Denote $\boldsymbol{\beta}=(\beta_{0,-1},...,\beta_{1,2})$ and let $\boldsymbol{\beta}^{\dagger}=(\beta^{\dagger}_{0,-1},...,\beta^{\dagger}_{1,2})$ be an indicator vector of the non-zero elements of $\boldsymbol{\beta}$, fixing the intercept and treatment terms as non-zero, i.e. $\beta^{\dagger}_{0,-1}=\beta^{\dagger}_{1,-1}=\beta^{\dagger}_{1,2}=1$. 

We characterize the specific objectives of our variable selection as:
\begin{objective} \label{O1}
For each time-point, select variables into the treatment model at time $t$ that have corresponding non-zero coefficients $\boldsymbol{\beta}$ in the model for $q_t^{\overline{a}}$. We estimate the coefficients of the propensity scores
\begin{align}
    & m_t(\boldsymbol{\overline{L}}_t,\overline{A}_{t-1}; \boldsymbol{\alpha}^{\dagger}), \quad t=0,1, \label{targetTM}
\end{align}
where $\boldsymbol{\alpha}^{\dagger}$, of the same length as $\boldsymbol{\alpha}$, has fixed elements equal to zero corresponding to the zero items in $\boldsymbol\beta^{\dagger}$. The optimal value of $\boldsymbol\alpha^{\dagger}$, denoted as $\boldsymbol\alpha_0^{\dagger}$ (with the same elements fixed at zero), minimizes the risk under the logistic quasi-log-likelihood loss function.
\end{objective}
 
 This specific variable selection criterion can be motivated as removing covariates from the adjustment set $\boldsymbol{\overline{L}}_T$ that are not associated with the potential outcome $Y^{\overline{a}}$ conditional on the remaining variables in $\boldsymbol{\overline{L}}_T$, and the past treatment $\overline{A}_{T-1}$. In particular, these covariates are not relevant for  sequential conditional exchangeability (i.e. are not confounders). However, the criterion retains variables that are conditionally associated with the potential outcome, regardless of whether they are confounders. This is an operationalization of the identification of a covariate subset that, when removed from the adjustment set, leads to a reduction in estimation variance in the nonparametric model. \cite{RotnitzkySmucler,LongEffAdenyo} \Cref{O1} also corresponds with the recommendations in~\cite{Lefebvre} to only adjust for variables that  affect the outcome through pathways that do not include treatment.

 \subsection{Longitudinal Outcome Adaptive Lasso (LOAL)}
In order to write out the estimator, we expand the notation of the possibly multivariate covariates. First, we use $\tau=0,1$ to index the propensity score model for $A_{\tau}$ (i.e. Models~(\ref{strattxmod0}) and (\ref{strattxmod1})). We use $t=0,1$ to index the covariates $\boldsymbol{L}_t$ as before, where $\boldsymbol{L}_0 \in \mathbb{R}^{p_0}$ and $\boldsymbol{L}_1 \in \mathbb{R}^{p_1}$ such that $p_t$ is the number of covariates of $\boldsymbol{L}_t$. The $k^{\text{th}}$ component  of $\boldsymbol{L}_t$ is denoted $L_{t,k}; k=1,\dots,p_{t}$.   Denote the set $\mathcal{J}=\{(0,1,\mathcal{J}_{0,0}),(1, 1,\mathcal{J}_{1,0}), (1, 2, \mathcal{J}_{1,1})\}$ as the 3-dimensional indices of the coefficients $\boldsymbol{\alpha}$ being shrunk, where the set $\mathcal{J}_{\tau,t}$ indexes the specific covariates in $\boldsymbol{L}_t$ being shrunk within propensity score model $\tau$. Note that overlapping indices in $\mathcal{J}_{0,0}$ and $\mathcal{J}_{1,0}$ index coefficients for the same covariates in different models. For example, $(0,0,1)$ and $(1,0,1)$ refer to the coefficients of covariate $L_{0,1}$ in Models~(\ref{strattxmod0}) and (\ref{strattxmod1}), respectively, or the equivalent in Model~(\ref{PM}). Also note that  the intercept coefficients (indices (0,-1) and (1,-1) in Model (\ref{PM})), as well as the coefficients associated with treatment (index (1,2) in Model (\ref{PM})) are not candidates for shrinking and so are excluded from $\mathcal{J}$. The indices in $\mathcal{J}$ are similarly used to refer to the corresponding coefficients $\boldsymbol{\beta}$ in  Models~(\ref{SM1}) and (\ref{SM2}).

Suppose that we have estimates $\hat{\boldsymbol{\beta}}$ of $\boldsymbol{\beta}$ in Models~(\ref{SM1}) and (\ref{SM2}) that are $\sqrt{n}$-consistent where $n$ is the sample size. \textcolor{black}{For instance, we might estimate the functions $q_0^{\bar{a}}$ and $q_1^{\bar{a}}$ with correctly specified parametric models and then regress these quantities onto $L_0$ and $\overline{L}_1$, respectively.} Given a regularization parameter $\lambda_{n} \geq 0$,  an outcome-adaptive LASSO estimator of $\boldsymbol{\alpha}^{\dagger}$ in the pooled Model~(\ref{PM}) as defined in \cref{O1} is given as
\begin{align}
    \hat{\boldsymbol\alpha}(\lambda_n) =\argmin_{\alpha}\sum_{\tau=0}^1\sum_{i=1}^n & \left[ a_{\tau,i} \log\{m_{\tau}(\overline{\boldsymbol{l}}_{\tau,i},\overline{a}_{\tau-1,i}; \boldsymbol\alpha)\} 
    +(1-a_{\tau,i})\log\{1-m_{\tau}(\overline{\boldsymbol{l}}_{\tau,i},\overline{a}_{\tau-1,i};\boldsymbol\alpha)\}\right] + \lambda_n\sum_{j\in \mathcal{J}} \hat{\omega}_j \lvert\alpha_j\rvert, \notag
\end{align}
where $\hat{\omega}_j=\lvert \hat{\beta}_j \rvert^{-\gamma}$ for all $j\in \mathcal{J}$, with tuning parameter $\gamma>1$.  By the results in, \cite{SE} this estimator is asymptotically normal and consistent for the selection of covariates in the Model~(\ref{targetTM}) if we assume that $\lambda_n/\sqrt{n} \rightarrow 0$ and $\lambda_n n^{\gamma/2-1}\rightarrow \infty$. Note that $\gamma>2$ is needed for the second convergence requirement.

For implementation purposes, this regularized regression can be run using a transformation of the pooled data, setting $V_{0,-1}=1-\mathbb{I}(t=1)$, $\boldsymbol{V}_{0,0}=\boldsymbol{L}_0-\mathbb{I}(t=1)\boldsymbol{L}_0$, $V_{1,-1}=\mathbb{I}(t=1)$, $\boldsymbol{V}_{1,0}=\mathbb{I}(t=1)\boldsymbol{L}_0$, $\boldsymbol{V}_{1,1}=\mathbb{I}(t=1)\boldsymbol{L}_1$, and $V_{1,2}=\mathbb{I}(t=1)A_0$ with respectively corresponding coefficients $\alpha_{0,-1},...,\alpha_{1,2}$ in Model~(\ref{PM}). Then, the adaptive LASSO is run with pooled outcome $A_t$ on covariates $V_{0,-1},...,V_{1,2}$, without an intercept term, using weights $\hat{\omega}_j=\lvert \hat{\beta}_j \rvert^{-\gamma}; \forall j\in \mathcal{J}$.

\subsection{\texorpdfstring{Estimation of $q^{\overline{a}}_t$ to estimate $\boldsymbol\beta$}{Estimation of q to estimate beta}}\label{Subsection: Estimation of beta}
The proposed variable selection for the propensity scores is based on the estimated $\beta$ parameters in Models~(\ref{SM1}) and~(\ref{SM2}), which need to be estimated at $\sqrt{n}$ rates. However, this requires preliminary estimates of $q_1^{\overline{a}}$ and $q_0^{\overline{a}}$. To get these, we could first use a flexible regression method to estimate $q_1^{\overline{a}}$ by regressing $Y$ on $\overline{\boldsymbol{L}}_1$ and $\overline{A}_1$. We then generate predictions from this model for each pattern of interest $\overline{a}_1=(a_0,a_1)$. 
We then run a pooled regression of the stacked vector ${q}^s_1$ on the covariates  $\overline{\boldsymbol{L}}_1$ and $a_0$ where $a_0$ takes the value zero or one depending on the pattern $\overline{a}_1$, corresponding to the working structural Model~(\ref{SM2}). This results in estimates of the coefficients in that model, denoted by $\hat{\beta}_{1,-1},...,\hat{\beta}_{1,2}$. 

For each pattern $\overline{a}\in\mathcal{A}$ where $\mathcal{A}$ is the set of all possible patterns, use a flexible regression method to regress $q_1^{\overline{a}}$ on $\boldsymbol{L}_0$ and $A_0$. We then use this model to make predictions setting $A_0=a_0$ to obtain $q_0^{\overline{a}}=\E(q_1^{\overline{a}}\mid \boldsymbol{L}_0,a_0)$.
We then run a pooled regression of the stacked vector $q^s_0$ on the covariate $\boldsymbol{L}_0$ according to the structural Model~(\ref{SM1}) to obtain estimates of $\beta_{0,-1}$ and $\beta_{0,0}$, which we will denote $\hat{\beta}_{0,-1}$ and $\hat{\beta}_{0,0}$, respectively. \textcolor{black}{For illustration, each estimation step is presented in the Figure \ref{fig: estimate beta}.}


\begin{figure}
    \centering
  \begin{tikzpicture}\color{black}
     \tikzset{state/.style={circle, draw, minimum size=0.8cm, inner sep=0}, double state/.style={circle, draw, double, minimum size=0.8cm, inner sep=0, line width=1pt}, double arrow/.style={double, line width=0.5pt, double distance=1pt} }
\node (foreach) at (-3.7,10) [align=left]{\textbf{For each $\bar{a}=(a_0, a_1)$},};
\node (yregress) at (-4,9) [align=left]{{$Y\sim \bar{L}_1+ \bar{A}_1$}};
\node (setA1) at (-3.6,8) [align=left]{{Set $\bar{A}_1=\bar{a}_1$}};
\node (predQ1) at (-1,7) [align=left]{{Predict $q_1^{\bar{a}}$}};
\node (q1regress) at (2.5, 9) [align=left]{\textbf{$q_1^{\bar{a}}\sim L_0+ A_0$}};
\node (setA0) at (3.2, 8) [align=left]{{Set $A_0=a_0$}};
\node (predQ0) at (6,7) [align=left]{{Predict $q_0^{\bar{a}}$}};

\node (stackQ1) at (-1,5.5) [align=left]{{Stack $q_1^{s}=\{q_1^{\bar{a}}; \bar{a}\in\mathcal{A}\}$}};
\node (regstackQ1) at (-1,5) [align=left]{{Regress $q_1^{s}\sim \bar{L}_1 +\bar{a}_1$}};
\node (estbeta1) at (-1,4.5) [align=left]{{Obtain $\hat{\beta}_{1,-1}, \hat{\beta}_{1,0}, \hat{\beta}_{1,1}, \hat{\beta}_{1,2}$}};
\node (stackQ0) at (6,5.5) [align=left]{{Stack $q_0^{s}=\{q_0^{\bar{a}}; \bar{a}\in\mathcal{A}\}$}};
\node (regstackQ0) at (6,5) [align=left]{{Regress $q_0^{s}\sim L_0 +a_0$}};
\node (estbeta0) at (6,4.5) [align=left]{{Obtain $\hat{\beta}_{0,-1}, \hat{\beta}_{0,0}$}};

\draw[>=stealth, ->, line width=0.5pt, scale=1] (yregress) -- (predQ1);
\draw[>=stealth, ->, line width=0.5pt, scale=1] (q1regress) -- (predQ0);
\draw[>=stealth, ->, line width=0.5pt, scale=1] (predQ1) -- (q1regress);
\draw[>=stealth, ->, line width=0.5pt, scale=1] (predQ1) -- (stackQ1);
\draw[>=stealth, ->, line width=0.5pt, scale=1] (predQ0) -- (stackQ0);

\node [draw, dashed, thick, inner sep=0.2cm, fit=(yregress) (setA1) (predQ1) (q1regress) (setA0) (predQ0) ] {};
\node [draw, thick, inner sep=0.1cm, fit=(stackQ1) (regstackQ1) (estbeta1) ] {};
\node [draw, thick, inner sep=0.1cm, fit=(stackQ0) (regstackQ0) (estbeta0)] {};
\end{tikzpicture}
\caption{\textcolor{black}{Diagram illustrating the steps required to estimate $q_t^{\bar{a}}$ and $\boldsymbol\beta$, as described in Section~\ref{Subsection: Estimation of beta}.}}
\label{fig: estimate beta}
\end{figure}

\subsection{Selection of the tuning parameters}\label{Selection of the tuning parameters}

To select values for the two tuning parameters of the LOAL, we first fix the value of $\gamma$ to a value slightly larger than 2 (we used 2.5 in the simulation and application) in order to ensure the required divergence of $\lambda_nn^{\gamma/2-1}$. \textcolor{black}{We allow for a wide range of candidate $\lambda_n$ values such that the largest value shrinks all coefficients to zero.} We propose to select $\lambda_n$ using a one-dimensional extension of the weighted absolute mean difference proposed in Shortreed and Ertefaie, \cite{SE} corresponding to a summary of a longitudinal balancing metric  over covariates and times. See Web Appendix A for details. \textcolor{black}{Note that in their simulation study, Shortreed and Ertefaie~\cite{SE} selected $\gamma$ such that $\lambda_nn^{\gamma/2-1}=n^2$ and set the range of values $\lambda_n=(n^\aleph)$ with $\aleph<0.5$ to ensure both convergence requirements. In either approach, $\gamma>2$ and the balance criterion used to select $\lambda_n$ produces a tradeoff between covariate balance and model fit.}

\section{Selective Fusion}\label{Section: LOAL Fusion}

In this section, we describe a second approach to dimension reduction which adaptively pools related coefficients across the time-point-specific treatment models using the fused LASSO.

 \subsection{Fusion goal}

The Model~(\ref{targetTM}) may not be sufficiently parsimonious in the following situation: suppose that, for some $k$, $\beta_{0,0,k}$ and $\beta_{1,0,k}$, the coefficients of covariate $L_{0,k}$ in the two structural models, are both large. This will mean that little penalty will be placed on the coefficients $\alpha_{0,0,k}$ and $\alpha_{1,0,k}$ in the LOAL procedure. In the situation where, for some $k\in \mathcal{J}_{0,0} \cap \mathcal{J}_{1,0}$, there is little difference in log-odds between ${L}_{0,k}$ and $A_1$ relative to ${L}_{0,k}$ and $A_0$, conditional on other terms in the model, we want the LOAL to fuse the terms $\alpha_{0,0,k}$ and $\alpha_{1,0,k}$. That is, we want to set $\alpha_{0,0,k}=\alpha_{1,0,k}$ or equivalently, have a single time-independent coefficient for ${L}_{0,k}$. 
This has the effect of smoothing over time and is the finite-sample \cref{O2}.

\begin{objective} \label{O2}
In finite samples,  fuse coefficients for common covariates across treatment models at different time points if it improves the pooled treatment model fit.
\end{objective}
By reducing the number of degrees of freedom, and avoiding potential overfitting of the propensity score models, we expect that this objective will lead to more efficient estimation of $\boldsymbol{\mu}$ in finite samples and avoid non-data-driven smoothing decisions under data sparsity.

\bmsubsection{Estimation}\label{Estimation}

In order to achieve \cref{O2}, we first obtain the estimates $\hat{\boldsymbol{\alpha}}^{\text{refit}}(\lambda_n)$ from the LOAL and a refitted logistic regression, and define $\boldsymbol\alpha^{*}$ as the parameter vector of the same length as $\boldsymbol\alpha$ that is set to zero at the indices of the zero-elements of $\hat{\boldsymbol{\alpha}}^{\text{refit}}(\lambda_n)$. Then we use a generalized Adaptive Fused LASSO ~\citep{Viallon_StatsComp}
\begin{align*}
    \argmin_{\boldsymbol{\alpha}^{*}}\sum_{\tau=0}^1&\sum_{i=1}^n \left[ a_{\tau,i} \log\{m_{\tau}(\overline{l}_{1,i},a_{0,i};\boldsymbol\alpha^{*})\} +(1-a_{\tau,i})\log\{1-m_{\tau}(\overline{l}_{1,i},a_{0,i};\boldsymbol\alpha^{*})\}\right]
+ \lambda_{1,n}\sum_{k\in \mathcal{J}^*_{0,0}\cap \mathcal{J}^*_{1,0},}\frac{\lvert \alpha^{*}_{1,0,k}-\alpha^{*}_{0,0,k} \rvert 
}{{\lvert \hat{\alpha}_{1,0,k}^{\text{refit}}(\lambda_n)-\hat{\alpha}_{0,0,k}^{\text{refit}}(\lambda_n) \rvert}^{\gamma_1}},
 \end{align*}
with $\gamma_1>0$ and where $\mathcal{J}^*_{0,0}\subset \mathcal{J}_{0,0}$ and $\mathcal{J}^*_{1,0}\subset \mathcal{J}_{1,0}$ represent the  indices of the selected covariates at $\tau=0$ and 1, respectively. We propose to select the tuning parameters $\gamma_1$ and $\lambda_{1,n}$ by Bayesian information criterion (BIC). \citep{Viallon_StatsComp} We omit a sparsity-inducing penalty for two reasons: our variable selection was performed in the separate first step, and the variable selection and fusion objectives have different statistical goals (covariate balance vs. model selection, respectively). 

It is important to note that, due to non-collapsibility and collinearity, the values of the coefficients in the pooled treatment models depend on the other covariates in the model.  So, we first need to identify the covariate set before being able to statistically determine whether or not two coefficients should be fused. 
Thus, our application of the adaptive fused LASSO with a logistic regression model involves a purposeful misspecification of the pooled treatment model where we have already potentially marginalized over covariates in the previous step. We define a graph over the remaining covariates indicating which we allow to fuse. 
 The oracle results of Viallon et al. (2016)~\cite{Viallon_StatsComp} apply relative to the model marginalized over the covariates removed in the first step, assuming that the treatment follows a Bernoulli distribution with a mean that is logit-linear in the remaining covariates. We also need that $\hat{\boldsymbol{\alpha}}^{\text{refit}}(\lambda_n)$ converges to $\boldsymbol \alpha^{\dagger}$ at a $\sqrt{n}$-rate, which is supported by the LOAL theory.  Our application of the adaptive fused LASSO does not include a sparsity component in the objective function (i.e. does not include a LASSO penalty for the sizes of the coefficients). Through the same arguments, the oracle results then hold for the fusion of equal parameters $\alpha_k$ that are connected in the graph. A formal statement is given in Web Appendix B.


\bmsubsection{\texorpdfstring{Fusion with $T>1$}{Fusion with T>1}}

This procedure can be expanded when the number of time-points is greater than 2. The Fused LASSO requires the user to define a graph indicating which coefficients are allowed to fuse.~\citep{Viallon_StatsComp} This graph should represent the maximum smoothing of the model through data pooling, corresponding perhaps to how propensity score models are typically pooled. For baseline covariates, a ``clique graph'' may be used where we allow the coefficients of common  variables to fuse between times $\tau=0,...,T$. Alternatively, we may use a ``chain graph'' where each chain connects the coefficient of the same baseline variable at successive time points. For common time-updated covariates, one may fuse coefficients of variables with the same lag relative to the treatment time $\tau$ across times $\tau$, for which we could use a chain graph for successive fusing or a clique graph for fusing between any two time-points.   We illustrate the usage of clique graphs with lagged time-dependent variables in the application.

\section{Simulations}


In this simulation study, we estimated the coefficients in the MSM (\ref{MSM}) where $L_0^1$ was always defined as the first confounder in the dataset. We applied LOAL \textcolor{black}{as described in Section~\ref{Section: LOAL}} and the two-step fused LOAL \textcolor{black}{(LOAL followed by the fusion step described in Section~\ref{Section: LOAL Fusion})} to estimate the three target parameters in the MSM~(\ref{MSM}) \textcolor{black}{using IPTW}. For benchmarks, we also ran the sequential G-computation~\citep{Bang} with all covariate main terms, IPTW with treatment models stratified by time-point and including all covariate main terms (``full IPTW''), and IPTW with pooled treatment models excluding all unwanted covariates (``IPTW oracle select'') and further with correctly fused coefficients for common terms (``IPTW oracle select and fuse''). For fair comparisons, the specification of the models for estimating each $q_t$ was common across methods that used these quantities.

The LOAL was implemented using adaptive weights in \texttt{glmnet}.~\cite{friedman2010regularization} We set $\gamma=2.5$ (which allows for the convergence of $\lambda_nn^{\gamma/2-1}$) and a very broad range for candidate $\lambda$ values and then selected the optimal $\lambda_n$ value according to the balance criterion (Web Appendix A). We performed the fusion step using the archived \texttt{FusedLasso} package~\cite{tibshirani2011solution} which implements the coordinate-wise optimization algorithm of H\"ofling, Binder, and Schumacher (2010),~\cite{Hoefling} implementing adaptive weights for fusion \cite{Viallon_StatsComp} but setting the adaptive weights for the main terms to zero so that no additional sparsity would be induced. We set $\gamma_1=2.5$ and searched over a very broad range for $\lambda_{1}$, selecting the optimal $\lambda_{1,n}$ using the BIC.

\begin{figure}[ht]
\centering
\includegraphics[width=0.48\textwidth]{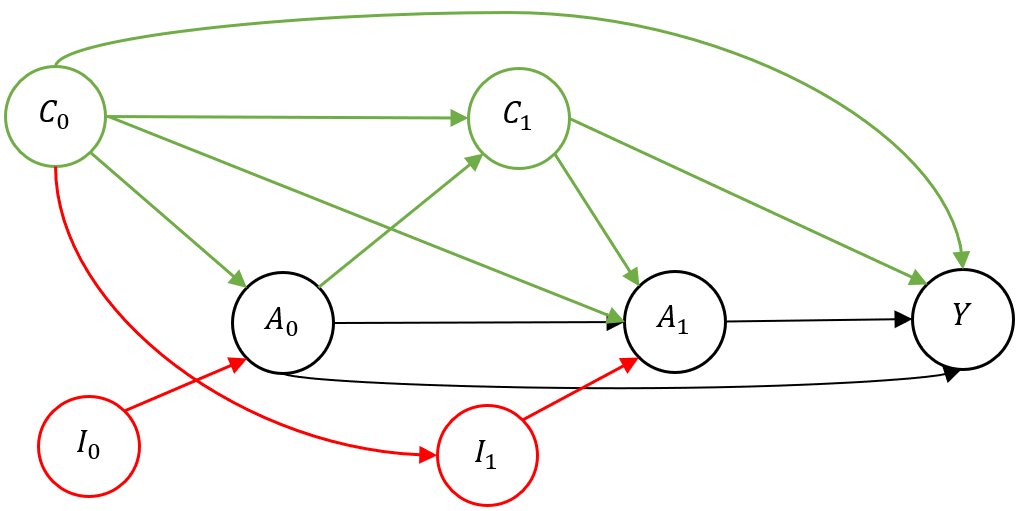} 
\caption{DAG representing the data generation in Scenarios 1. The target variable selection retained all covariates labeled $C$. \label{DAGs}}
\end{figure}

\bmsubsection{\textcolor{black}{Evaluation in a simple scenario}}\label{Scenario 1: low dimensional with two time-points}
In the simple scenario \textcolor{black}{(Scenario 1)}, we generated i.i.d. data $O=(C_0,I_0,A_0,C_1,I_1,A_1, Y)$ according to the left DAG in Figure~\ref{DAGs}, where $C_0$, and $I_0$ were independently generated from a standard normal distribution and $A_0$ and $A_1$ were Bernoulli distributed. Variables $C_1$ and $I_1$ were Gaussian-distributed with means $(A_0+C_0)$ and $(C_0)$, respectively. The instruments $I_0$ and $I_1$ only affected the treatment probabilities. Notably, $A_0$ was equal to one with probability $\logit^{-1}(1.515C_0+I_0)$ and $A_1$ with probability $\logit^{-1}(-0.5+0.5C_0+0.25C_1+0.5A_0+I_1)$. This made it so that in the marginal treatment models without instruments, the coefficients of $C_0$ in each model were both equal to 1.28. No other coefficients were equal. All four covariates were standardized to zero mean and unit standard deviation.

In order to evaluate the robustness of our method to misspecification of the working structural models, the Gaussian-distributed outcome $Y$ had a mean specified in three ways: a) as a function of the main terms $C_0$, $A_0$, $C_1$, and $A_1$, b) the same as (a) but with an added interaction between $C_0$ and $C_1$ (covariate interaction) and c) the same as (a) but with an added interaction between $A_0$ and $C_1^2$ (effect modification). We specified the linear model for $q_1$ as containing only the main terms of all covariates; the linear model for $q_0$ also contained interactions between $C_0$ and $A_0$ and a squared term for $I_0$. The complete data generating mechanism is given in Web Appendix C.1.

We varied the sample sizes from $n=200,500,1000$. Table~\ref{Scen1abc} reports $\sqrt{n}$ times the absolute bias and $n$ times Monte Carlo mean squared error (MSE) over 1,000 draws for each estimator for these three outcome generating models. The variable selection results for the proposed estimators are given in Table~\ref{Scen1abc_varselect}.
 In Scenario 1(a) where the models used for the outcome process were correctly specified, the G-computation estimator was unbiased with the lowest standard errors. All IPTW estimators were unbiased since they adjusted for all confounders; the oracle IPTWs had lower MSE than the IPTW adjusting for all covariates (``full IPTW'') though there was no difference between the two oracles.  LOAL performed as well as the oracles, as did the fused LOAL. In Scenario 1(b), all outcome models were misspecified and G-Computation was the most biased. The full and oracle IPTW estimators were consistent but held some bias at these sample sizes; the oracle IPTW estimators had similar MSEs that were smaller than the full IPTW. LOAL and fused LOAL had lower MSE than the oracle IPTWs and comparable bias. In Scenario 1(c), the G-computation was highly biased but the full and oracle IPTW estimators were consistent with little bias. The oracles had the lowest MSE. The LOAL and fused LOAL were more biased than the oracle IPTWs and had higher MSE, but were still less biased than the G-computation and had lower MSE than the full IPTW.

\begin{sidewaystable} 
\centering
\caption{Scenario 1:  $\sqrt{n}$ times the absolute value of bias ($n$ times mean squared error) of methods estimating the parameters in the MSM of equation~\ref{MSM}. IPTW oracle fits a glm of the target propensity score model with correctly selected variables and fused coefficients. The Fused LOAL uses the estimates of LOAL for the adaptive weights. \label{Scen1abc} }
\resizebox{1.2\width}{!}{%
\begin{tabular}{lccccccccc}
\toprule
Method\textbackslash{}Scenario & \multicolumn{3}{c}{\begin{tabular}[c]{@{}l@{}}a) Outcome model with\\ main terms\end{tabular}}   & \multicolumn{3}{c}{\begin{tabular}[c]{@{}l@{}}b) Outcome model with\\ covariate interaction\end{tabular}} & \multicolumn{3}{c}{\begin{tabular}[c]{@{}l@{}}c) Outcome model with\\ effect modification\end{tabular}} \\
                               & $\mu_0$ & $\mu_1$ & $\mu_2$ &                   $\mu_0$ & $\mu_1$ & $\mu_2$                                &           $\mu_0$ & $\mu_1$ & $\mu_2$                                           \\ 
True values   & -1.5  & 1.5   & 1.25  & 1 &  2.75  &   1.25  &         -1.5 &   4   &  5  \\ \midrule
\textbf{n=200}       &     &     &    &     &     &      &      &     &  \\
G-comp main terms  & 0.0(37)&0.1(24)&0.0(21)&7.9(183)&0.2(185)&0.9(110)&29.4(461)&20.2(362)&12.0(232)\\
IPTW full main terms  &0.3(55)&0.4(43)&0.2(40)&5.7(224)&0.1(261)&0.2(145)&1.9(256)&1.8(330)&3.4(188)\\
IPTW oracle select &0.2(46)&0.3(36)&0.2(33)&4.7(202)&0.2(229)&0.0(135)&1.0(186)&1.5(259)&2.7(152)\\
IPTW oracle select and fuse&0.1(46)&0.4(36)&0.1(33)&4.7(202)&0.1(231)&0.1(135)&0.9(185)&0.9(266)&2.7(153)\\
LOAL &0.2(46)&0.5(37)&0.2(34)&6.0(190)&0.2(209)&0.5(123)&4.9(228)&5.6(284)&5.7(166)\\
Fused LOAL  &0.1(46)&0.6(36)&0.2(34)&6.0(191)&0.2(213)&0.5(123)&4.9(228)&5.5(285)&5.6(166)\\ \midrule
\textbf{n=500  }  &  &   &  &   &   &  &   & &  \\
G-comp main terms  &0.1(55)&0(39)&0.1(33)&11.8(342)&0.8(292)&2.1(183)&45.5(1061)&32.1(798)&18.1(473) \\
IPTW full main terms  &0.5(97)&0.4(82)&0.4(76)&6.1(478)&0.1(563)&0.3(303)&3.0(516)&4.2(673)&3.9(354)\\
IPTW oracle select   &0.1(77)&0.2(67)&0.1(60)&5.3(395)&1.1(464)&0.3(280)&1.7(352)&3.6(503)&3.3(283) \\
IPTW oracle select and fuse&0.1(78)&0.2(66)&0.1(61)&5.3(392)&0.9(459)&0.3(279)&1.7(342)&3.2(495)&3.3(281)\\
LOAL &0.1(80)&0.4(63)&0.1(63) &7.2(375)&0.5(432)&0.3(255)&6.9(381)&7.0(537)&7.5(304)\\
Fused LOAL  &0.0(81)&0.4(63)&-0.1(64)&7.3(372)&0.4(424)&0.3(253)&6.9(382)&6.8(542)&7.4(304)\\ 
\midrule
\textbf{n=1000 }     &  &   &  &  &  &  & &  & \\
G-comp main terms  &0.1(81)&0.0(53)&0.0(48)&16.6(606)&1.5(411)&2.9(252)&63.9(2063)&43.4(1459)&25.5(871) \\
IPTW full main terms  &0.3(148)&0.3(121)&0.2(117)&5.3(741)&0.0(1042)&0.2(517)& 3.2(753)&2.6(1081)&4.7(546)\\
IPTW oracle select  &0.1(115)&0.4(95)&0.1(87)&5.3(613)&0.9(778)&0.8(439)&1.1(524)&2.1(793)&3.7(420) \\
IPTW oracle select and fuse &0.0(116)&0.4(94)&0.1(88)&5.2(612)&0.9(784)&0.8(439)&1.1(524)&1.7(810)&3.6(424)\\
LOAL &0.1(117)&0.4(94)&0.1(91)&6.7(571)&0.2(726)&0.2(394)&8.1(590)&11.1(930)&8.9(479)\\
Fused LOAL  &0.1(118)&0.4(94)&0.0(92)&6.8(568)&0.2(727)&0.2(394)&8.2(592)&10.8(936)&8.8(481)\\ 
\bottomrule \\
\end{tabular}}
\end{sidewaystable}

In Table~\ref{Scen1abc_varselect}, we give the proportion of each selected variable and fused coefficient for each of the proposed methods. At the top of the table, we see that, when the oracle estimates of ${\alpha}$ (i.e. estimated with correctly selected propensity score models) are used to weight the fused LASSO, the $L_0$ coefficients correctly fused 99\% of the time at all sample sizes. In Scenario 1(a), LOAL correctly selected the covariates with non-zero $\beta$ coefficients in the working structural models  $79-100\%$ of the time  for all sample sizes, with greater true positive rates for  larger $\beta$ values. The method also correctly omitted the instruments with almost no false positives by $n=500$. The success of fusion notably depended on the success of the variable selection, so that the $L_0$ covariates fused when the true model was selected in the first phase. In Scenarios 1(b) and 1(c) where the models to estimate $q_t$ were misspecified, the convergence of the covariate selection was slower. In Scenario 1(c), the $\hat{\beta}$ did not converge to zero for the instruments, making it so that de-selection of instruments was not consistent (though the selection of confounders was consistent, but with slow convergence). Since fused LASSO relies on the correct covariate selection, the proportion of fusion was lower than for the other scenarios.

\begin{table}\center
\caption{Scenario 1: Proportion selection of each covariate into each treatment model and  fusion of the coefficients of common terms across the two models, out of 1000 runs. $C_0$ and $C_1$ are true confounders and $I_0$ and $I_1$ are both instruments. The coefficients for $C_0$ are common across the two models under the target adjustment set. The true $\beta$s are the coefficients in the working models corresponding to equations~(\ref{SM1}) and~(\ref{SM2}). The $\lim_{n\rightarrow\infty}\hat{\beta}$ are the converging values of the estimates of $\beta$ which depend on the specification of the $q$ models. Fused LOAL was implemented in two ways: first, taking initial values of $\hat{\alpha}$ as estimated in a variable selection oracle model and secondly taking initial values from the LOAL (the former procedure being independent of the outcome).  \label{Scen1abc_varselect}}
\resizebox{1.2\width}{!}{%
\begin{tabular}{llllllllll}
\toprule
$n$&Method&\multicolumn{6}{l}{Selected covariates}&\multicolumn{2}{l}{Fused non-zero}\\
   &                  & \multicolumn{2}{l}{$A_0$ model}         & \multicolumn{4}{l}{$A_1$ model} & \multicolumn{2}{l}{coefficients} \\
   &                  & $C_0$ & $I_0$       & $C_0$  & $I_0$  & $C_1$ & $I_1$ &  $C_0$     & $I_0$     \\
                     \midrule
 \multicolumn{10}{l}{\emph{Fusion results with oracle $\hat{\alpha}$ (independent of scenario)}}\\
  \textbf{200}& Fused LASSO&1.00&0.00&1.00&0.00&1.00&0.00&0.99&0.00\\
\textbf{500}& Fused LASSO& 1.00&0.00&1.00&0.00&1.00&0.00&0.99&0.00\\
\textbf{1000} &Fused LASSO & 1.00&0.00&1.00&0.00&1.00&0.00&0.99&0.00\\ 
\midrule
\multicolumn{10}{l}{\emph{a) Outcome generating model with main terms}}\\
   \textbf{200}& Fused LOAL & 1.00&0.12&0.79&0.00&0.94&0.03&0.77&0.00\\ 
 \textbf{500}& Fused LOAL & 1.00&0.03&0.90&0.00&0.99&0.00&0.89&0.00\\ 
\textbf{1000}& Fused LOAL &1.00&0.01&0.95&0.00&1.00&0.00&0.94&0.00\\ \\
 
 &\textbf{True $\beta$s} & 1.50 &0.00 &0.50&0.00 &1.41 &0.00& \\
   &\textbf{$\lim_{n\rightarrow\infty}\hat{\beta}$}&1.50&0.00&0.50&0.00&1.65&0.00\\
 \midrule
 \multicolumn{10}{l}{\emph{b) Outcome generating model with covariate interaction}}\\
     \textbf{200}& Fused LOAL &0.96&0.09&0.79&0.03&0.56&0.17&0.75&0.00\\
   \textbf{500}&  Fused LOAL &1.00&0.08&0.93&0.02&0.71&0.12&0.84&0.00\\
 \textbf{1000}&  Fused LOAL &1.00&0.06&0.98&0.01&0.84&0.14&0.83&0.00\\\\
 & \textbf{True $\beta$s} &2.75&0.00&1.75&0.00&1.41&0.00  \\
   & \textbf{$\lim_{n\rightarrow\infty}\hat{\beta}$}&2.70&0.00&1.73&0.00&1.64&-0.04\\
   \midrule
 \multicolumn{10}{l}{\emph{c) Outcome generating model with effect modification}}\\
    \textbf{200}& Fused LOAL &1.00&0.16&0.24&0.03&1.00&0.21&0.22&0.00\\
    \textbf{500}& Fused LOAL &1.00&0.13&0.32&0.02&1.00&0.17&0.28&0.00\\
     \textbf{1000}& Fused LOAL &1.00&0.11&0.43&0.02&1.00&0.17&0.34&0.00\\  
    \textbf{5000}& Fused LOAL &1.00&0.06&0.78&0.01&1.00&0.16&0.60&0.00\\ \\
   &   \textbf{True $\beta$s} &4.00&0.00&0.50&0.00&4.95&0.00\\
 &        \textbf{$\lim_{n\rightarrow\infty}\hat{\beta}$}&5.40 &  0.04 &0.61&  0.05 & 7.82&  0.09 \\
\bottomrule \\
\end{tabular}}
\end{table} 

\bmsubsection{\textcolor{black}{Higher-dimensions, more time-points, and varied instrument strengths, comparisons across estimators, and inference}}\label{Additional simulation scenarios and results}
We additionally ran a higher dimensional scenario \textcolor{black}{(Scenario 2)} which demonstrated the good performance of the LOAL and Fused LOAL in terms of estimation, variable selection, and fusion, with 30 covariates and two time points. We ran a third scenario with five time-points, demonstrating primarily that the smoothing by Fused LOAL approaches the oracle estimation and can positively impact estimation variance. Finally, for the first two scenarios, we compared the performance of LTMLE, implemented with and without LOAL, \textcolor{black}{LTMLE with covariate screening and stacking (we used the superlearner \texttt{R} package including main terms logistic regressions, logistic regressions with main terms and interactions, and variable screeners, with all screening applied separately to the treatment and outcome models, respectively),} and C-LTMLE (a comparator variable selection method). While C-LTMLE was slightly better than LOAL in terms of MSE when the outcome models were correctly specified, LTMLE with LOAL performed better when they were not. A major benefit of LOAL over C-LTMLE is that it dramatically lowers computational complexity. \textcolor{black}{LTMLE with variable screening unsurprisingly did not perform as well as the causal variable selection methods because variable screening applied directly to the treatment models will prioritize the selection of variables correlated with treatment, including instruments.} See Web Appendix C.2-4 for details. \textcolor{black}{In addition, to further investigate the potential for inference of the LOAL, we applied the m-out-of-n bootstrap to Scenarios 1(a) and 2. This bootstrap method is designed to improve bootstrap inference in settings where the standard bootstrap fails, including in covariate selection settings. \cite{bickel1997mboot, politis1994large, bickel2008choice} However, we observed undercoverage of the bootstrap-based confidence intervals in our setting. Full details are provided in Web Appendix Section C.5. Finally, we compared the performance of LOAL versus the full model IPTW while firstly varying the strength of one instrument and secondly shifting all the propensity scores from lower to higher values. LOAL had far superior performance in terms of $n$ times MSE across all scenarios.
}

\section{Example: The Nicotine Dependence in Teens (NDIT) study}\label{Example: The Nicotine Dependence in Teens (NDIT) study}

We now illustrate the application of our proposed methodology using data from the Nicotine Dependence in Teens (NDIT) study to estimate the effect of the timing of alcohol initiation during adolescence on depressive symptoms in early adulthood. The NDIT study is a prospective longitudinal study initiated in 1999-2000, comprising 1,294 grade seven students recruited from 10 high schools in Montréal, Canada. \citep{Loughlin2015} Self-report questionnaires were administered at three-month intervals, resulting in a total of 20 cycles from 1999 to 2005. An additional post-high school survey was conducted in 2007 or 2008. Data were collected from repeated assessments of a wide range of sociodemographic, substance use, psychosocial, lifestyle, and physical and mental health variables. We consider data from 1,231 students who were in grade seven in September 1999 and who were not previous regular (at least weekly use) alcohol users. 

The baseline variables (cycle 1) included in our analysis were reported sex (female vs. male), mother's education (less than university vs. at least some university), single-parent home, French spoken at home, country of birth (outside Canada vs. Canada), self-esteem, impulsivity, and novelty-seeking. The time-varying covariates $\boldsymbol{L}$  considered were current depressive symptoms
, participation in team sports, family-related stress, other type of stress, worry about weight, and ever smoked. The exposure $A_t$ was the indicator of initiation of regular alcohol use at or before cycle $t$. 
Note that if $A_t = 1$ at a given time, then $A_{k} = 1$ at all times $k > t$ by our definition. We considered data from cycles 1 to 5, spanning calendar years 1999 to 2000, for the time-varying covariates and exposure. The outcome $Y$ was depressive symptoms experienced within the past two weeks as measured using the Major Depressive Inventory (MDI) in 2007 or 2008 \citep{Bech} when participants were age 20.4 years on average (i.e., approximately two years after cycle 20). The outcome is a continuous score ranging from 0 to 50, with higher scores indicating more severe symptoms. Since not all participants initially recruited took part in all cycles of the study, we denote loss to follow-up (i.e. censoring) by cycle $t$ as $C_t=1$, and $C_t = 0$ otherwise. The observed data structure is written as $O = (\boldsymbol{L}_1, A_1, \boldsymbol{L}_2, C_2, A_2,\cdots, A_5, \boldsymbol{L}_{6}, C_{6},  Y).$ Note that $\boldsymbol{L}_1$ contains the baseline covariates in addition to the time-varying covariates at the first time. 

We denote an arbitrary exposure pattern as $\overline{\boldsymbol{a}}=(a_1, \cdots, a_5)$. Define $\mathcal{D}$ as the treatment regimen space, corresponding to the 6 possible treatment patterns: initiation  at time 1, 2, 3, 4, or 5, or no initiation at any time point. For example, initiation at time 2 is represented as $(0,1,1,1,1)$. The parameters of interest were defined through the working MSM  
 \begin{align}
     &\mathbb{E}[Y^{\overline{\boldsymbol{a}}}|\text{Sex}]=\mu_0+\mu_{1}\text{Sex}+ \mu_2 cum(\overline{\boldsymbol{a}})+{\mu}_{3}\{\text{Sex}\times cum(\overline{\boldsymbol{a}})\},\label{ParameterITT}
\end{align}
where $cum(\overline{a})$ gives the number of exposed time points in $\overline{a}$. 

We extended out methods to incorporate the simultaneous presence of time-dependent treatment and censoring. We considered \textcolor{black}{the following} implementations of IPTW and LTMLE:
\begin{itemize}
    \item IPTW full: Fit stratified models for the probability of being exposed (and censored, respectively) at each given time according to all previous covariates' main terms among participants who were previously unexposed and uncensored.
    \item IPTW LOAL and IPTW fused LOAL: Included selected variables in the treatment and censoring models using the LOAL and fused LOAL procedures, respectively. 
    \item LTMLE full: Included all covariates in the stratified treatment and censoring models. 
    \item LTMLE LOAL and LTMLE fused LOAL: Included only the covariates selected using the LOAL and fused LOAL procedures, respectively, in the treatment and censoring models.
    \item \textcolor{black}{LTMLE SL: LTMLE with superlearner screening for stratified treatment, censoring and outcome models. The library contains ``SL.mean'', ``SL.glm'', ``SL.gam'',  ``SL.gam, screen.randomForest'', ``SL.glm.interaction'', ``SL.glm.interaction, screen.randomForest'', ``SL.earth'', ``SL.earth, screen.randomForest''.}
\end{itemize}
For all implementations, the outcome models included the main terms of the baseline and time-varying covariates, current and lagged exposure terms, and the first-order interactions of sex and exposures for uncensored participants.

For LOAL and fused LOAL, we performed variable selection and fusion for the treatment model and censoring model separately, but the penalization parameters $\lambda_n^a$ (for treatment) and $\lambda_n^c$ (for censoring) were selected jointly by minimizing the sum of two longitudinal balancing metrics over covariates and times with respect to treatment and censoring at each time point. For fused LOAL, the penalty graph connected common baseline variables across time points, as well as common time-varying variables with the same lag across time points (e.g., corresponding $\boldsymbol{L}_{t-1}$ variables are connected together when modeling $A_t$, and when modeling $C_t$). 

 
We fixed the tuning parameter $\gamma$ at 2.5 and used 20 candidate values for the tuning parameters $\lambda^a$ with the range $(e^{-4}, e^{8})$ and $\lambda^c$ with the range $(e^{-5}, e^{10})$, with values increasing evenly on a log scale, and which, at the extremes, included both null and complete variable selection and fusing. 

The full treatment model included 135 parameters (including five intercepts) and was reduced to 37 parameters by LOAL, where the variables sex, and current depressive symptoms were selected  in each time period. The fusion step further reduced the number of parameters to 23, a reduction of $83\%$ in the number of parameters as compared to the full model (see \Cref{selfusion trt}). The full censoring model included 180 parameters (including five intercepts and 15 coefficients of past treatments), of which 112 remained after LOAL and 55 after fused LOAL, representing a reduction of $69\%$ of the number of parameters. The variables selected in the censoring model included sex, country of birth, current depressive symptoms, ever smoked, family-related stress, other stress, participation in team sports and worry about weight (see \Cref{selfusion cen}). 

\begin{table}\center
\caption{Selected and fused parameters in the treatment model. `CurDep' represents current depressive symptoms; `FamStress' represents family stress; `TeamSport' represents participation in team sports; `WorWeight' represents worry about weight; `NA' represents that the time varying variable is not applicable at the given time. A blank space means that the variable was not selected by the LOAL in the first step. The values are color coded such that common colors indicate fused parameters.}
\resizebox{1.2\width}{!}{%
\begin{tabular}{lrrrrr}
  \toprule
Variable $\backslash$ Time &  1 &  2 &  3 &  4 &  5 \\ 
  \midrule
Intercept & -4.067 & -3.475 & -3.898 & -3.609 & -1.469 \\ 
Sex & \textcolor{BlueGreen}{-0.313} & \textcolor{BlueGreen}{-0.313} & \textcolor{BlueGreen}{-0.313} & \textcolor{BlueGreen}{-0.313} & \textcolor{BlueGreen}{-0.313} \\
CountryBirth & \textcolor{Plum}{-0.838} & \textcolor{Plum}{-0.838}  &  & \textcolor{Plum}{-0.838} & \textcolor{Plum}{-0.838}\\ 
MotherEducation &  &  &  &  & 0.491 \\ 
  CurDep$_{t=1}$ & \textcolor{Salmon}{1.493} &  &  &  &  \\ 
  CurDep$_{t=2}$ & NA & \textcolor{Salmon}{1.493}  &  &  &  \\ 
  CurDep$_{t=3}$ & NA & NA & \textcolor{Salmon}{1.493} & \textcolor{BrickRed}{0.127} &  \\ 
  CurDep$_{t=4}$ & NA & NA & NA & \textcolor{Salmon}{1.493}  & \textcolor{BrickRed}{0.127} \\ 
  CurDep$_{t=5}$ & NA & NA & NA & NA & \textcolor{Salmon}{1.493} \\ 
  EverSmoke$_{t=3}$ & NA & NA & 1.349 & 1.455 & 0.916 \\ 
  FamStress$_{t=2}$ & NA &  &  &  & -0.208 \\ 
  FamStress$_{t=4}$ & NA & NA & NA &  & 0.072 \\ 
  OtherStress$_{t=5}$ & NA & NA & NA & NA & 0.081 \\ 
  TeamSport$_{t=3}$ & NA & NA &  & -0.020 & 0.147 \\ 
  TeamSport$_{t=5}$ & NA & NA & NA & NA & 0.117 \\ 
  WorWeight$_{t=1}$ &  & \textcolor{Green}{0.277} &  & 0.410 & 0.074 \\ 
  WorWeight$_{t=3}$ & NA & NA & \textcolor{Magenta}{-0.119} &  &  \\ 
  WorWeight$_{t=4}$ & NA & NA & NA & \textcolor{Magenta}{-0.119} & \textcolor{Green}{0.277} \\ 
   \bottomrule \\
\end{tabular}}
\label{selfusion trt}
\end{table}

\begin{table}[!]\center
\caption{Selected and fused parameters in the censoring model. `CurDep' represents current depressive symptoms; `FamStress' represents family stress; `TeamSport' represents participation in team sports; `WorWeight' represents worry about weight; `NA' represents that the time-dependent variable is not applicable at the given time. A blank space means that the variable was not selected by the LOAL in the first step. The values are color coded according to fused clique.}
\resizebox{1.2\width}{!}{%
\begin{tabular}{lrrrrr}
  \toprule
Variable $\backslash$ Time & 2 & 3 & 4 & 5 & 6 \\ 
  \midrule
Intercept & -4.980 & -4.106 & -2.445 & -1.064 & -1.443 \\ 
A1 & 2.458 &  &  & 0.423 & -0.810 \\ 
  A2 & NA &  & -1.395 & 0.037 & -1.429 \\ 
  A3 & NA & NA & 0.840 & -0.594 & 3.187 \\ 
  A4 & NA & NA & NA & 0.709 & -1.946 \\ 
  A5 & NA & NA & NA & NA & -0.097 \\ 
 Sex & \textcolor{LimeGreen}{-0.336} & \textcolor{LimeGreen}{-0.336} & \textcolor{LimeGreen}{-0.336} & \textcolor{LimeGreen}{-0.336} & \textcolor{LimeGreen}{-0.336} \\ 
SelfEsteem & \textcolor{Plum}{-0.133}& \textcolor{Plum}{-0.133} & \textcolor{Plum}{-0.133} & \textcolor{Plum}{-0.133} &  \\ 
  CountryBirth & \textcolor{Bittersweet}{0.760} & \textcolor{Bittersweet}{0.760} & \textcolor{Bittersweet}{0.760} & \textcolor{Bittersweet}{0.760} & \textcolor{Bittersweet}{0.760} \\ 
  MotherEducation &  &  & \textcolor{Green}{-0.255} & \textcolor{Green}{-0.255} &  \\ 
  CurDep$_{t=1}$ &  &  & -0.823 & \textcolor{NavyBlue}{-0.360} & -0.263 \\ 
  CurDep$_{t=2}$ & \textcolor{Cerulean}{-0.242} & \textcolor{Purple}{-0.167} & \textcolor{Magenta}{0.358} &  & \textcolor{NavyBlue}{-0.360}\\ 
  CurDep$_{t=3}$ & NA & \textcolor{Cerulean}{-0.242} & \textcolor{Purple}{-0.167} & \textcolor{Magenta}{0.358} &  \\ 
  CurDep$_{t=4}$ & NA & NA & \textcolor{Cerulean}{-0.242} & \textcolor{Purple}{-0.167} & \textcolor{Magenta}{0.358} \\ 
  CurDep$_{t=5}$ & NA & NA & NA & \textcolor{Cerulean}{-0.242} & \textcolor{Purple}{-0.167} \\ 
  CurDep$_{t=6}$ & NA & NA & NA & NA & \textcolor{Cerulean}{-0.242} \\ 
  EverSmoke$_{t=1}$ &  & \color{red}0.235 & \color{teal}0.264 & \color{olive}-0.001 & 0.568 \\ 
  EverSmoke$_{t=2}$ & & \color{violet}-0.192 & \color{red}0.235 & \color{teal}0.264 & \color{olive}-0.001 \\ 
  EverSmoke$_{t=3}$ & NA & \color{orange}0.506 & \color{violet}-0.192 & \color{red}0.235 &  \\ 
  EverSmoke$_{t=4}$ & NA & NA & \color{orange}0.506 & \color{violet}-0.192 &  \\ 
  EverSmoke$_{t=5}$ & NA & NA & NA & \color{orange}0.506 &  \\ 
  EverSmoke$_{t=6}$ & NA & NA & NA & NA & \color{orange}0.506 \\ 
  FamStress$_{t=2}$ &  \textcolor{Turquoise}{0.022} & \textcolor{Maroon}{0.044}  & \textcolor{PineGreen}{-0.039} & 0.329 & -0.246 \\ 
  FamStress$_{t=3}$ & NA &  & \textcolor{Maroon}{0.044}  & \textcolor{PineGreen}{-0.039} &  \\ 
  FamStress$_{t=4}$ & NA & NA &  & \textcolor{Maroon}{0.044} & \textcolor{PineGreen}{-0.039}\\ 
  FamStress$_{t=5}$ & NA & NA & NA & \textcolor{Turquoise}{0.022} & \textcolor{Maroon}{0.044}  \\ 
  FamStress$_{t=6}$ & NA & NA & NA & NA &  \textcolor{Turquoise}{0.022} \\ 
 OtherStress$_{t=1}$ & \color{brown}0.070 & \color{gray}-0.219 & 0.353 & 0.150 & -0.054 \\ 
  OtherStress$_{t=3}$ & NA &  &  & \color{gray}-0.219 &  \\ 
  OtherStress$_{t=4}$ & NA & NA & \color{cyan}-0.025 & \color{brown}0.070 &  \\ 
  OtherStress$_{t=5}$ & NA & NA & NA & \color{cyan}-0.025 & \color{brown}0.070 \\ 
  TeamSport$_{t=2}$ & \textcolor{Aquamarine}{-0.105} & \textcolor{BrickRed}{-0.185} & \color{pink}0.128 & \color{violet}0.004 & 0.191 \\ 
  TeamSport$_{t=3}$ & NA &  & \textcolor{BrickRed}{-0.185} & \color{pink}0.128 & \color{violet}0.004 \\ 
  TeamSport$_{t=4}$ & NA & NA & \textcolor{Aquamarine}{-0.105} & \textcolor{BrickRed}{-0.185} &  \\ 
  TeamSport$_{t=5}$ & NA & NA & NA & \textcolor{Aquamarine}{-0.105} & \textcolor{BrickRed}{-0.185} \\ 
  TeamSport$_{t=6}$ & NA & NA & NA & NA & \textcolor{Aquamarine}{-0.105} \\ 
  WorWeight$_{t=1}$ & \color{teal}-0.018 & \color{orange}0.053 & 0.044 & -0.388 & 0.012 \\ 
  WorWeight$_{t=2}$ & \color{magenta}0.145 &  &  &  &  \\ 
  WorWeight$_{t=3}$ &  NA& \color{magenta}0.145 &  &  &  \\ 
  WorWeight$_{t=4}$ & NA & NA & \color{magenta}0.145 & \color{teal}-0.018 & \color{orange}0.053 \\ 
  WorWeight$_{t=5}$ & NA & NA & NA & \color{magenta}0.145 & \color{teal}-0.018 \\ 
  WorWeight$_{t=6}$ & NA & NA & NA & NA & \color{magenta}0.145 \\ 
   \bottomrule \\
\end{tabular}}
\label{selfusion cen}
\end{table}

Estimated coefficients and standard errors are presented in \Cref{ndit results}. The standard errors of the \textcolor{black}{four} LTMLE estimates were obtained through the influence functions without accounting for variable selection and are thus not valid post-inference standard errors. For IPTW, we applied the robust sandwich variance estimator. All methods consistently indicated that females had more severe depressive symptoms compared to males. Point estimates from IPTW full, LTMLE full, \textcolor{black}{LTMLE SL,} LTMLE LOAL, and LTMLE fused LOAL suggested that early alcohol initiation was associated with more depressive symptoms during young adulthood among male participants. All IPTW implementations suggested that alcohol initiation in female participants was associated with less severe depressive symptoms; however, the LTMLE implementations concluded null or harmful impacts of earlier drinking initiation for both sexes. 
Notably, the incorporation of propensity scores limited to covariates selected by LOAL dramatically reduced standard error estimates in both the IPTW and LTMLE analyses. In addition, LOAL plus fusion more than halved the estimated standard error of the LTMLE estimator compared to LTMLE with only LOAL. \textcolor{black}{LOAL was able to limit the extreme cumulative probabilities of treatment and censoring used in the analysis; see Appendix Table~\ref{ps_ndit}.}

 \begin{table}[!]
\centering
\caption{Estimates of the MSM parameters for the NDIT study application. Estimated standard errors are presented in brackets. IPTW [LTMLE] full represents IPTW [LTMLE] with pooled treatment models including all covariate main terms and pooled censoring models including all covariate main terms and treatment terms; IPTW [LTMLE] LOAL represents IPTW [LTMLE] with pooled treatment and censoring models after covariate selection by LOAL; IPTW [LTMLE] fused LOAL represents IPTW [LTMLE] with pooled treatment and censoring models with both selection by LOAL and coefficient fusion\textcolor{black}{; LTMLE SL represents LTMLE with superlearner screening for treatment, censoring, and outcome models.}}
\label{ndit results}
\resizebox{1.3\width}{!}{%
\begin{tabular}{lcccc}
  \toprule
Method$\backslash$Coefficient & Intercept & Female sex & cum$(\overline{\boldsymbol{a}})$ & Female sex$\times $cum$(\overline{\boldsymbol{a}})$\\ 
  \midrule
 IPTW full & 7.397(1.119) & 5.708(2.177) & 0.583(0.646) & -1.833(1.564) \\ 
  IPTW LOAL & 8.203(0.975) & 5.774(1.898) & -0.224(0.438) & -0.352(0.844) \\ 
  IPTW fused LOAL & 8.553(0.943) & 4.223(1.452) & -0.047(0.636) & -0.455(1.038) \\ 
  LTMLE full & 7.361(0.337) & 3.569(0.573) & 0.072(0.253) & 0.008(0.504) \\ 
  LTMLE LOAL & 7.570(0.104) & 3.479(0.182) & 0.005(0.085) & 0.028(0.154) \\ 
  LTMLE fused LOAL & 7.712(0.028) & 3.500(0.045) & 0.002(0.042) & -0.008(0.076) \\ 
   \textcolor{black}{LTMLE SL} &  \textcolor{black}{7.571(0.279)} & \textcolor{black}{3.477(0.436)} & \textcolor{black}{0.064(0.172)} & \textcolor{black}{0.005(0.360)} \\ 
   \bottomrule \\
\end{tabular}}
\end{table}
Complete details of the application and results are given in Web Appendix D.

\section{Discussion}
\textcolor{black}{Many causal confounder selection methods have been proposed for point treatment settings but very few for longitudinal data. Despite this, in practice, variable selection methods are highly utilized; a 2019 descriptive review found that 69/299 (~24\%) of articles published in epidemiology journals explicitly used data-driven variable selection methods.\cite{talbot2019descriptive}} In this paper, we extended the Outcome Adaptive LASSO propensity score variable selection approach of Shortreed and Ertefaie\cite{SE} to the setting with time-varying treatment over discrete time points. We first estimate regularized coefficients of the time-saturated propensity score models. We then fuse the resulting nonzero coefficients using a generalized adaptive fused LASSO. Allowing for sparse model identification can avoid forcing a Markov-type assumption where we assume a priori that treatment can only depend on the most recent values of the time-updated covariates and baseline covariates.  Oracle properties of the \textcolor{black}{outcome-adaptive LASSO~\cite{SE} and the} generalized adaptive fused LASSO~\citep{Viallon_StatsComp} guarantee oracle performance of these estimators \textcolor{black}{in larger samples}. In our setting, this means that 
the fused \textcolor{black}{LOAL will select then} fuse the coefficients correctly according to the marginal pooled treatment model as sample size increases. 

Our simulation studies show that implementation of our method can improve estimation by IPTW and LTMLE compared to the same estimators without variable selection. 
The success of the selection relied on the specification of the outcome model to the extent that the estimates of the $\beta$s in the working model converged correctly to either zero or non-zero; however, the success of variable selection also depended on the variance of the estimators of these $\beta$s. The fusion was highly successful whenever the correct covariates were selected in the LOAL step.

The application demonstrated the usage of our method in a realistically complex longitudinal study in epidemiology where the interest was in estimating the effect of the initiation time of regular alcohol consumption on depression symptoms in young adulthood. We extended our method to select adjustment variables in both the treatment and censoring models, and used the balance criterion to jointly select the tuning parameter for the two models. The reduction of covariates and fusion of coefficients in the treatment and censoring models both led to apparent major gains in efficiency. 

An important limitation of all frequentist covariate selection methods for the propensity score that exclude instruments is that \textcolor{black}{uniformly valid inference is not  available.~\cite{LeebPotscher, nonuniformCausal} }
This means that in practice, it may be best to avoid covariate selection in combination with IPTW and LTMLE if possible. \textcolor{black}{LOAL and fused LOAL may still be used to identify the relevant low-dimensional set of confounding variables needed and appropriateness of pooling over time for analyses on a separate observational dataset.} \textcolor{black}{However, C-TMLE has been shown to be asymptotically linear under certain conditions, \cite{van2011targeted} relying on cross-validation to select the optimal number of selection steps. While we do not expect this version of LOAL implemented with IPTW to be asymptotically linear, doubly robust inference approaches that also target residual bias terms due to incorrectly estimated treatment functions are a potential avenue for valid inference.~\cite{DRinferencevanderLaan2014,JuCHAL} Bootstrap methods valid under very general conditions, such as m-out-of-n and multiplier bootstrap, are other possibilities.\cite{jones2024causal,bickel1997mboot}}   
Another limitation of our method, as currently proposed \textcolor{black}{with working parametric models}, is that it cannot handle nonlinearities or interactions between covariates in the propensity score models. 
Potential extensions of our method may incorporate nonparametric approaches used in the single time-point setting such as causal ball screening~\cite{TangCausalBall} for high-dimensional covariates and outcome highly adaptive lasso,~\cite{JuCHAL} the latter of which has valid closed-form expressions for confidence intervals. To conclude, we consider this work as \textcolor{black}{a step forward in the} development of nonparametric shrinkage estimators that trade-off bias and variance in the MSM parameter estimation for longitudinal treatments while allowing for valid inference.

\bmsection*{Acknowledgments}
The authors extend their thanks to Vivian Viallon who kindly shared his code implementing the archived package \texttt{FusedLasso}, to Linbo Wang for helpful discussions, and an anonymous reviewer who identified a valid approach for inference.

\bmsection*{Financial disclosure}
This project was supported in part by a Discovery Grant from the Natural Sciences and Engineering Research Council of Canada to MES; a Canada Research Chair to MES; R33NS120240, R01ES034021 and R01DA048764 from the National Institutes of Health to AE; and a Fonds de recherche du Qu\'ebec research career award to DT (\#312198).

\bmsection*{Conflict of interest}

The authors declare no potential conflict of interests.

 \bmsection*{Supporting information}

The R codes used for the simulations are available on github [to be uploaded, currently attached to submission].

\bibliography{sample.bib}

\newpage
\appendix

\bmsection{Balance metric for variable selection\label{appA}}
\vspace*{12pt}
Our tuning parameter selection for the LOAL evaluates the balance of every covariate or function of covariates at each time point between units with different current treatment values. It is similar to a summary of the longitudinal balancing metric in \cite{Jackson_diagnostic} (Diagnostic 3). However, unlike,~\cite{Jackson_diagnostic} we do not subset on treatment history as this will result in vanishing data support for greater numbers of time points; our definition of the weights in  the main manuscript, which does not stabilize conditional on past treatments, produces independence between past treatment and current covariates.~\cite{adenyo2024evaluation}  
Each covariate $L_{t,k}$ considered at each time point $\tau\geq t$ will be weighted by the corresponding structural model coefficient ($\hat{\beta}_{\tau,t,k}$) divided by its standard error ($\sigma_{\hat{\beta}_{\tau,t,k}}$). We let $\hat{\boldsymbol{\alpha}}^{refit}(\lambda_n)$ represent the estimated values of the $\boldsymbol{\alpha}$ parameters using a logistic regression on the covariates selected by LOAL under tuning parameter value $\lambda_n$, where the value is defined as zero if the corresponding coefficient was not selected. Then, define the LOAL-weight for subject $i$ at time $\tau$  as $\hat{w}^{\lambda_n}_{\tau,i}(\bar{a}_{\tau})=w_{\tau}\{\bar{a}_{\tau}, \bar{\boldsymbol{l}}_{\tau,i};\hat{\boldsymbol{\alpha}}^{refit}(\lambda_n)\}$ corresponding to the weights defined in the main manuscript. 
For the simple example, we use the metric
\begin{align*}
  &\sum_{k=1}^{p_0}\frac{\mid  \hat{\beta}_{0,0,k} \mid }{\sigma_{\hat{\beta}_{0,0,k}}}
 \left| \frac{\sum_{i=1}^n{a_{0,i}l_{0,k,i}}\hat{w}^{\lambda_n}_{0,i}(1)}{\sum_{i=1}^n{a_{0,i}}\hat{w}^{\lambda_n}_{0,i}(1)} -
  \frac{\sum_{i=1}^n{(1-a_{0,i})l_{0,k,i}}\hat{w}^{\lambda_n}_{0,i}(0)}{\sum_{i=1}^n{(1-a_{0,i})}\hat{w}^{\lambda_n}_{0,i}(0)}
 \right| \\
 &\qquad+\sum_{k=1}^{p_0}
\frac{\mid\hat{\beta}_{1,0,k}\mid }{\sigma_{\hat{\beta}_{1,0,k}}}
 \left| \frac{\sum_{i=1}^n a_{1,i}l_{0,k,i}\hat{w}^{\lambda_n}_{1,i}(a_{0,i},1)}{\sum_{i=1}^na_{1,i}\hat{w}^{\lambda_n}_{1,i}(a_{0,i},1)} -
  \frac{\sum_{i=1}^n{(1-a_{1,i})l_{0,k,i}}\hat{w}^{\lambda_n}_{1,i}(a_{0,i},0)}{\sum_{i=1}^n{(1-a_{1,i})}\hat{w}^{\lambda_n}_{1,i}(a_{0,i},0)}
 \right| \\
 &\qquad \qquad +\sum_{k=1}^{p_1} \frac{\mid\hat{\beta}_{1,1,k}\mid }{\sigma_{\hat{\beta}_{1,1,k}}}
 \left| \frac{\sum_{i=1}^n a_{1,i}l_{1,k,i}\hat{w}^{\lambda_n}_{1,i}(a_{0,i},1)}{\sum_{i=1}^na_{1,i}\hat{w}^{\lambda_n}_{1,i}(a_{0,i},1)} -
  \frac{\sum_{i=1}^n{(1-a_{1,i})l_{1,k,i}}\hat{w}^{\lambda_n}_{1,i}(a_{0,i},0)}{\sum_{i=1}^n{(1-a_{1,i})}\hat{w}^{\lambda_n}_{1,i}(a_{0,i},0)}
 \right|.
\end{align*}

\bmsection{Asymptotics of the Generalized Adaptive Fused LASSO\label{appB}}%
Here we state a theorem on the convergence of the estimates of the generalized adaptive fused LASSO that does not include a sparsity-inducing penalty and is performed after variable selection. The result is directly connected to the main result in Viallon et al.~\cite{Viallon_StatsComp} We present this statement with the purpose of giving details about the theoretical properties of the second stage of our procedure.
 
 Recall that the parameter $\boldsymbol{\alpha}=(\boldsymbol{\alpha}_{0,-1},\boldsymbol{\alpha}_{0,0},\boldsymbol{\alpha}_{1,-1},\boldsymbol{\alpha}_{1,0},\boldsymbol{\alpha}_{1,1},\boldsymbol{\alpha}_{1,2})$ is defined according to the model (4) in the main manuscript. 
 Define $\boldsymbol{\alpha}^{\dagger}=(\alpha_{0,-1},\boldsymbol{\alpha}^{\dagger}_{0,0},\alpha_{1,-1},\boldsymbol{\alpha}^{\dagger}_{1,0},\boldsymbol{\alpha}^{\dagger}_{1,1},\alpha_{1,2})$, a parameter vector potentially containing fixed zeros among the elements of $\boldsymbol{\alpha}^{\dagger}_{0,0}$, $\boldsymbol{\alpha}^{\dagger}_{1,0}$, and $\boldsymbol{\alpha}^{\dagger}_{1,1}$. The elements that are not fixed zeros are free parameters. 
 Suppose that the marginalized distribution of the treatments $A_t; t=0,1$  corresponds to Bernoulli distributions with probability of success $m_{\tau}(\overline{L}_1,A_0;\boldsymbol{\alpha}_0^{\dagger})$ where $\boldsymbol{\alpha}^{\dagger,*}$ are defined as the true parameter values under maximum likelihood. More specifically, the distribution of $A_0$ conditional on the elements of $L_0$ corresponding to non-fixed-zero components of $\boldsymbol{\alpha}^{\dagger}_{0,1}$ is Bernoulli with probability of success $m_0(\overline{L}_1,A_0;\boldsymbol{\alpha}^{\dagger,*})$; the distribution of $A_1$ conditional on $A_0$ and the elements of $(L_0, L_1)$ corresponding to non-fixed-zero components of $(\boldsymbol{\alpha}^{\dagger}_{1,0},\boldsymbol{\alpha}^{\dagger}_{1,1})$ is Bernoulli with probability of success $m_1(\overline{L}_1,A_0;\boldsymbol{\alpha}^{\dagger,*})$.
 
 Let $\mathcal{J}^{\dagger}_{0,1}$ and $\mathcal{J}^{\dagger}_{1,1}$ denote the indices of the covariates corresponding to the non-zero elements of $\boldsymbol{\alpha}^{\dagger}_{0,0}$ and $\boldsymbol{\alpha}^{\dagger}_{1,0}$, respectively. 
 We define a graph $\mathcal{G}=(V,E)$ with vertices $V=\{(0,0,k_0),(1,0,k_1); k_0\in \mathcal{J}^{\dagger}_{0,0} \text{ and } k_1\in \mathcal{J}^{\dagger}_{1,0}\}$ and edges $E$ that connect all (pairs of) corresponding indices for $k \in\mathcal{J}^{\dagger}_{0,0}\cap\mathcal{J}^{\dagger}_{1,0}$. This is the graph that will be used to run the adaptive fused LASSO.

 Define the estimator $\hat{\boldsymbol{\alpha}}^{\dagger}$ as the minimizer of
 \begin{align}
    &\sum_{\tau=0}^1\sum_{i=1}^n \left[ a_{\tau,i} \log\{m_{\tau}(\bar{l}_{1,i},a_{0,i};\boldsymbol{\alpha}^{\dagger})\} +(1-a_{\tau,i})\log\{1-m_{\tau}(\bar{l}_{1,i},a_{0,i};\boldsymbol{\alpha}^{\dagger})\}\right]\notag\\
& + \lambda_{1,n}\sum_{k\in \mathcal{J}^{\dagger}_{0,0}\cap \mathcal{J}^{\dagger}_{1,0},}\frac{\lvert \alpha^{\dagger}_{1,0,k}-\alpha^{\dagger}_{0,0,k} \rvert 
}{{\lvert \tilde{\alpha}^{\dagger}_{1,0,k}-\tilde{\alpha}^{\dagger}_{0,0,k} \rvert}^{\gamma_1}}\label{AGFL}
 \end{align}
 where $\tilde{\alpha}^{\dagger}_{1,0,k}$ and $\tilde{\alpha}^{\dagger}_{0,0,k}$ are $\sqrt{n}$-consistent estimates of $\alpha^{\dagger,*}_{1,0,k}$ and $\alpha^{\dagger,*}_{0,0,k}$, respectively.
 
 Now following \cite{Viallon_StatsComp} with adaptations to our setting,  we define  $\mathcal{J}^*=\{k\in \mathcal{J}^{\dagger}_{0,1}\cap \mathcal{J}^{\dagger}_{1,1}: \alpha^{\dagger,*}_{0,1,k} = \alpha^{\dagger,*}_{1,1,k}\}$, i.e. the set of indices where fusing should occur. Furthermore, let $\mathcal{B}=\{(0,1,k),(1,1,k) \in E : j \in \mathcal{J}^*\}\subseteq E$, which is a set of connected indices where the true values of the parameters are equal.    Define the graph $\mathcal{G}_{\mathcal{B}}=(V,\mathcal{B})$ as the one containing the complete set of vertices of $\mathcal{G}$ but with edges only between the connected vertices of $E$ where the corresponding parameters values are equal. 
 Define $s_0$ as the number of connected components of the graph $\mathcal{G}_{\mathcal{B}}$.
 
 Now define $\alpha_{\mathcal{B}}^{*}=(\alpha_{0,0}^{*},\alpha^{\dagger,*}_{0,1},\alpha_{1,0}^*,\alpha^*_{1,1,\mathcal{J}_{1,1}\setminus \mathcal{J}^*},\alpha_{1,2}^{\dagger,*},\alpha_{1,3}^{*})^T$, which are the same as $\alpha^{\dagger,*}$ after removal of the redundant terms in $\alpha_{1,1}$ that are equal to their connected terms in $\alpha_{0,1}$; and let $\hat{\alpha}_{\mathcal{B}}$ be its estimate by the adaptive fused LASSO in (\ref{AGFL}). Finally, define $\mathcal{B}_n=\{(0,1,k),(1,1,k)\in E: \hat{\alpha}^{\dagger}_{0,1,k}=\hat{\alpha}^{\dagger}_{1,1,k}\}$, the edges that fused in the procedure.
 
 \begin{theorem}If $\lambda_{1,n}/\sqrt{n} \rightarrow 0$ and $\lambda_{1,n}n^{(\gamma_1-1)/2}\rightarrow \infty$, then under mild assumptions, the minimizer of (\ref{AGFL}) satisfies $P(\mathcal{B}_n = \mathcal{B})\rightarrow 1$ as $n\rightarrow \infty$ and $\sqrt{n}\left(\hat{\alpha}_{\mathcal{B}}-\alpha_{\mathcal{B}}^{*}\right)$ converges in distribution to a Gaussian distribution of dimension $s_0+3$ with mean zero.
 \end{theorem}
 
 The mild assumptions are given explicitly in Viallon et al. \cite{Viallon2013} as AL1 and AL2. The proof of our theorem follows exactly the steps of their proof of Theorem 2 excluding the sparsity element.

\bmsection{Simulation study details and extended results\label{appc}}%

\bmsubsection{Scenario 1 data generating mechanisms\label{appC.1}}

\begin{figure}[h!]
\centering
\includegraphics[width=0.48\textwidth]{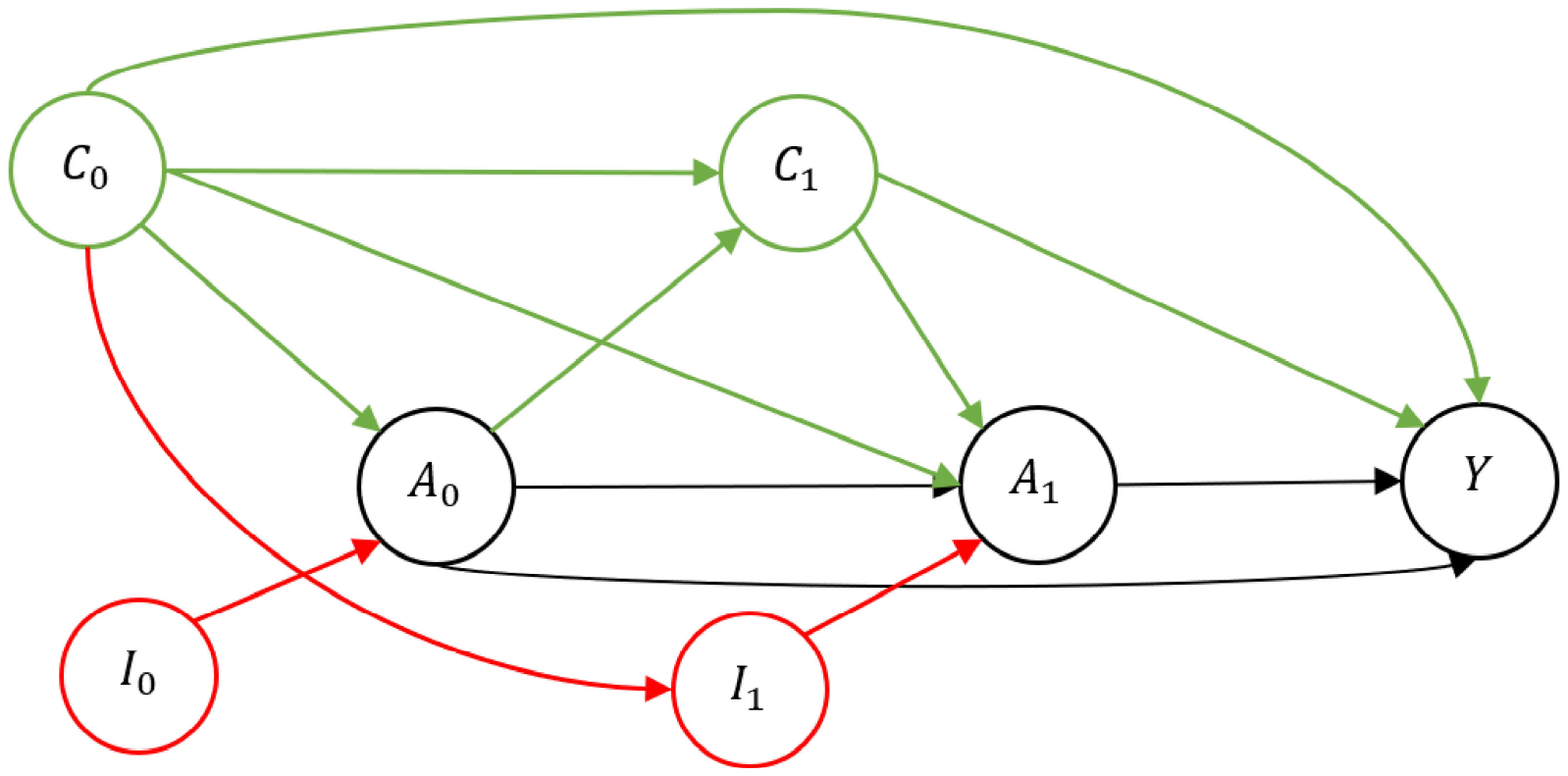} 
\includegraphics[width=0.49\textwidth]{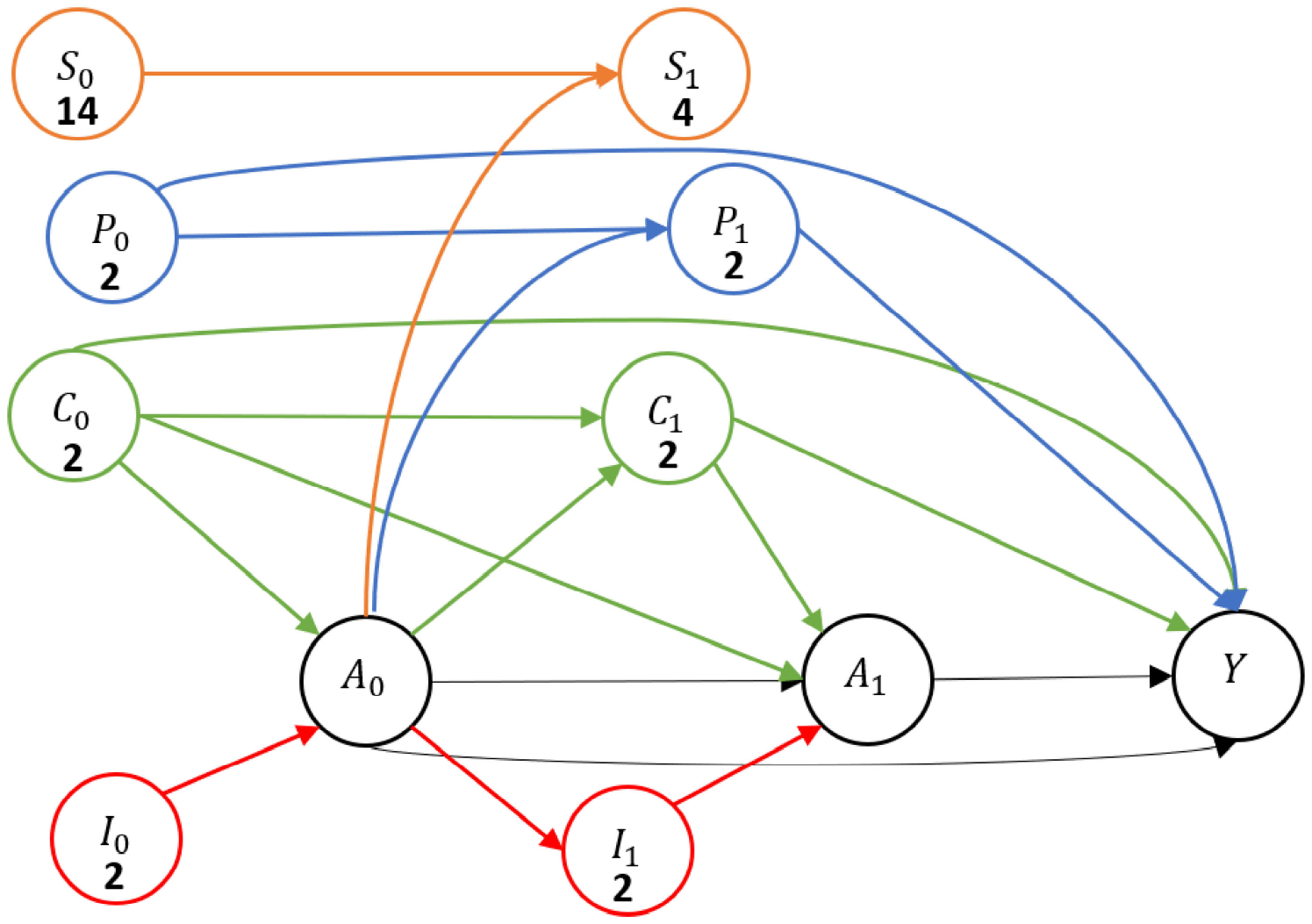}
\caption{DAGs representing the data generation in Scenarios 1 (left) and 2 (right). The target variable selection retained all covariates labeled $C$ and $P$ (the latter only for Scenario 2 in both models). \label{DAGs2}}
\end{figure}

The data in Scenario 1 were generated according to the left-hand DAG in Figure~\ref{DAGs2} and more specifically from the mechanisms presented in Table~\ref{Table simulation scenario 1}. 

\begin{table} 
\centering
\caption{\normalsize Simulation Scenario 1 data generating mechanism}
\label{Table simulation scenario 1}
 \resizebox{1.2\width}{!}{%
 \begin{tabular}{ll}
 \\[-1.2em]
\toprule\\[-1.2em]
\textbf{Variable} & \textbf{Generating Mechanism}\\ \\[-1.2em]
\midrule\\[-1.2em]
$C_0$    & $\sim N(\text{mean}=0,\text{sd}=1)$\\
$I_0$    & $\sim  N(\text{mean}=0,\text{sd}=1)$\\
$A_0$    & $\sim \text{Bernoulli(logit}(p)=1.515C_0+I_0)$\\
\midrule
$C_1$    & $\sim N(\text{mean}=C_0+A_0,\text{sd}=1)$\\
$I_1$    & $\sim  N(\text{mean}=C_0,\text{sd}=1)$\\
$A_1$    & $\sim \text{Bernoulli(logit}(p)=-0.5+0.5C_0+0.25C_1+0.5A_0 +I_1)$\\
\midrule
$Y$ for Scenario 1(a)   & $\sim N(\text{mean}=-1.5+0.5C_0+0.5A_0+C_1+A_1,\text{sd}=0.5)$\\
$Y$ for Scenario 1(b)   & $\sim N(\text{mean}=-1.5+0.5C_0+0.5A_0+C_1+A_1+2.5C_0C_1,\text{sd}=0.5)$\\
$Y$ for Scenario 1(c)   & $\sim N(\text{mean}=-1.5+0.5C_0+0.5A_0+C_1+A_1+2.5A_0C_1^2,\text{sd}=0.5)$\\
\bottomrule
\end{tabular}}
\end{table}

\bmsubsection{Scenario 2: higher dimensional covariates with two time-points\label{appC.2}}

In this scenario, we generated 20 independent covariates at time 0 and 10 covariates at time 1, two treatments and a continuous outcome, according to the right-hand DAG in Figure~\ref{DAGs2} and more specifically the complete data generating mechanism in Table~\ref{Table simulation scenario 2}. At time 0 there were two confounders, jointly denoted $C_0$; two pure causes of the outcome, $P_0$; two instruments, $I_0$, and 14 spurious covariates $S_0$. At time 1 there were two confounders, $C_1$; two pure causes of the outcome, $P_1$, two instruments, $I_1$, and four spurious covariates $S_1$. All of the time 1 covariates were affected by the corresponding covariate at time 0 and also by the previous treatment $A_0$. The data were generated in such a way that the coefficients of both confounders in $C_0$ at the two time-points were equal in the marginal pooled treatment model that excluded instruments and spurious covariates. The coefficients of the two variables $P_0$ were equal to zero in this same model. The outcome was Gaussian with mean linear in the main terms of $C$, $P$, $A_0$, and $A_1$. The model to estimate $q_1$ conditioned on all main terms (and thus contained the truth) for all methods; the models for $q_0$ were linear in the main terms.

\begin{table} 
\centering
\captionsetup{width=1\textwidth}
\caption{\normalsize Simulation Scenario 2 data generating mechanism}\label{Table simulation scenario 2}
\centering
 \resizebox{1.2\width}{!}{%
 \begin{tabular}{l l}
 \\[-1.2em]
\toprule\\[-1.2em]
\textbf{Variable} & \textbf{Generating Mechanism}\\ \\[-1.2em]
\midrule\\[-1.2em]
$C_{0,j}$ for $j=(1,2)$   &  $\sim N(\text{mean}=0,\text{sd}=1)$\\
$P_{0,j}$ for $j=(1,2)$    &  $\sim N(\text{mean}=0,\text{sd}=1)$\\
$I_{0,j}$ for $j=(1,2)$   &  $\sim  N(\text{mean}=0,\text{sd}=1)$\\
$S_{0,j}$ for $j=(1,...,14)$   &  $\sim  N(\text{mean}=0,\text{sd}=1)$\\
$A_0$    & $\sim \text{Bernoulli(logit}(p)=C_{0,1}+C_{0,2}+I_{0,1}+I_{0,2})$\\
\midrule
$C_{1,1}$  & $\sim N(\text{mean}=0.5C_0+0.5A_0,\text{sd}=1)$\\
$C_{1,2}$  & $\sim N(\text{mean}=0.2C_0-A_0,\text{sd}=1)$\\
$P_{1,1}$  & $\sim N(\text{mean}=0.5C_0+0.5A_0,\text{sd}=1)$\\
$P_{1,2}$  & $\sim N(\text{mean}=0.2C_0-A_0,\text{sd}=1)$\\
$I_{1,1}$  & $\sim N(\text{mean}=-0.5A_0,\text{sd}=1)$\\
$I_{1,2}$  & $\sim N(\text{mean}=A_0,\text{sd}=1)$\\
$S_{1,j}$ for $j=(1,...,4)$    &  $\sim  N(\text{mean}=0.5C_0+0.2A_0,\text{sd}=1)$\\
$A_1$    & $\sim \text{Bernoulli(logit}(p)=1.026C_{0,1}+0.987C_{0,2}+0.5A_0+C_{1,1}+C_{1,2}+I_{1,1}+I_{1,2})$\\
\midrule
$Y$    & $\sim N(\text{mean}=1+0.6C_{0,1}+0.6C_{0,2}+0.6P_{0,1}+0.6P_{0,2}+0.6C_{1,1}+0.6C_{1,2}$\\
&~~~~~ $+0.6P_{1,1}+0.6P_{1,2}+0.5A_0+A_1,\text{sd}=1)$\\
\bottomrule
\end{tabular}}
\end{table}

\begin{figure}[h!]
\centering
\includegraphics[width=0.49\textwidth]{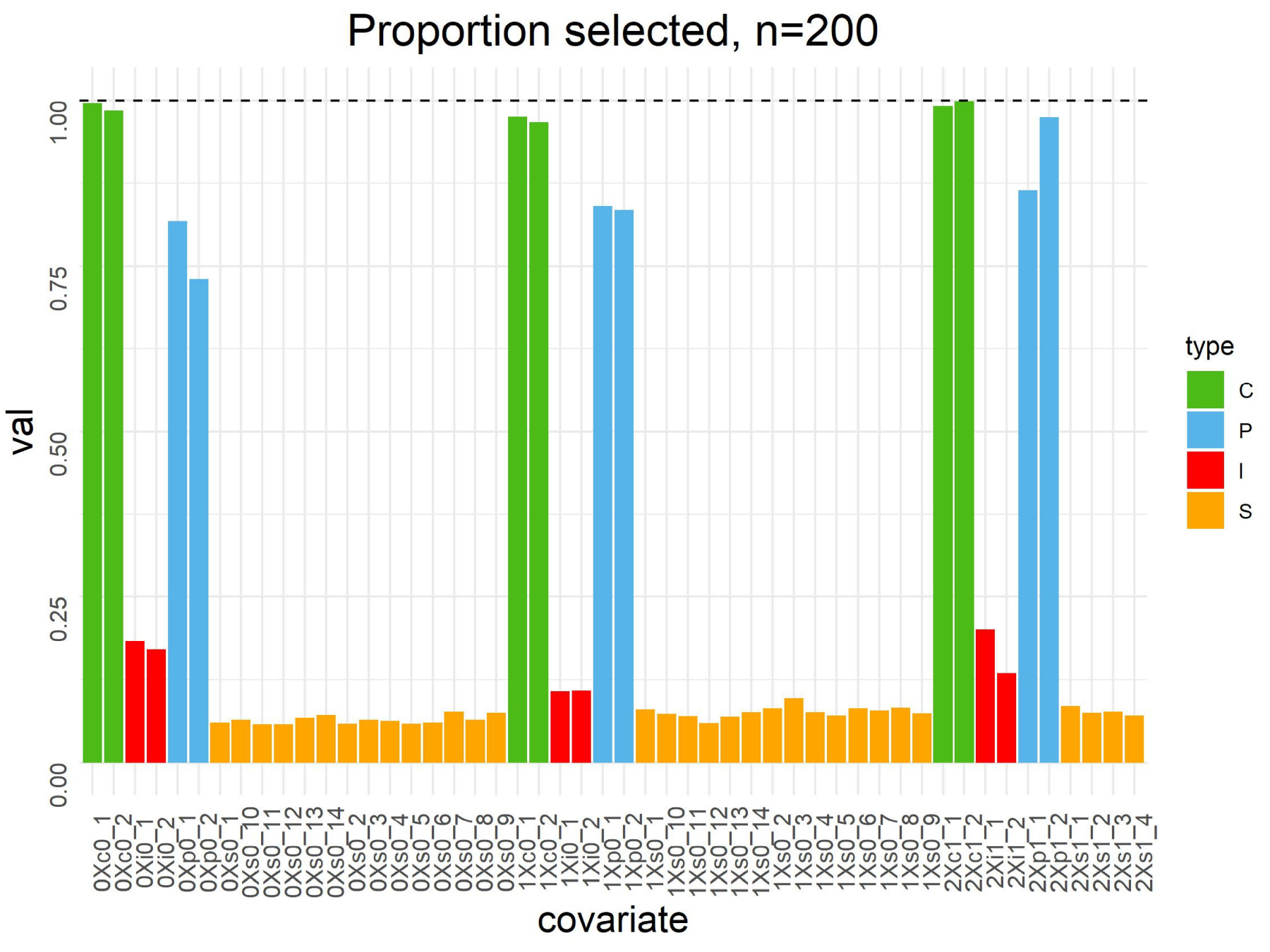} 
\includegraphics[width=0.49\textwidth]{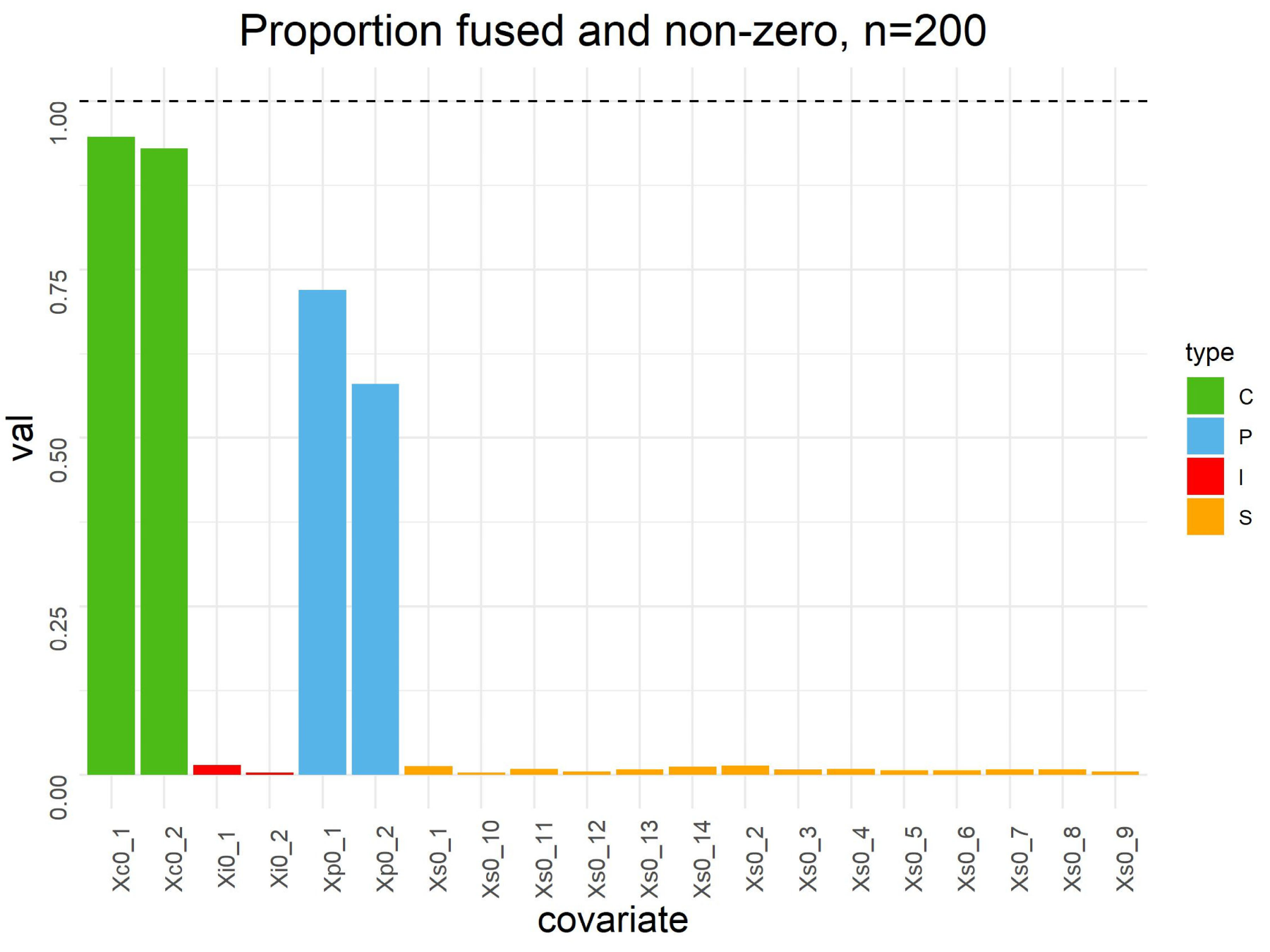}
\includegraphics[width=0.49\textwidth]{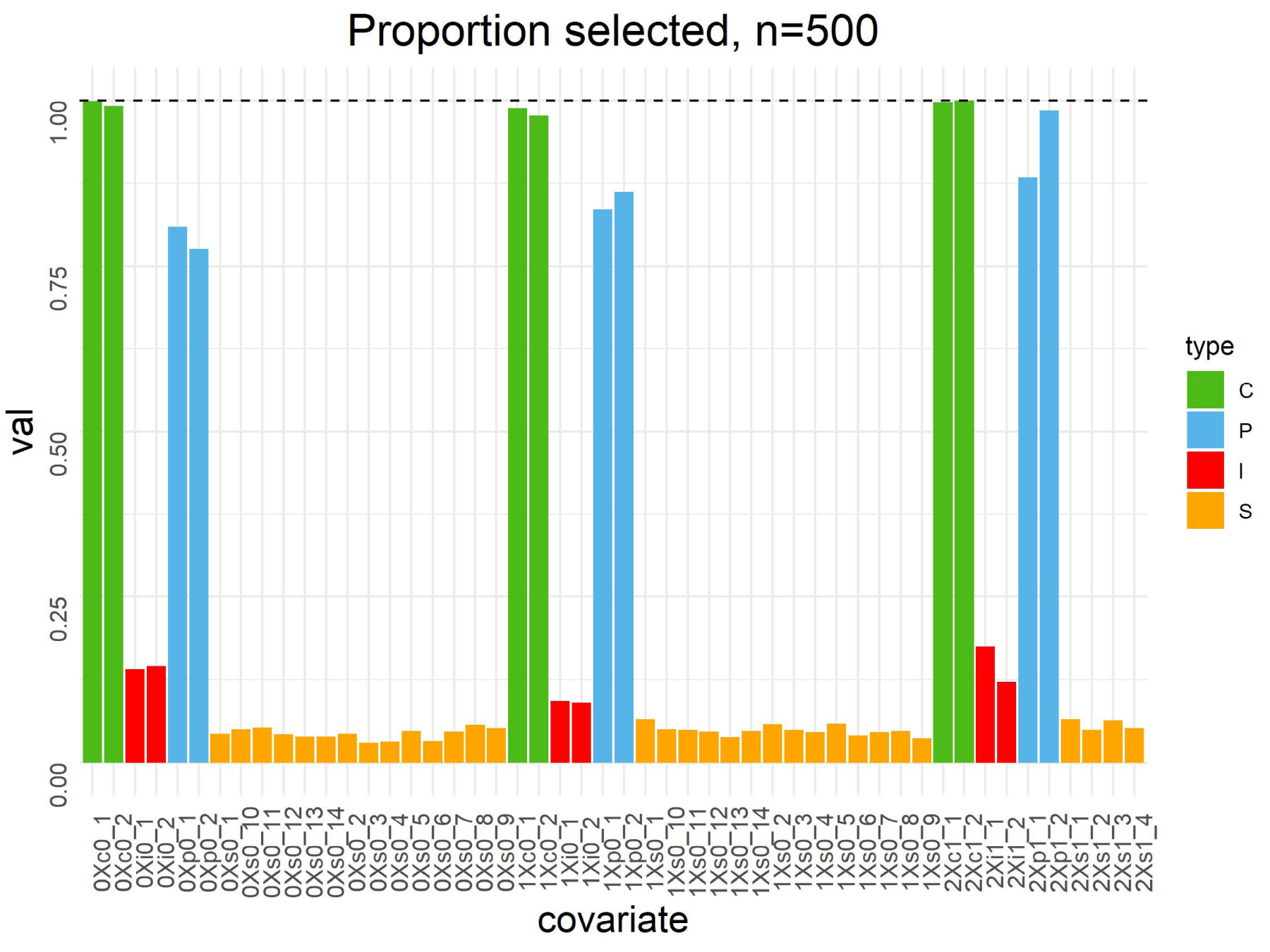} 
\includegraphics[width=0.49\textwidth]{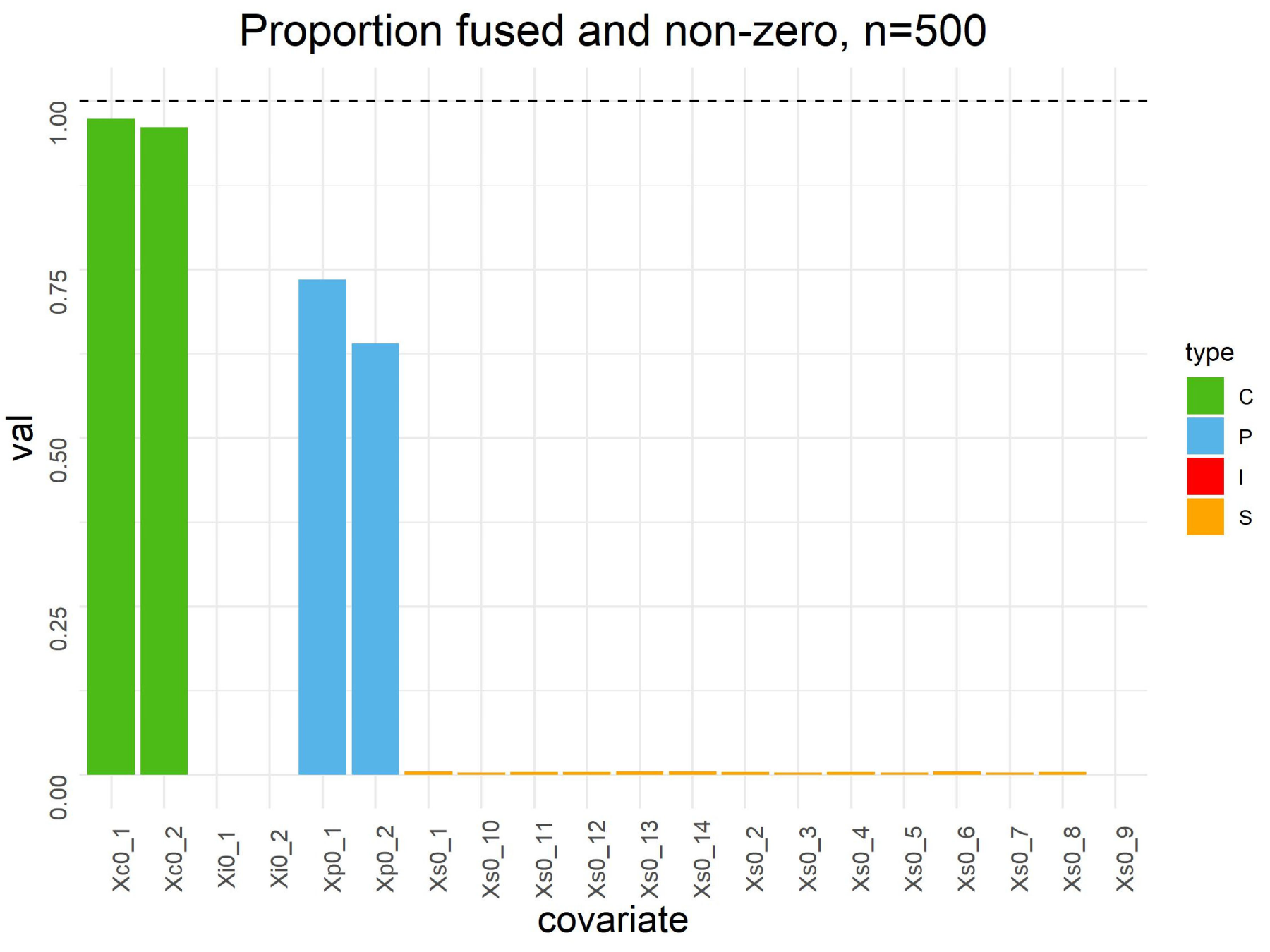}
\includegraphics[width=0.49\textwidth]{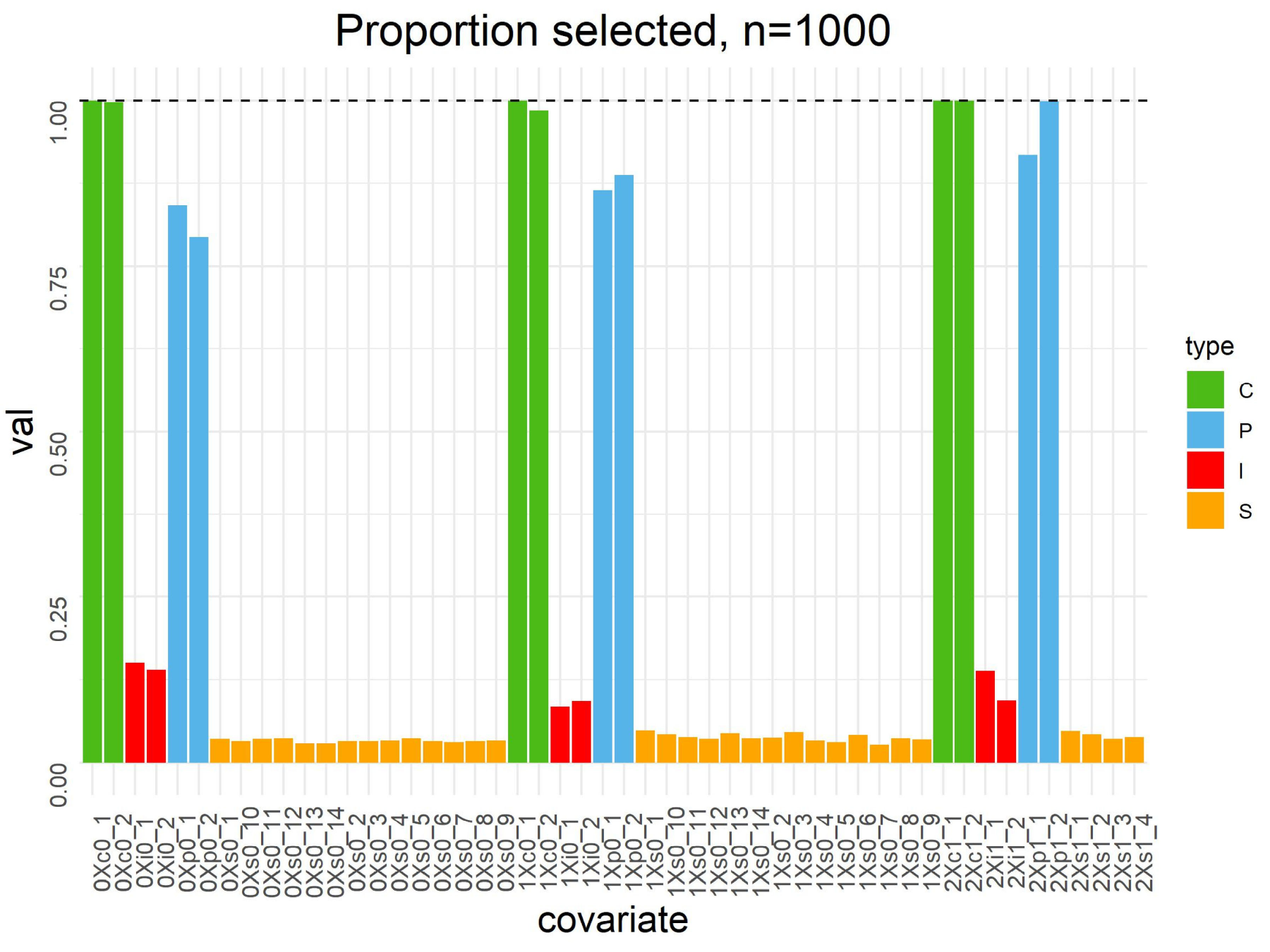} 
\includegraphics[width=0.49\textwidth]{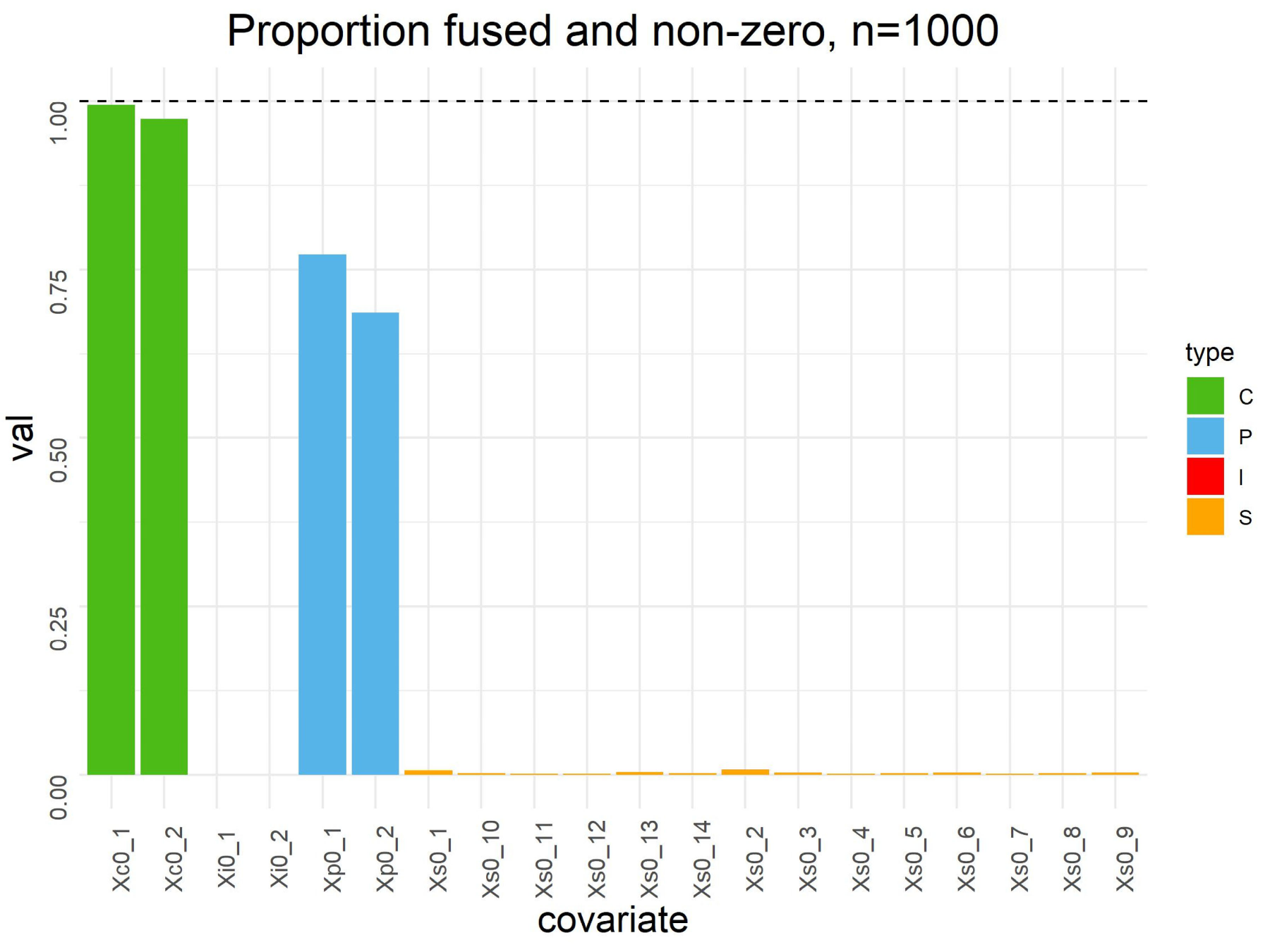}
\caption{Proportion variable selection (left) and fusion (right) for $n=200$ (top), $n=500$ (middle), and $n=1000$ (bottom) in Scenario 2. \textcolor{black}{Labels: $C$ are the confounders (should be selected), $P$ are the pure causes of the outcome (should be selected), $I$ are the instruments, and $S$ are noise variables.}\label{ScenGen2tpSelectandFuseresults}}
\end{figure}

The $\sqrt{n}$-bias and $n$-MSE are given in the first three data columns of Table~\ref{Scen23}. G-computation was unbiased with the lowest MSE as it was approximately correctly specified. The oracle IPTWs produced lower bias and MSE than the full IPTW. While the LOAL and fused LOAL had higher bias than their oracle counterparts, they had lower MSE. Figure~\ref{ScenGen2tpSelectandFuseresults} gives the proportion selection for each covariate at each time point and proportion fused (and non-zero) for corresponding baseline covariates between the two time points. While the confounders $C$ were selected nearly 100$\%$ of the time at all sample sizes, the selection of $P$ varied between roughly 75-100$\%$, and appeared to be slowly converging. The selection of instruments $I$ varied between  10-20$\%$ and appeared to be slowly converging to zero. Spurious covariates $S$ were selected less often (below 10$\%$) and converged close to zero by $n=1000$. The fusion of both $C_0$ variables quickly converged to almost 100$\%$ by $n=1000$. The terms $P_0$ often fused when they were both selected and non-zero, about 60-75$\%$ of the time.

\bmsubsection{Scenario 3: five time-points with only baseline covariates\label{appC.3}}

Finally, in order to demonstrate the potential usefulness of fusing, we developed a scenario with five treatments over time. Using the same approach as in Scenario 2, we simulated 20 baseline covariates (two confounders $C_0$, two pure predictors of the outcome $P_0$, two instruments $I_0$, and 14 spurious covariates $S_0$). To facilitate the construction of this scenario, we did not generate covariates past the baseline. The data generation was done in such a way that, in the marginal pooled treatment model, all 5 of the coefficients for each confounder were equal, and all 10 of the outcome predictor coefficients were equal to zero. Thus, variable selection should reduce the number of parameters (including 5 intercepts and 10 coefficients of treatment) from 115 to 35, and fusion should further reduce the number of parameters to 19. See Table~\ref{Table simulation scenario 3} for the complete data generating mechanism.

\begin{table} 
\centering
\captionsetup{width=1\textwidth}
\caption{\normalsize Simulation Scenario 3 data generating mechanism
\label{Table simulation scenario 3}}
 \resizebox{1.2\width}{!}{%
 \begin{tabular}{l l}
 \\[-1.2em]
\toprule\\[-1.2em]
\textbf{Variable} & \textbf{Generating Mechanism}\\ \\[-1.2em]
\midrule\\[-1.2em]
$C_{0,j}$ for $j=(1,2)$   & \multirow{5}{*}{\makecell{$\sim$ Multivariate normal distribution with mean 0 and the covariance matrix \\  $\Sigma_{(20\times 20)}=\left[\begin{array}{ccccc}0.64&0.192&0.192&\cdots&0.192\\0.192&0.64&0.192&\cdots&0.192\\ \vdots&\vdots&\vdots&\vdots&\vdots\\0.192&0.192&0.192&\cdots&0.64\end{array}\right]$}}\\
$P_{0,j}$ for $j=(1,2)$    &\\
$I_{0,j}$ for $j=(1,2)$   & \\
$S_{0,j}$ for $j=(1,...,14)$   & \\
   & \\  & \\
$A_0$    & $\sim \text{Bernoulli(logit}(p)=0.5C_{0,1}+C_{0,2}-0.5I_{0,1}-0.5I_{0,2})$\\\midrule
$A_1$    & $\sim \text{Bernoulli(logit}(p)=0.542C_{0,1}+1.075C_{0,2}-0.545I_{0,1}-0.545I_{0,2}-0.5A_0)$\\
$A_2$    & $\sim \text{Bernoulli(logit}(p)=0.568C_{0,1}+1.142C_{0,2}-0.565I_{0,1}-0.569I_{0,2}-0.5A_1)$\\
$A_3$    & $\sim \text{Bernoulli(logit}(p)=0.615C_{0,1}+1.23C_{0,2}-0.61I_{0,1}-0.61I_{0,2}-0.5A_2)$\\
$A_4$    & $\sim \text{Bernoulli(logit}(p)=0.66C_{0,1}+1.322C_{0,2}-0.655I_{0,1}-0.655I_{0,2}-0.5A_3)$\\
\midrule
$Y$    & $\sim N(\text{mean}=0.6C_{0,1}+0.6C_{0,2}+0.6P_{0,1}+0.6P_{0,2}+0.5A_0+0.5A_1+0.5A_2$\\
&~~~~~~~~~~~~~~~~ $+0.5A_3+0.5A_4,\text{sd}=1)$\\
\bottomrule
\end{tabular}}
\end{table}

The $\sqrt{n}$-bias and $n$-MSE results are given in the last three columns of Table~\ref{Scen23}. G-computation was approximately correctly specified and unbiased. The full IPTW was more biased. IPTW with oracle variable selection decreased the bias and MSE relative to the full IPTW. IPTW with oracle variable selection and fusing produced  smaller  MSE  compared to IPTW with oracle variable selection. Fused LOAL produced  reductions in bias for $\mu_1$ and MSE reductions for $\mu_0$ and $\mu_1$ compared to LOAL. 

\begin{figure}[h!]
\centering
\includegraphics[width=0.49\textwidth]{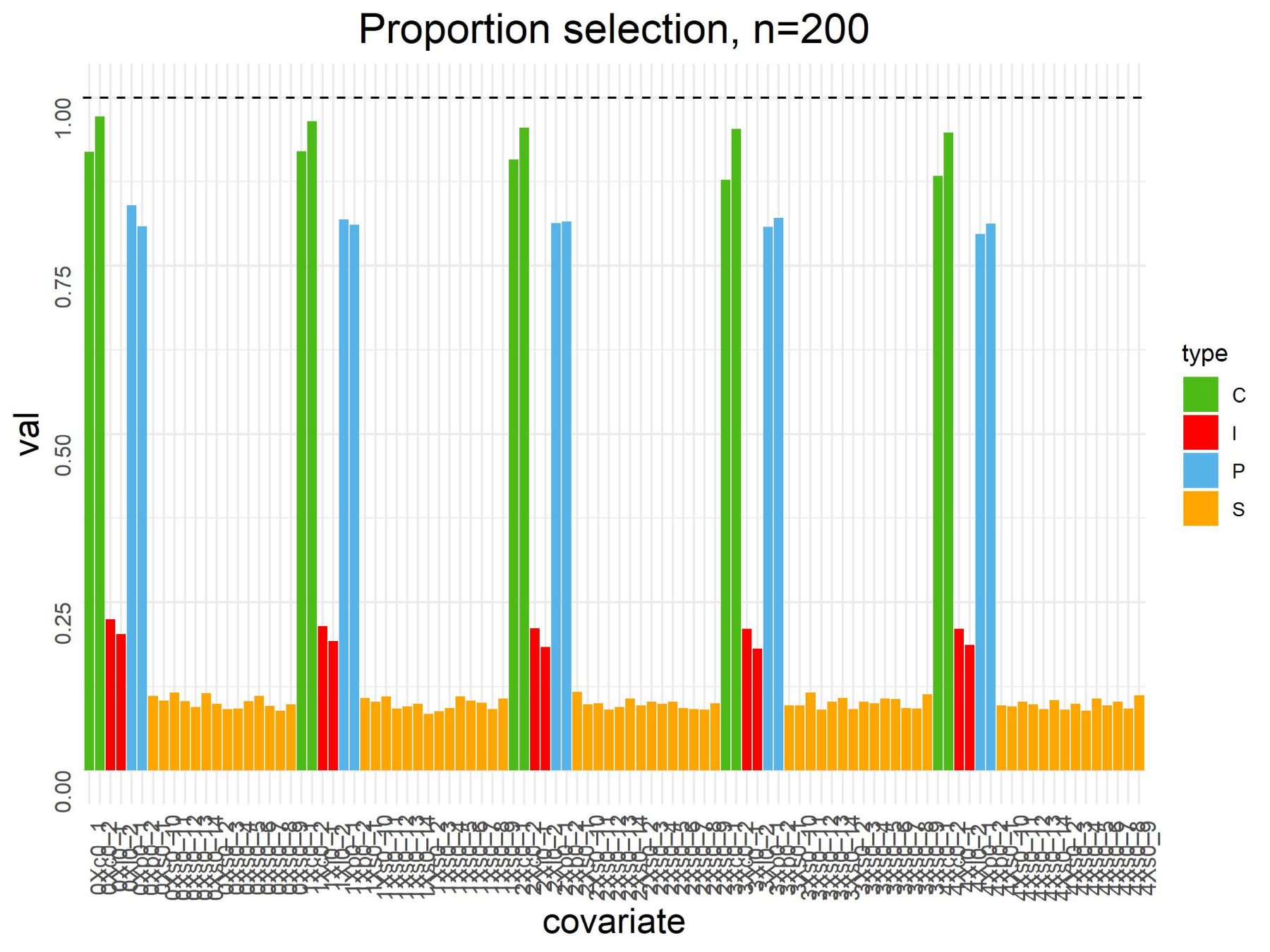}
\includegraphics[width=0.49\textwidth]{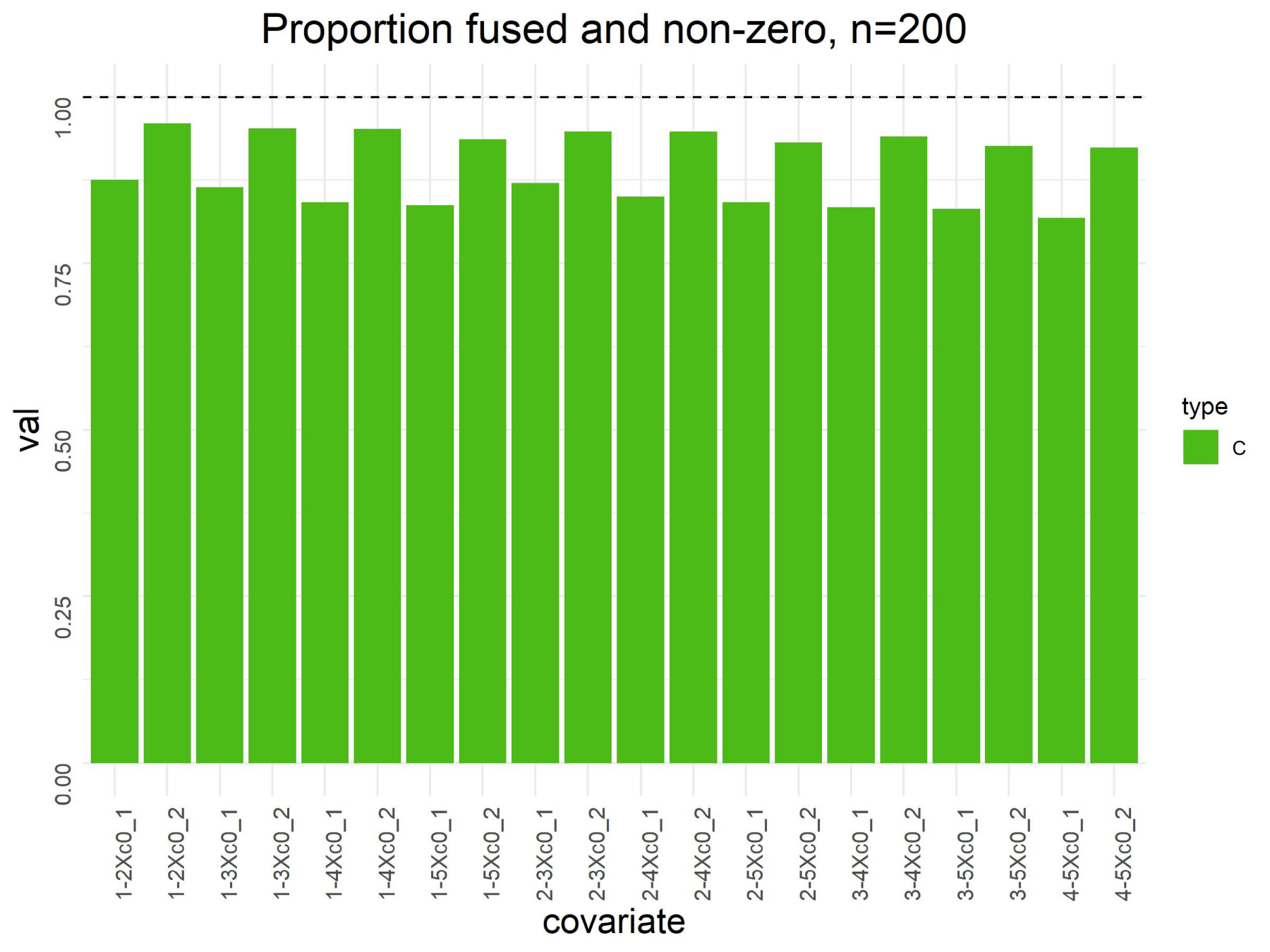}
\includegraphics[width=0.49\textwidth]{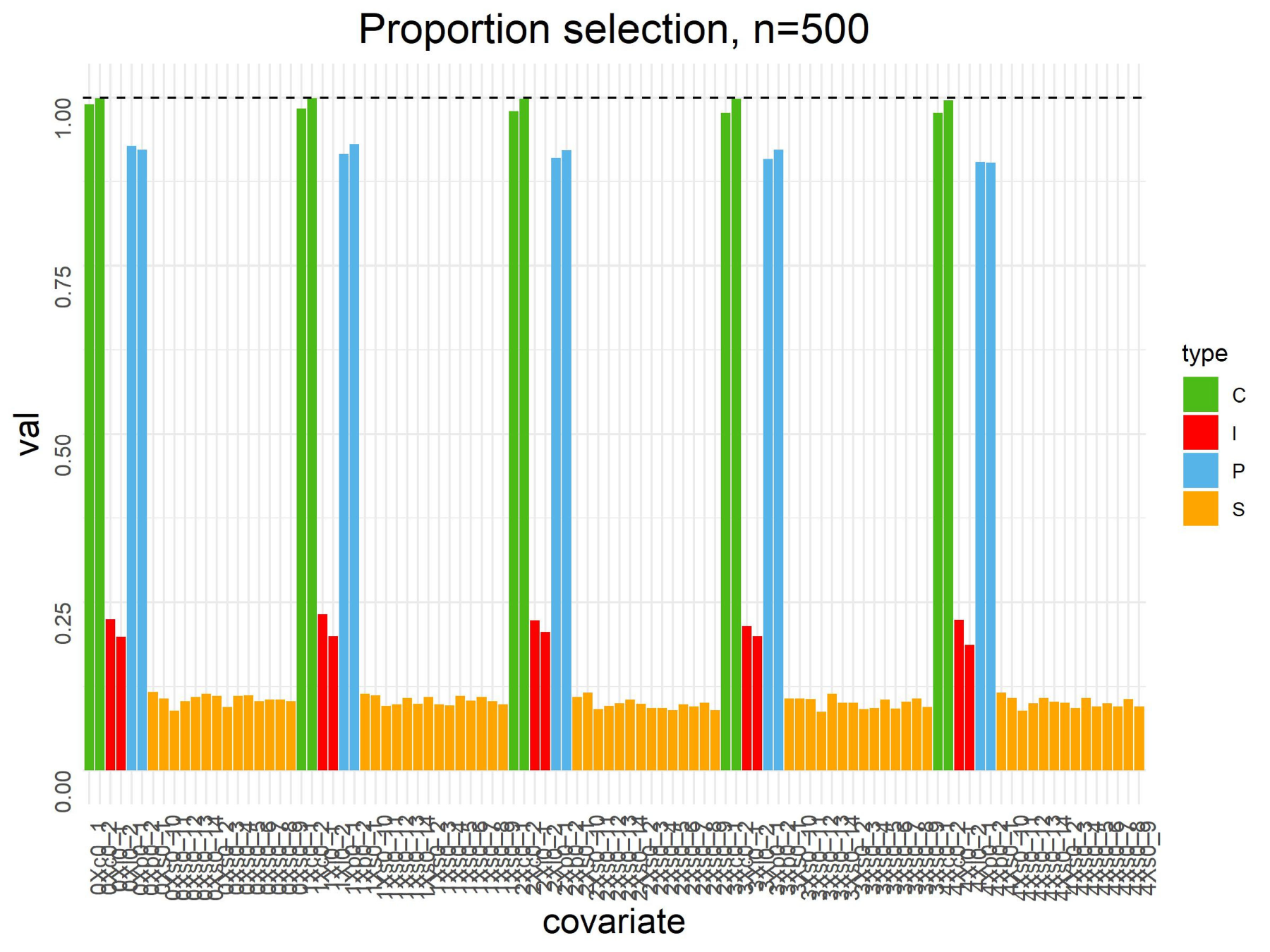}
\includegraphics[width=0.49\textwidth]{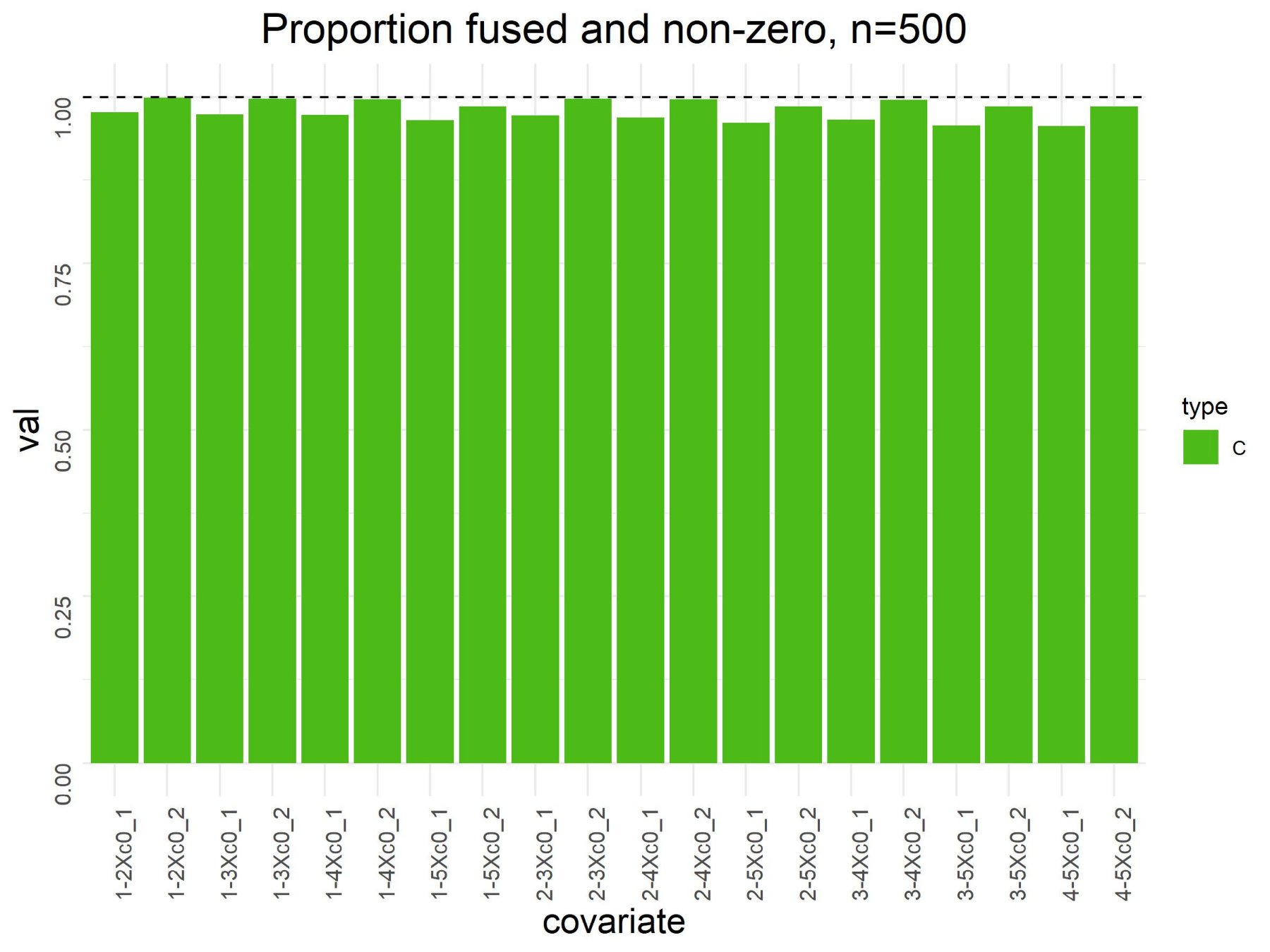}
\includegraphics[width=0.49\textwidth]{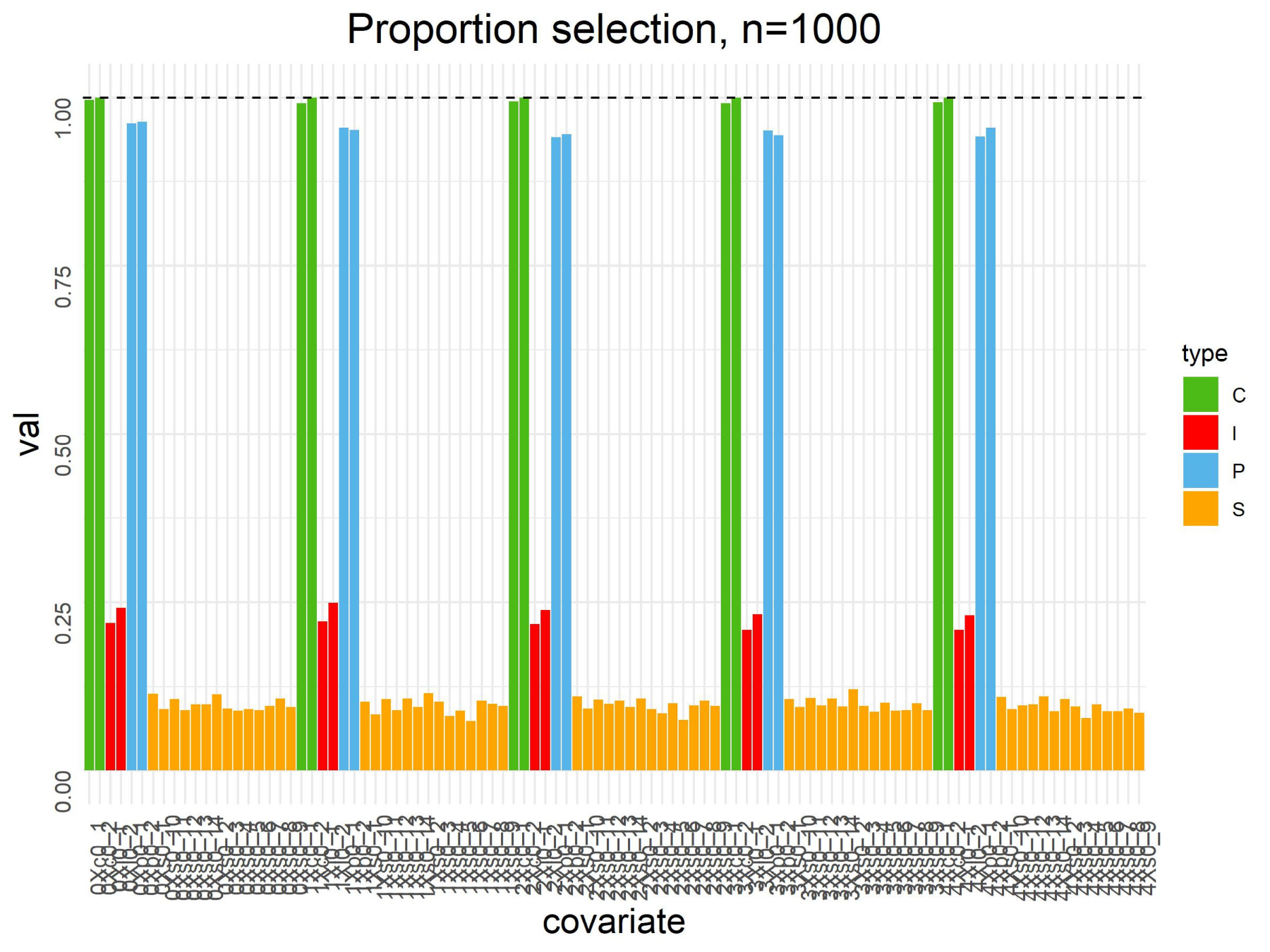} 
\includegraphics[width=0.49\textwidth]{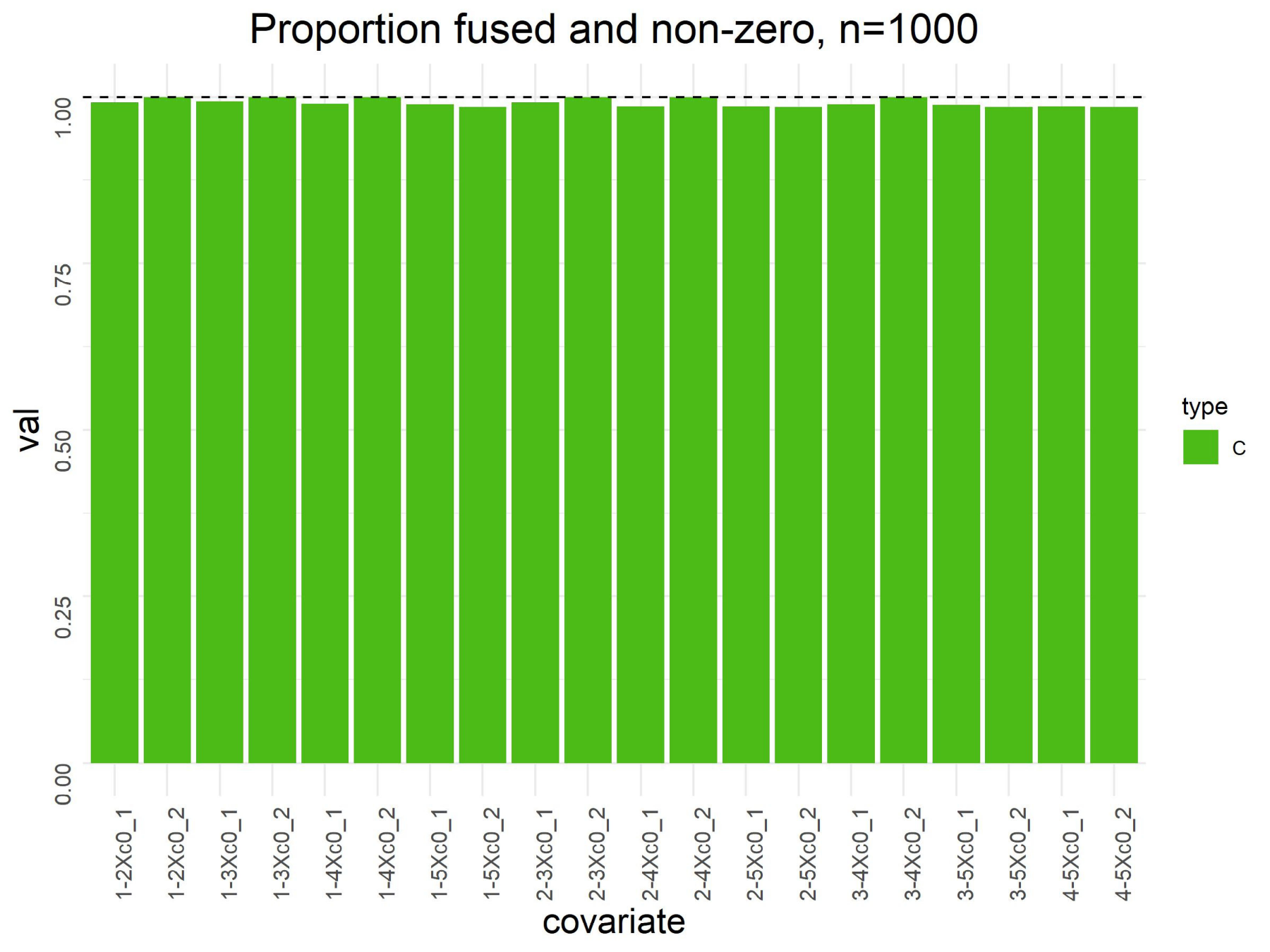}
\caption{Proportion variable selection (left) and fusion of the confounders (right) for $n=200$ (top), $n=500$ (middle), and $n=1000$ (bottom) in Scenario 3. Each bar in the right-hand plots represents the proportion of fusion of each confounder at each pair of time-points. \textcolor{black}{Labels: $C$ are the confounders (should be selected), $P$ are the pure causes of the outcome (should be selected), $I$ are the instruments, and $S$ are noise variables.}\label{ScenlongSelectandFuseresults}}
\end{figure}

The selection and fusion results are given in Figure~\ref{ScenlongSelectandFuseresults}. The confounder variable selection was again close to ideal. The selection of pure causes of the outcome increased to above 90$\%$ as $n$ increased. Instrument and spurious covariate selection was again low. We also verified that the fusion of non-zero coefficients of the confounders matched the selection of confounders and was near perfect for $n=1000$.

\begin{table} 
\centering
\captionsetup{width=1\textwidth}
\caption{Scenarios 2 and 3 $\sqrt{n}$ times the absolute value of bias ($n$ times mean squared error) of methods estimating the parameters in the marginal structural model of equation~(see Equation 1 in Section 2.1 of the manuscript). The Fused LOAL uses the estimates of LOAL for the adaptive weights. \label{Scen23} }
 \resizebox{1.2\width}{!}{%
\begin{tabular}{lcccccc}
\toprule 
Method\textbackslash{}Scenario & \multicolumn{3}{c}{\begin{tabular}[c]{@{}l@{}} Scenario 2, two time-points,\\$dim(L_0)=20$, $dim(L_1)=10$ \end{tabular}}   & \multicolumn{3}{c}{\begin{tabular}[c]{@{}l@{}} Scenario 3, five time-points,\\$dim(L_0)=20$ \end{tabular}}  \\  & $\mu_0$ & $\mu_1$ & $\mu_2$ &                   $\mu_0$ & $\mu_1$ & $\mu_2$  \\ 
True values  & 1.00    & 0.88   & 0.45    &   0.0  &    1.14    &  0.5  \\ \midrule 
\textbf{n=200}    &     &     &      &     &    &   \\
G-comp main terms  & 0.3(11)&0.2(5)&0.4(7)&0.1(9)&0.1(4)&0.0(1) \\
IPTW full main terms  &3.5(118)&1.5(40)&3.7(65)&2.1(54)&0.4(16)&0.8(8)\\
IPTW oracle select &2.2(41)&1.6(18)&2.0(27)&0.6(22)&0.2(8)&0.2(3)\\
IPTW oracle select and fuse   &2.2(40)&1.5(17)&2.0(27)&0.6(20)&0.1(7)&0.2(3)\\
LOAL &2.8(36)&1.8(17)&2.6(25)&1.1(23)&0.4(8)&0.4(3)\\
Fused LOAL  &2.9(36)&1.7(16)&2.6(25)&1(22)&0.3(7)&0.4(3)\\ \midrule
\textbf{n=500  }            \\
G-comp main terms  &0.6(9)&0.2(5)&0.8(6.5)&0.3(8)&0(3)&0.1(1) \\
IPTW full main terms  &4.0(153)&1.7(51)&3.9(90)&2.1(73)&0.7(18)&0.9(11)\\
IPTW oracle select &2.1(59)&1.4(23)&2.1(41)&0.6(23)&0.3(9)&0.3(4)\\
IPTW oracle select and fuse   &2.1(59)&1.4(23)&2.1(41)&0.6(22)&0.3(8)&0.3(3)\\
LOAL &3.1(45)&2.0(20)&3.0(30)&1.1(25)&0.6(9)&0.5(4)\\
Fused LOAL  &3.2(46)&1.9(20)&3.1(31)&1(24)&0.4(9)&0.5(4)\\ 
\midrule
\textbf{n=1000 }      \\
G-comp main terms  & 0.7(9)&0.3(5)&0.8(7)&0.1(8)&0(3)&0(1)\\
IPTW full main terms  & 4.4(182)&2.5(70)&4.3(114)&1.5(80)&0.6(20)&0.7(13)\\
IPTW oracle select &2.8(75)&2.0(30)&2.4(48)&0.3(28)&0.3(10)&0.2(4)\\
IPTW oracle select and fuse   &2.8(75)&2.0(30)&2.4(48)&0.3(26)&0.3(9)&0.2(4)\\
LOAL &3.7(59)&2.5(26)&3.3(38)&0.8(30)&0.5(10)&0.4(5)\\
Fused LOAL  &3.7(59)&2.5(26)&3.3(39)&0.8(29)&0.3(9)&0.4(5)\\ 
\bottomrule 
\end{tabular}}
\end{table}

\bmsubsection{Comparison with C-LTMLE \textcolor{black}{and LTMLE with Super Learner with variable screening}\label{appC.4}}

We used the simulation Scenarios 1 and 2 to contrast the estimation of $\mathbb{E}(Y^{1,1})$ by LTMLE using variable selection with LOAL to \textcolor{black}{variable selection with SL screening methods and to} an implementation of C-LTMLE. The C-LTMLE was implemented for two time-points to estimate the interventional mean in previous work.~\cite{SchnitzerCLTMLE} This C-LTMLE procedure uses initial estimates of $q_t;t=1,0$ to select covariates into the two propensity score models in a greedy fashion, creating a sequence of nested models with corresponding TMLE-updated estimates of $q_t;t=0,1$ with uniformly decreasing risk. Cross-validation is then used to select the point at which to stop, and this number of selections is used to determine the chosen propensity score models. (See \cite{GruberCTMLE} and \cite{SchnitzerCLTMLE} for more details.) Using the same linear main-terms models to estimate $q_t$ throughout, we fit 1) LTMLE using propensity scores conditional on the full set of covariates, \textcolor{black}{2) LTMLE using Super Learner with library ``SL.mean'', ``SL.glm'', ``SL.glm, screen.glmnet'', ``SL.glm.interaction'', ``SL.glm.interaction, screen.glmnet'', 3)} C-LTMLE, and 4) LTMLE using propensity scores with the covariates selected by LOAL.

The results in Table~\ref{CLTMLEvTMLEOAL} show that both C-LTMLE and LTMLE with LOAL generally improved the bias and MSE over LTMLE implemented with all covariates \textcolor{black}{or with superlearner}. In Scenarios 1(a) and 2 when the outcome models were nearly correctly specified, \textcolor{black}{both C-LTMLE and LTMLE with LOAL generally performed similarly in terms of bias, while C-LTMLE had lower MSE in higher-dimensional Scenario 2.} However, with incorrect model specifications (Scenarios 1(b) and 1(c)), LTMLE LOAL had less bias and MSE than all other approaches compared here. 
Compared to LTMLE with LOAL, C-LTMLE \textcolor{black}{was substantially slower, while LTMLE with superlearner screening had a similar runtime}; for example, in Scenario 2 with $n=1000$, C-LTMLE had a 90 second runtime while \textcolor{black}{LTMLE with superlearner with variable screening took 9 seconds, and }LTMLE with LOAL took only 3 seconds. However, the C-LTMLE procedure may be made more scalable with a preliminary ordering of the variables rather than the greedy procedure we employed.~\cite{Juscalable}

\begin{table}[ht]
\centering
\captionsetup{width=.68\textwidth}
\caption{$\sqrt{n}$ times absolute value of bias ($n$ times mean squared error) for the estimation of $E(Y^{(1,1)})$ of C-LTMLE, LTMLE with propensity score covariates selected by LOAL, LTMLE with no variable selection, \textcolor{black}{and LTMLE with superlearner (SL)}.}.  \label{CLTMLEvTMLEOAL} 
 \resizebox{1.2\width}{!}{%
\begin{tabular}{lcccc}
\toprule
Method\textbackslash{}Scenario & 1(a)   & 1(b) &1(c) &   2\\ 
$E(Y^{(1,1)})$ true values     & 1.0&3.5&8.5 &1.9 \\ \midrule
\textbf{n=200}          \\
LTMLE full& 0.6(48)& 10.2(317)& 13.0(1619)& 4.0(316)\\
\textcolor{black}{LTMLE SL}&\textcolor{black}{0.6(13)} & \textcolor{black}{13.0(308)} & \textcolor{black}{15.4(839)} & \textcolor{black}{2.2(26)}\\
C-LTMLE  &0.3(21)&6.7(268)&10.4(1125) &0.6(18)\\
LTMLE LOAL &0.4(17)& 5.2(215)& 8.6(706)& 1.9(71)\\
\midrule
\textbf{n=500  }              \\
LTMLE full&0.8(53)& 13.2(617)& 20.2(2520)& 4.1(287)\\
\textcolor{black}{LTMLE SL}& \textcolor{black}{0.8(14)} & \textcolor{black}{22.3(678)} & \textcolor{black}{26.0(1422)} & \textcolor{black}{2.2(28)}\\
C-LTMLE  & 0.3(19)&10.8(409)&17.6(1444)&1.2(23)\\
LTMLE LOAL & 0.4(19)&9.2(348) & 11.7(1126)& 2.0(63)\\
\midrule
\textbf{n=1000 }                      \\
LTMLE full& 0.8(65)& 17.0(844)& 29.6(4516)& 3.8(320)\\
\textcolor{black}{LTMLE SL}& \textcolor{black}{0.5(13)} & \textcolor{black}{35.8(1478)} & \textcolor{black}{43.9(2961)} & \textcolor{black}{2.1(27)}\\
C-LTMLE  & 1.3(21)&14.0(630)&24.7(2022)&1.9(33)\\
LTMLE LOAL &0.4(22) & 12.0(489)& 17.4(1956)& 2.0(70)\\
\bottomrule 
\end{tabular}}
\end{table}

\bmsubsection{\textcolor{black}{Inference using the m-out-of-n bootstrap\label{appC.5}}}

\color{black}

The m-out-of-n bootstrap is a resampling method designed to improve inference in settings where the standard bootstrap fails, especially when the parameter of interest is non-smooth, such as when it involves an extrema. \cite{bickel1997mboot, politis1994large, bickel2008choice} Instead of resampling $n$ observations from a dataset of size $n$, this method draws only $m$ observations with replacement where it requires that $m/n\rightarrow 0~$and$~m\rightarrow \infty$. \cite{bickel1997mboot} A key challenge is choosing an appropriate $m$: too small leads to high variance; too large reintroduces bias. Bickel and Sakov (2008) \cite{bickel2008choice} proposed a data-driven solution by selecting $m$ based on the stability of the bootstrap distribution across multiple values of $m$, using metrics like the Kolmogorov–Smirnov distance to detect when the distribution ``stabilize''. In other words, we want to find an optimal $m^*$ such that the limiting distribution of the bootstrap approximates the true generative distribution.

In order to perform the m-out-of-n bootstrap, the covergence rate $\tau_n^2$ of the estimator needs to be known or estimated. If $\tau_n$ is assumed to be of the form $\tau_n=n^{\epsilon}$, we can estimate $\epsilon$ by running a linear regression of $log[\hat{Var}(\hat{\boldsymbol{\mu}}^m)]$ on $-2log(m)$ where the variance $var(\hat{\boldsymbol{\mu}}^m)$ can be estimated by sampling with multiple subsampling sizes $m$. \cite{bertail1999subsampling} We present the m-out-of-n bootstrap algorithm in Table \ref{bootstrap algorithm}. Based on the findings of Chakraborty et al., \cite{chakraborty2013inference} we define $q=0.95$ and $K=14$, which implies that the minimum value of $m$ is approximately half the total sample size.

\begin{table}[]
    \centering
    \caption{\textcolor{black}{ m-out-of-n bootstrap algorithm} \label{bootstrap algorithm} }\color{black}
   \begin{tabular}{llll}
\toprule
\textbf{Step} & \multicolumn{3}{l}{\textbf{Description}} \\
\midrule
1 & \multicolumn{3}{l}{Estimate the parameters of the marginal structural model (MSM) using the full sample. Denote this estimate by $\hat{\boldsymbol{\mu}}=(\hat{\mu}_0,\hat{\mu}_1,\hat{\mu}_2)$.  }\\
\midrule \vspace{2mm}
2 & \multicolumn{3}{l}{For each \( j = 0, \dots, K \)} \\ \vspace{2mm}
 & 2.1 & \multicolumn{2}{l}{\makecell[l]{Set the bootstrap sample size as $m_j = \lfloor q^j \cdot n \rfloor$ where $\lfloor x\rfloor$ denotes the largest integer smaller than $x$ and $q$ is a predefined tuning parameter.}}\\
\vspace{2mm}
 &  2.2 & \multicolumn{2}{l}{For \( b = 1, \dots, B \), draw a sample of size \( m_j \) with replacement from the original data of size $n$.} \\\vspace{2mm}

&  2.3 & \multicolumn{2}{l}{Estimate the target parameter ${\boldsymbol{\mu}}$ from this sample, and denote it \( \hat{\boldsymbol{\mu}}[{jb}]=(\hat{\mu}_0[jb],\hat{\mu}_1[jb],\hat{\mu}_2[jb]) \).} \\\vspace{2mm}

& 2.4 & \multicolumn{2}{l}{For each parameter $\mu_l \in \boldsymbol{\mu}, l=0,1,2$, estimate the variance of each bootstrap resampled estimate, denoted $var^j(\mu_l)=var(\hat{\mu}_l[jb])$.}\\
\midrule \vspace{2mm}
3 & \multicolumn{3}{l}{For each parameter $\mu_l \in \boldsymbol{\mu}, l=0,1,2$,} \\ \vspace{2mm}
 & \raisebox{1.5ex}{3.1}  & \multicolumn{2}{l}{\makecell[l]{For each $j$, compute the empirical cumulative distribution function of
$\sqrt{m_j} (\hat{{\mu}}_l[{jb}]- \hat{{\mu}}_l)$, and denote it by\\ $R_{m_j}(x, \hat{\mu}_l)=\frac{1}{B}\sum_{b=1}^B\mathbb{I}\{\sqrt{m_j} (\hat{{\mu}}_l[{jb}]- \hat{{\mu}}_l)\leq x\}$ .}}\\\vspace{2mm}
  & \raisebox{4ex}{3.2}  & \multicolumn{2}{l}{\makecell[l]{Data-adaptively select an optimal $m_j$, denoted $m^{*}$, which produces  the minimum value of the sup-norm of the successive differences\\ between the bootstrap empirical distribution functions, i.e., \\
$m^* = \arg\min_{m_j} \biggl\{\sup_x \left| R_{m_j}(x, \hat{\mu}_l) - R_{m_{j+1}}(x, \hat{\mu}_l) \right|\biggr\}$ and $j^*$ refers to the corresponding index such that $m^*=m_{j^*}$. }}\\\vspace{2mm}
 & 3.3 & \multicolumn{2}{l}{Run a simple linear regression of $log[var^j(\mu_l)]$ on $-2log(m_j)$ to obtain the coefficient $\epsilon_l$.}\\
 & 3.4 & \multicolumn{2}{l}{Compute the 95\% confidence intervals by calculating $\hat{\mu}\pm 1.96(\frac{m^*}{n})^{\epsilon_l} \sqrt{var^{j^*}(\hat{\mu}_l})$.}\\

\bottomrule
\end{tabular}
\label{tab:my_label}
\end{table}


We report the coverage rates and mean width of confidence intervals for Scenario 1(a) (low dimensional covariates with two time-points) with sample size of 200, 500, and 1000, and Scenario 2 (high dimensional covariates with two time-points) with sample size of 500 and 1000 in Table \ref{Scen1aS2_BOOTSTRAP}. We also evaluate the performance of the naive bootstrap (n-out-of-n bootstrap). Due to computational burden, a total of 200 replicates were conducted for each scenario and each sample size, and within each replicate, the m-out-of-n bootstrap was applied with 200 bootstrap samples. 
The results demonstrate that the m-out-of-n bootstrap generally achieves higher coverage rates than the naive bootstrap across both scenarios and all sample sizes although confidence intervals via the naive bootstrap have the smaller average width. For example, in Scenario 2 with n=1000, coverage rates under the m/n bootstrap range from 0.83 to 0.84, compared to 0.71 to 0.78 under the naive method.

\begin{table} 
\centering
\captionsetup{width=0.8\textwidth}
\caption{\color{black} Coverage rates and mean width of 95\% confidence intervals for the parameters estimated using LOAL in 200 simulations of Scenario 1(a) (low dimensional with two time points) and of Scenario 2 (high dimensional with two time points) \label{Scen1aS2_BOOTSTRAP} }
 \resizebox{1.2\width}{!}{%
 \color{black}
\begin{tabular}{lccccc}
\toprule 
 Secenario & Parameters &  \makecell[c]{Coverage rate in\\naive bootstrap}  & \makecell[c]{Coverage rate in\\m/n bootstrap}&  \makecell[c]{CI width in\\naive bootstrap} & \makecell[c]{CI width in\\m/n bootstrap}\\\midrule
\multirow{13}{*}{S1(a)} & \textbf{n=200 }  & & &  &      \\
&  $\mu_0$ & 0.86 & 0.88 & 0.62 & 0.76 \\ 
&  $\mu_1$& 0.85 & 0.88 & 0.49 & 0.60 \\ 
&  $\mu_2$ & 0.94 & 0.96 & 0.49 & 0.60 \\ 
 \cmidrule(lr){2-6}
& \textbf{n=500}      \\
 &  $\mu_0$  & 0.82 & 0.88 & 0.42 & 0.51 \\ 
 &  $\mu_1$ & 0.82 & 0.89 & 0.32 & 0.39 \\ 
 &  $\mu_2$ & 0.90 & 0.92 & 0.34 & 0.41 \\ \cmidrule(lr){2-6}
& \textbf{n=1000 }      \\
 & $\mu_0$  & 0.85 & 0.89 & 0.31 & 0.38 \\ 
 & $\mu_1$  & 0.83 & 0.90 & 0.24 & 0.29 \\ 
 & $\mu_2$ & 0.87 & 0.90 & 0.26 & 0.31 \\ 
\midrule
\multirow{9}{*}{S2} & \textbf{n=500 } & & & & \\
 &  $\mu_0$  &0.81 & 0.88 & 0.80 & 0.97 \\ 
 &  $\mu_1$ & 0.81 & 0.87 & 0.52 & 0.64 \\
 &  $\mu_2$ & 0.78 & 0.82 & 0.61 & 0.75 \\ 
\cmidrule(lr){2-6}
& \textbf{n=1000 }      \\
 & $\mu_0$  & 0.77 & 0.84 & 0.63 & 0.73 \\ 
 & $\mu_1$  & 0.78 & 0.83 & 0.41 & 0.47 \\ 
 & $\mu_2$ & 0.71 & 0.83 & 0.48 & 0.56 \\ 
\bottomrule 
\end{tabular}}
\end{table}

\bmsubsection{Evaluating positivity violations \label{appC.6}}

In order to show the extent of practical positivity violations in Scenarios 1 and 2, in Table \ref{ps_S1S2} we provide summaries of the cumulative products of the treatment probabilities used for the estimation of the MSM parameters. These probabilities were estimated first using a logistic regression conditional on all terms (``full model") and then conditional on just those selected by LOAL (``LOAL"), with data corresponding to 200 draws of sample size $n=500$. Given that the data generating mechanism for treatment is identical in Scenarios (a-c), the corresponding results were the same in the absence of selection. The full model yielded minimum cumulative treatment probabilities of 0.000 in both Scenarios 1 and 2, indicating severe practical positivity violations. In contrast, LOAL mitigated this issue to some extent, with minimum scores ranging from 0.002 to 0.003 across all scenarios. LOAL yielded slightly lower quantile values (10\%–90\%), compared to the full model across all scenarios, which indicates that LOAL shifted the bulk of the cumulative probabilities downwards.

\begin{table}[ht]
\centering
\captionsetup{width=1\textwidth}
\caption{\color{black}Summary of the cumulative probabilities of treatment using all covariates (``full model'') vs selection by LOAL in Scenarios 1 and 2 with 200 draws of $n=500$ \label{ps_S1S2} }
\resizebox{1.2\width}{!}{%
\color{black}
\begin{tabular}{llrrrrrrr}
  \toprule
Scenario & Method & Min. & 10\% & 25\% & 50\% & 75\% & 90\% & Max. \\ 
  \midrule
\multirow{2}{*}{S1(a)}& Full & 0.000 & 0.134 & 0.262 & 0.494 & 0.757 & 0.897 & 1 \\ 
 & LOAL & 0.003 & 0.135 & 0.202 & 0.403 & 0.677 & 0.830 & 0.998 \\ \midrule
\multirow{2}{*}{S1(b)} & Full & 0.000 & 0.134 & 0.262 & 0.494 & 0.757 & 0.897 & 1 \\ 
& LOAL & 0.003 & 0.136 & 0.203 & 0.402 & 0.675 & 0.827 & 0.997 \\ \midrule
 \multirow{2}{*}{S1(c)} & Full & 0.000 & 0.134 & 0.262 & 0.494 & 0.757 & 0.897 & 1 \\ 
 & LOAL & 0.002 & 0.124 & 0.206 & 0.406 & 0.646 & 0.797 & 0.998 \\ \midrule
 \multirow{2}{*}{S2} & Full & 0.000 & 0.163 & 0.339 & 0.591 & 0.813 & 0.928 & 1 \\ 
  & LOAL & 0.003 & 0.142 & 0.245 & 0.417 & 0.640 & 0.802 & 0.999 \\ 
   \bottomrule
\end{tabular}}
\end{table}

To investigate the impact of practical positivity violations in the simulation study, we focused on Scenario 1(a) and varied the intercept and the coefficient of $I_0$ in the probability function of $A_0$. More specifically,  $A_0$ follows the Bernoulli distribution with probability $\text{logit}(p)=\nu_0+1.515C_0+\nu_II_0$ where we set the intercept $\nu_0$ to 21 values over the range $[-1.5, 1.5]$ in increments of 0.15, and the coefficient of $I_0$, $\nu_I$, to 21 values in $[0, 2]$ in increments of 0.1. Therefore, increasing $\nu_I$ amplifies the effect of $I_0$ on the probability of $A_0$, inducing greater variability across units and pushing propensity scores closer to the boundaries of 0 and 1. Varying the intercept $\nu_0$ shifts the overall probability of $A_0$.

\begin{figure}[ht]
\centering
\includegraphics[width=0.9\textwidth]{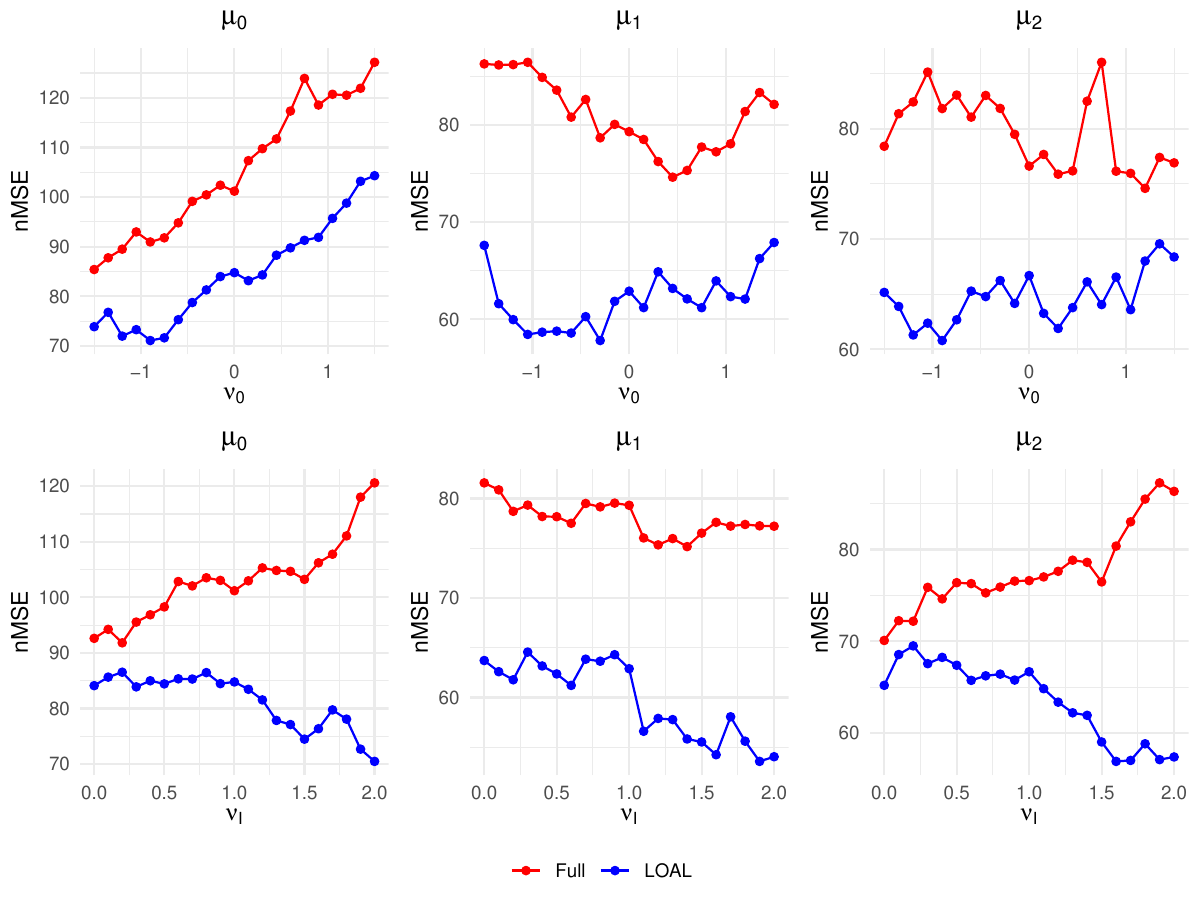} 
\caption{\color{black}$n$ times MSE for parameters over 200 draws when varying treatment data generating model intercept $\nu_0$ (top row) and instrument coefficient $\nu_I$ (bottom row) in Scenario 1(a), $n=500$. Results are shown using full model (red) and model selected by LOAL (blue). 
\label{fig_mse_S1a}}
\end{figure}

Figure \ref{fig_mse_S1a} displays the $n$ times the mean squared error (nMSE) over 200 draws for the estimated parameters $\mu_0, \mu_1, \mu_2$ using full model and model selected by LOAL across the full range of $\nu_0$ and $\nu_I$, the intercept and coefficient of the instrument $I_0$ in the treatment generating model respectively, in Scenario 1(a). As the strength of the instrument ($\nu_I$) increased, the performance of full model deteriorated, particularly for $\mu_0, \mu_2$ as indicated by rising nMSE. LOAL, in contrast, reduced in nMSE  exhibiting robustness in the presence of varying degrees of positivity violations due to instruments. When varying the treatment generating model intercept $\nu_0$, LOAL had consistently better performance than using full model, though its performance was hampered by larger positive values of $\nu_0$.

\color{black}
\FloatBarrier
\bmsection{Complete report on the NDIT study analysis\label{appD}}%

The Nicotine Dependence in Teens (NDIT) study is a prospective longitudinal study of 1,294 grade 7 students recruited from 10 Montréal-area (Canada) high schools in 1999-2000. \cite{Loughlin2015} 
Self-report questionnaires were administered from grade 7 to 11 at each of the 10 schools every three months for a total of 20 cycles from 1999 to 2005 (i.e., during the five years of high school). Mail or in-person questionnaires were administered in 2007/2008 (cycle 21) when participants were age 20.4 years on average.
The data collected include repeated measures of a wide range of socio-demographic, substance use, psychosocial, lifestyle, and physical and mental health variables.

\bmsubsection{NDIT data\label{appD.1}}

\bmsubsubsection{Exposure\label{Exposure}}
Participants were asked ``During the past three months, how often did you drink alcohol (beer, wine, hard liquor)?'' We considered a participant exposed to regular alcohol  use if the participant answered ``once or a couple of times a week'' or ``usually every day'' (alternatives were ``never,'' ``a bit to try'' or ``once or a couple of times a month''). Therefore, ``alcohol use'' in this paper refers to ``at least weekly use''. In defining the population of interest, we excluded all participants reporting  regular alcohol use at time zero. We denoted the binary exposure over time as $A_t$ for time $t$.

\bmsubsubsection{Censoring\label{Censoring}}

We denote the censoring indicators as $C_t$ for each time $t$. A participant was censored by time $t$, denoted $C_t=1$, when they were lost to follow-up or when they skipped more than one entire year of follow-up; otherwise, $C_t=0$.

\bmsubsubsection{Covariates}\label{covars}

\textbf{Baseline covariates}
As baseline variables, we included socio-demographic characteristics including sex (with male as the reference sex), mothers' education (no university vs. some university), whether the participant lived in a single-parent home, if the participant spoke French at home, and country of birth (outside Canada vs. Canada), which were assessed in the first data collection cycle. We also included: self-esteem, impulsivity, and novelty-seeking (a genetic tendency to feel intense excitement and actively explore new or potentially rewarding experiences, while also avoiding monotony and possible punishment \cite{cloninger1987systematic}). While these three variables were measured in the 12-th cycle, because they are considered personality traits and unlikely to vary considerably over time, they were included as baseline covariates \cite{liuNDIT}. Self-esteem was measured using Rosenberg’s Self-Esteem Scale \cite{Rosenberg}; higher values indicate higher self-esteem. \cite{Racicot} Impulsivity was measured with an abbreviated version of the Eysenck Impulsivity Scale, \cite{Eysenck} which was previously validated among adolescents \cite{Wills}; higher scores indicate greater impulsivity. \cite{Racicot} Novelty-seeking was assessed using nine items based on Cloninger’s Tridimensional Personality Questionnaire \cite{Otter}; high scores indicate greater novelty-seeking. \cite{Racicot}

\textbf{Time-varying covariates}
The time-varying covariates $\boldsymbol{L}_t$ were measured for time $t$ and included: Current depressive symptoms; participation in team sports; family-related stress (validated 4-point scale) with higher values indicating more stress; other type of stress (validated 4-point scale); worry about weight;  and ever smoked. Unlike the outcome, current depressive symptoms 
were measured with a validated six-item symptoms scale \cite{Chaiton, Escobedo}; 
higher scores indicated higher levels of depressive symptoms. Family stress was measured using a validated scale over the past three months; 
higher values indicate higher levels of stress.~\cite{Racicot} Other stress referred to the past three months with higher values indicating higher levels of stress.

\bmsubsubsection{Outcomes}\label{Outcome}

The outcome $Y$, depression symptoms, was measured using the Major Depressive Inventory (MDI) in 2007/2008. \cite{Bech1997,Bech} Participants asked how much time in the past two weeks they had: 1) felt low in spirit; 2) lost interest in or could no longer enjoy their daily activities; 3) felt a lack of energy and strength; 4) felt less confident; 5) had a bad conscience or feelings of guilt; 6) felt life was not worth living; 7) had difficulty concentrating; 8) felt very restless; 9) felt subdued or slowed down; 10) had trouble sleeping at night or waking up too early; 11) suffered from reduced appetite; and, 12) suffered from increased appetite. A score of four or more for items 1) and 2), and a score of three or more for the other items indicated a diagnostic demarcation for the depression symptom. For items 8) and 9), the highest score was retained for scoring, and similarly for items 11) and 12). Based on a 6-point scale ranging from 0 (\emph{no time}) to 5 (\emph{all the time}) for each item, responses were summed to generate a continuous score from  0 to 50 with higher scores indicating more severe symptoms. \cite{Chaiton, Bech} This scale measures depression symptoms over the past two weeks. 

\bmsubsection{Handling covariate missingness in the analysis}\label{appD.2}
    
We addressed missing values in covariates that were unrelated to censoring using imputation methods (For cases where missing data were due to censoring, we applied IPTW or LTMLE with censoring weights). To impute missing values in time-dependent covariates, we employed the Last Observation Carried Forward method which was applied for no more than one full academic year of follow-up after the last measured value. For handling missing data in baseline and the remaining time-varying covariates, we utilized multiple imputations by chained equations (MICE), mice \texttt{R} package, \cite{Van} maintaining time-ordering of the follow-up variables throughout this process. Our imputation process involved a single database, and subsequent analyses were conducted to derive the estimates.

\bmsubsection{Target trial}\label{appD.3}
We aim to study the effect of time of initiation of drinking during early high school  on depression in young adulthood. 
To do so, we analyzed data from the NDIT study collected over a span of the first five cycles from 1999 to 2000 and the 21$^{st}$ cycle in 2007/2008. We define the target trial with corresponding intention-to-treat parameters of interest. This hypothetical trial recruits participants who have not yet initiated regular drinking at the beginning of grade 7, e.g. $A_0=0$ for all participants. The target trial randomizes drinking initiation to one of the follow-up time points and conducts an analysis of the correlation between time of initiation and depression symptoms in young adulthood $Y$.

\bmsubsection{Parameter of interest}\label{appD.4}
To perform the intention-to-treat analysis with our observational data, we defined the exposure variable such that once an individual was exposed to regular alcohol use, they were considered exposed for the duration of the study unless they were lost to follow-up. Specifically, if an individual had a value of 1 for $A_t$ at any $t=1,\cdots, 5$, we coded the variable to set all subsequent time points, $A_{t+k}$, to 1 for $k= 1, \cdots, 6-t$. Define $\mathcal{D}$ as the regimen space for the intention-to-treat analysis, then $\mathcal{D}$ represents 6 treatment patterns where the initiation time varies between 1 and 5, and no initiation for all time points, i.e. 

$$\mathcal{D}= \left\{\begin{array}{ccccc}(1, &1, & 1, & 1, & 1)\\(0, &1, & 1, &1, & 1 )\\(0,&0, & 1, & 1, & 1 )\\ (0,&0, & 0, & 1, & 1 )\\ (0,&0, & 0, & 0, & 1 ) \\(0,&0, & 0, & 0, & 0)\end{array}\right\}.$$

Then the parameters of interest can be defined through the working MSM,  
 \begin{align*}
    &\mathbb{E}[Y^{\bar{\boldsymbol{a}}}|\text{Sex}]=\mu_0+\mu_{1}\text{Sex}+ \mu_2\text{cum}(\bar{\boldsymbol{a}})+{\mu}_{3}\{\text{Sex}\times \text{cum}(\bar{\boldsymbol{a}})\}
\end{align*}
where $\mathbb{E}[Y^{\bar{\boldsymbol{a}}}|\text{sex}]$ represents the mean counterfactual outcome under some intervention pattern $\bar{\boldsymbol{a}}$ in a sex subgroup such that sex=1 denotes female, and $\text{cum}(\bar{\boldsymbol{a}})$ counts the cumulative exposures in the pattern. The true parameter values  ${\boldsymbol{\mu}}$ minimize the expectation of a squared error loss function, summing over all patterns in the intention-to-treat space $\mathcal{D}$, corresponding to the parameters estimated in the hypothetical target trials.

\bmsubsection{Model specification}\label{appD.5}

\bmsubsubsection{Data structure}\label{Data structure}

Given the above, the following represents the observed data structure:
 $$O = \{\boldsymbol{L}_1, 
 A_1, \boldsymbol{L}_2, C_2, A_2,\cdots, A_5, \boldsymbol{L}_{6}, C_{6},  Y\}.$$
Note that $\boldsymbol{L}_1$ contains the baseline covariates and the time-varying covariates at the first time and there is no censoring prior to the first exposure time. 

\bmsubsubsection{Outcome models}\label{Outcome models }

We use the notation $\bar{\boldsymbol{L}}_t$ to denote the history of baseline and time-dependent covariates up to time $t$ and likewise $\bar{\boldsymbol{A}}_t$ represents the history of the exposure $A_1, \cdots, A_t$. We rescaled the bounded continuous outcome $Y$ to be contained in $(0,1)$. Denote $T=6$ as the total number of time points. 
Starting with $q_{T+1}(\bar{\boldsymbol{a}}_{T+1}, \bar{\boldsymbol{L}}_{T+1}) = Y$, we  recursively define $$q_t(\bar{\boldsymbol{a}}_t, \bar{\boldsymbol{L}}_t)=\mathbb{E}\{q_{t+1}(\bar{\boldsymbol{a}}_{t+1}, \bar{\boldsymbol{L}}_{t+1})|  \bar{\boldsymbol{L}}_{t}, C_{t}=0,\bar{\boldsymbol{A}}_{t}=\bar{\boldsymbol{a}}_{t}\}, ~~~~t=T,\cdots,1.$$
For the NDIT data setting, the history of the exposures up to time $T$ is equal to the history of the exposures to time $T-1$. 
In order to obtain preliminary estimates  $q_{t,n}(\bar{\boldsymbol{a}}_t, \bar{\boldsymbol{L}}_t)$ of $q_{t}(\bar{\boldsymbol{a}}_t, \bar{\boldsymbol{L}}_t)$, we modeled the outcome for $t=T$ or most recent estimate of $q_{t+1}(\bar{\boldsymbol{a}}_t, \bar{\boldsymbol{L}}_t)$, conditional on main terms of the baseline and time varying covariates, exposure terms (current and lagged) and the first-order interactions of sex and exposure terms for uncensored participants. Then we generated predictions from this model for each pattern of interest $\bar{\boldsymbol{a}}$. We fit logistic regressions stratified on time $t$, corresponding to:\\
For $t=T$,
\begin{align*}
     Y\sim & \sum_{k=1}^t\boldsymbol{L}_{k}+\sum_{k=1}^{t-1}A_{k}+ \sum_{k=1}^{t-1}\left\{\text{Sex}\times A_{k}\right\}
\end{align*}
For $t=T-1, \cdots, 1$,
\begin{align*}
      q_{t+1,n}(\bar{\boldsymbol{a}})\sim & \sum_{k=1}^t\boldsymbol{L}_{k}+\sum_{k=1}^{t}A_{k}+ \sum_{k=1}^{t}\left\{\text{Sex}\times A_{k}\right\}.
\end{align*}


Taking as outcome a vector composed of stacked components $q_{t,n}(\bar{\boldsymbol{a}}_t, \bar{\boldsymbol{L}}_t)$ for each pattern $\bar{\boldsymbol{a}}$, we then run regressions according to the following working regression models:
\begin{align}
    &\hat{E}\{q_{1,n}\mid \bar{\boldsymbol{L}}_1, a_1\}=\beta_{1,0} + \boldsymbol{\beta}_{1,1} \boldsymbol{L}_1+\beta_{1,2} a_1, \notag\\
    &\hat{E}\{q_{2,n}\mid \bar{\boldsymbol{L}}_2, C_2=0,\bar{\boldsymbol{a}}_2\}=\beta_{2,0} + \boldsymbol{\beta}_{2,1} \boldsymbol{L}_1+ \boldsymbol{\beta}_{2,2}\boldsymbol{L}_2+ \beta_{2,3}a_1+\beta_{2,4} a_2 , \notag\\
    &\vdots\notag\\
     &\hat{E}\{q_{T,n}\mid \bar{\boldsymbol{L}}_T,C_T=0,\bar{\boldsymbol{a}}_{T} \}=\beta_{T,0} +\sum_{k=1}^{T}\boldsymbol{\beta}_{T,k} \boldsymbol{L}_k+ \sum_{k=1}^{T}\beta_{T, T+k}a_k  \notag\\ \label{outcome model}
\end{align}
with true parameter values $\boldsymbol{\beta}=\{\beta_{\tau, t};\tau=(1,\cdots,T),~t=(0, \cdots, 2\tau)\}$ minimizing the risk under a squared-error loss function.

\bmsubsubsection{Pooled treatment model and pooled censoring model}\label{Pooled treatment model and pooled censoring model}

As discussed in the manuscript, we define a ``full" model for the probability of treatment at all time-points that adjusts for the full covariate history. The treatment model was fit using those who had not yet initiated and were uncensored at each time point. 
A pooled logistic regression for the conditional probability of treatment, ${P}(A_t = 1\mid {A}_{t-1}=0, \bar{\boldsymbol{L}}_t, C_{t}=0)$, was specified as follows. We define $m_t(\bar{\boldsymbol{L}}_t;\boldsymbol\alpha)$ as the corresponding model for the probability of treatment at times $t=1,\cdots, T$.
\begin{align}
     &\text{logit} \left\{ m_t( \bar{\boldsymbol{L}}_{t};\boldsymbol{\alpha}) \right\} \notag \\
     =~& \mathbb{I}(t=1) \left(\alpha_{1,0}+ \boldsymbol{\alpha}_{1,1}\boldsymbol{L}_1 \right) +\notag \\
     & \mathbb{I}(t=2)  \left(\alpha_{2,0} +\boldsymbol{\alpha}_{2,1}\boldsymbol{L}_1 +\boldsymbol{\alpha}_{2,2} \boldsymbol{L}_2 \right) +\notag \\
     &\vdots\notag \\
      & \mathbb{I}(t=T-1)  \left(\alpha_{T-1,0} +\sum_{k=1}^{T-1} \boldsymbol{\alpha}_{T-1,k}\boldsymbol{L}_k \right).\notag
 \end{align}

In the above, $\boldsymbol{\alpha}= \{\alpha_{\tau, t};\tau=(1,\cdots,T-1),~t=(0, \cdots, \tau)\}$ are the coefficients of the covariates in the pooled propensity score model. Notably, the exposure does not appear in this model since it is fit on the subset of participants who had not yet initiated drinking (and so all past exposure is null).

In addition, the pooled censoring model adjusted for the history of covariates and treatments in order to estimate  $h_t(\bar{\boldsymbol{L}}_t,\bar{A}_{t-1};\boldsymbol\theta)={P}(C_t = 0\mid
     \bar{\boldsymbol{L}}_t, \bar{A}_{t-1},C_{t-1}=0); t=2,\cdots, T$. The model was specified as
\begin{align}
     &\text{logit} \left\{1-h_t(\bar{\boldsymbol{L}}_t,\bar{A}_{t-1},;\boldsymbol\theta)\right\} \notag \\
     =& 
     \mathbb{I}(t=2)  \left(\theta_{2,0} +\boldsymbol{\theta}_{2,1}\boldsymbol{L}_1+\boldsymbol{\theta}_{2,2}\boldsymbol{L}_2 +\theta_{2,3} A_1\right) +\notag \\
     &\vdots\notag \\
      &\mathbb{I}(t=T)  \left(\theta_{T,0} +\sum_{k=1}^{T}\boldsymbol{\theta}_{T,k}\boldsymbol{L}_k+ \sum_{k=1}^{T-1} \theta_{T,T+k}A_k \right).\notag
 \end{align}
In the above, $\boldsymbol{\theta}=\{\theta_{\tau, t};\tau=(2,\cdots,T),~t=(0, \cdots, 2\tau-1)\}$ are the coefficients of the covariates in the pooled censoring model.

\bmsubsection{Cumulative weights for treatment and censoring}\label{appD.6}
To define the weights used in estimation and the balance criteria, we must extend our definition of the model for the probability of exposure to be deterministic when exposure was already initiated in the past. Thus, we define
\begin{equation*}
m_t^*(  \bar{\boldsymbol{L}}_{t}, a_{t-1};\boldsymbol{\alpha})=\left\{\begin{array}{ll}m_t(\bar{\boldsymbol{L}}_{t};\boldsymbol{\alpha}) & for~a_{t-1}=0 \\1 & for~a_{t-1}=1. \end{array}\right.  
\end{equation*}

The cumulative weights for treatment at times $t=(1, \cdots, T-1)$ are
\newcommand{\fdfrac}[2]{\mbox{$\displaystyle\frac{#1}{#2}$}}
\begin{align}
&~~w^a_{t}(a_t,\bar{\boldsymbol{L}}_{t}; \boldsymbol{\alpha})=
 \fdfrac{\mathbb{I}(A_t=a_t)}{a_tm_t( \bar{\boldsymbol{L}}_{t};\boldsymbol{\alpha})+(1-a_t)[1-m_t( \bar{\boldsymbol{L}}_{t};\boldsymbol{\alpha})]} \quad\text{for}~t=1, \text{and}\notag\\
&~~w^a_{t}(a_t,\bar{\boldsymbol{a}}_{t-1},\bar{\boldsymbol{L}}_{t}; \boldsymbol{\alpha})=\fdfrac{\mathbb{I}(A_t=a_t,\bar{\boldsymbol{A}}_{t-1}=\bar{\boldsymbol{a}}_{t-1})}{ \prod_{k=1}^{t}a_km^*_k(a_{k-1},\bar{\boldsymbol{L}}_{k};\boldsymbol{\alpha})+(1-a_k)[1-m^*_k(a_{k-1}, \bar{\boldsymbol{L}}_{k};\boldsymbol{\alpha})]}
\quad \text{for}~t=2,\cdots, T-1. \label{weighta}
\end{align}
    The cumulative weight for censoring at time $t=(2, \cdots, T)$ is
 \begin{align}   
&~~w^c_{t}(\bar{\boldsymbol{L}}_{t}, \bar{\boldsymbol{a}}_{t-1};  \boldsymbol{\theta})=\prod_{k=2}^{t}\frac{\mathbb{I}(C_k=0)}{h_k(\bar{\boldsymbol{L}}_{k}, \bar{\boldsymbol{a}}_{k-1};\boldsymbol{\theta})}.\label{weightc}\\
   &\notag
\end{align}
Combinations of these weights are used in the balance criteria.

The cumulative weights used in IPTW are written as: $w_t^{iptw}=1/P(A_t=a_t\mid \boldsymbol{L}_t=\boldsymbol{l}_t)$ for time $t=1$  and $w_t^{iptw}=1/P(A_1=a_1\mid \boldsymbol{L}_1=\boldsymbol{l}_1) \prod_{k=2}^t w_k$ for time $t=(2,\cdots, T)$ where
{\begin{align*}
w_k=\left\{\begin{array}{ll}\fdfrac{\mathbb{I}(C_k=0)}{P(A_k=a_k\mid a_{k-1}=0, \bar{\boldsymbol{L}}_{k}=\bar{\boldsymbol{l}}_{k}, C_k=0)P(C_k=0\mid \bar{\boldsymbol{L}}_{k}=\bar{\boldsymbol{l}}_{k}, C_{k-1}=0, a_{k-1}=0)} & for~a_{k-1}=0 \\\fdfrac{\mathbb{I}(C_k=0)}{P(C_k=0\mid \bar{\boldsymbol{L}}_{k}=\bar{\boldsymbol{l}}_{k}, C_{k-1}=0, \bar{\boldsymbol{A}}_{k-1}=\bar{\boldsymbol{a}}_{k-1})} & for~a_{k-1}=1. \end{array}\right.  
\end{align*}}%

\bmsubsection{Longitudinal Outcome Adaptive Lasso} \label{appD.7}
We implemented LOAL to select variables for the treatment models and censoring models separately at each time that have corresponding non-zero coefficients $\boldsymbol{\beta}$ in the $q_t$ model fits.

\bmsubsubsection{LOAL for treatment}\label{LOAL for treatment}
Given a regularization parameter $\lambda_{n}^a \geq 0$, the pooled LOAL estimator for $\boldsymbol\alpha^{\dagger}$ is given as,  
\begin{align}
    \hat{\boldsymbol\alpha}(\lambda^a_n) &=\argmin_{\alpha}\sum_{\tau=1}^T\sum_{i=1}^n \left[ a_{\tau,i} \log\{m_{\tau}( \bar{\boldsymbol{l}}_{\tau,i}, \bar{\boldsymbol{a}}_{\tau-1,i}; \boldsymbol\alpha)\} \right.\notag\\
       & \left.+(1-a_{\tau,i})\log\{1-m_{\tau}(\bar{\boldsymbol{l}}_{\tau,i}, \bar{\boldsymbol{a}}_{\tau-1,i};\boldsymbol\alpha)\}\right] + \lambda_n^a\sum_{j\in \mathcal{J}^a} \hat{\omega}_j \lvert\alpha_j\rvert, \notag
\end{align}
where $\hat{\omega}_j=\lvert \hat{\beta}_j \rvert^{-\gamma}$ for $j\in \mathcal{J}^a$ and $\gamma=2.5$. Here $\mathcal{J}^a$ represents the indices of the coefficients $\boldsymbol{\alpha}$ being shrunk, i.e.,
\begin{align*}
    \mathcal{J}^a=\{&(1,1,\mathcal{J}^a_{1,1}),\\
    &(2,1,\mathcal{J}_{2,1}),(2,2,\mathcal{J}^a_{2,2}),\\
   &\vdots\\
    &(T-1,1,\mathcal{J}^a_{T-1,1}),(T-1,2,\mathcal{J}^a_{T-1,2}),\cdots,(T-1,T-1,\mathcal{J}^a_{T-1,T-1})\}
\end{align*}
where $\mathcal{J}^a_{\tau,t}$, for $\tau,t=(1,\cdots, T-1), t\leq \tau$, represents the indices of set of coefficients at each time. Specifically, $\mathcal{J}^a_{\tau,t}$ indexes the specific covariates in $\boldsymbol{L}_t$ being shrunk within propensity score model $\boldsymbol A_{\tau}$. 

This regularized regression for the treatment can be implemented by a transformation of the pooled data, setting 
\begin{align*}
    &V_{1,0}=\mathbb{I}(t=1), \boldsymbol{V}_{1,1}=\mathbb{I}(t=1)\boldsymbol{L}_1,  \\
    &V_{2,0}=\mathbb{I}(t=2), \boldsymbol{V}_{2,1}=\mathbb{I}(t=2)\boldsymbol{L}_1, \boldsymbol{V}_{2,2}=\mathbb{I}(t=2)\boldsymbol{L}_2, \\&\vdots \\
    &V_{T-1,0}=\mathbb{I}(t=T-1), \boldsymbol{V}_{T-1,1}=\mathbb{I}(t=T-1)\boldsymbol{L}_1, \cdots, 
    \boldsymbol{V}_{T-1,T-1}=\mathbb{I}(t=T-1)\boldsymbol{L}_{T-1}
\end{align*}
with respectively corresponding coefficients in $\boldsymbol{\alpha}$. Then, the adaptive LASSO is run with pooled outcome $A_{\tau}$ on these covariates $V_{1,0},...,\boldsymbol{V}_{T-1,T-1}$, without an intercept term, using weights $\hat{\omega}_j=\lvert \hat{\beta}_j \rvert^{-\gamma}; j\in \mathcal{J}^a$.

\bmsubsubsection{LOAL for censoring}\label{LOAL for censoring}
For the censoring, given the same $\hat{\omega}_j=\lvert \hat{\beta}_j \rvert^{-\gamma}$ for $j\in \mathcal{J}^c$ and $\gamma=2.5$, the pooled LOAL estimator $\boldsymbol\theta^{\dagger}_0$ on $\lambda_{n}^c \geq 0$ is,
\begin{align}
    \hat{\boldsymbol\theta}(\lambda_n^c)&=\argmin_{\theta}\sum_{\tau=2}^T\sum_{i=1}^n \left[(1- c_{\tau,i} )\log\{h_{\tau}(\bar{\boldsymbol{l}}_{\tau,i},\bar{\boldsymbol{a}}_{\tau-1,i}; \boldsymbol\theta)\} \right.\notag\\
       & \left.+ c_{\tau,i}\log\{1-h_{\tau}(\bar{\boldsymbol{l}}_{\tau,i},\bar{\boldsymbol{a}}_{\tau-1,i};\boldsymbol\theta)\}\right] + \lambda_n^c\sum_{j\in \mathcal{J}^c} \hat{\omega}_j \lvert\theta_j\rvert, \notag
\end{align}
where $\mathcal{J}^c$ represents the indices of the coefficients $\boldsymbol{\theta}$ being shrunk. $\mathcal{J}^c_{\tau,t}$ for $\tau=(2,\cdots, T),t=(1,\cdots, T), t\leq\tau$ indexes the specific covariates in $\boldsymbol{L}_t$ within the censoring models $C_{\tau}$. Note that the intercepts and the coefficients corresponding to treatments are not shrunk,
\begin{align*}
    \mathcal{J}^c=\{&(2,1,\mathcal{J}^c_{2,1}),(2,2,\mathcal{J}^c_{2,2}),\\&(3,1,\mathcal{J}^c_{3,1}),(3,2,\mathcal{J}^c_{3,2}),(3,3,\mathcal{J}^c_{3,3}),\\
    &\vdots\\
    &(T,1,\mathcal{J}^c_{T,1}),(T,2,\mathcal{J}^c_{T,2}),\cdots,(T,T,\mathcal{J}^c_{T,T})\}.
\end{align*}
Likewise, this regularization for the censoring can be applied through a transformation of the pooled data, but encompassing not only the variables of time and covariates but also treatment related variables, 
\begin{align*}
    &U_{2,0}=\mathbb{I}(t=2), U_{2,1}=\mathbb{I}(t=2)\boldsymbol{L}_1, U_{2,2}=\mathbb{I}(t=2)\boldsymbol{L}_2,  U_{2,3}=\mathbb{I}(t=2)A_1, \\ 
    &\vdots \\
    &U_{T,0}=\mathbb{I}(t=T), U_{T,1}=\mathbb{I}(t=T)\boldsymbol{L}_1, \cdots, U_{T,T}=\mathbb{I}(t=T)\boldsymbol{L}_{T},\\
    &\hspace{8mm}U_{T,T+1}=\mathbb{I}(t=T)A_1,\cdots, U_{T,2T-1}=\mathbb{I}(t=T)A_{T-1}
\end{align*}
with respectively corresponding coefficients in $\boldsymbol{\theta}$. Then, the adaptive LASSO is run with pooled outcome $C_t$ on these covariates $U_{2,0},...,U_{T,2T-1}$ without an intercept term.

\bmsubsection{Selection of \texorpdfstring{$\lambda_n^a$}{lambda\_n\^a} and \texorpdfstring{$\lambda_n^c$}{lambda\_n\^c}}\label{Selection of lambda}
The tuning parameter $\lambda_n^a$ and $\lambda_n^c$ were selected jointly by minimizing the sum of the balance metrics for treatment and censoring. The balance metric for treatment, $\mathcal{M}$, is a summary of weighted absolute mean differences (wAMDs) of the covariates between the exposure groups. \cite{SE}  Similarly, the balance metric for censoring, $\mathcal{N}$, is based on the wAMDs of the covariates between uncensored and censored groups. 

The weights involved are cumulative inverse probability weights for current treatment or censoring.
For balance across treatment groups at a given time, we only consider histories with no past exposure, since the only comparison to make is in people who initiated or did not initiate exposure. Let $\hat{\boldsymbol\alpha}^{refit}(\lambda_n^a)$ represent the estimates from a logistic regression of the treatment on the covariates selected by LOAL under tuning parameter $\lambda_n^a$ where the value is set to be zero if the corresponding coefficient was not selected. Similarly, $\hat{\boldsymbol\theta}^{refit}(\lambda_n^c)$ represent the estimates from a logistic regression of the censoring on the covariates selected by LOAL under tuning parameter $\lambda_n^c$. Based on equations (\ref{weighta}) and (\ref{weightc}), we define the weight for subject $i$ at time $t$ for the current treatment as 
$$w_{t,i}^{a}=w^a_{t}\{a_{t,i},\bar{\boldsymbol{a}}_{t-1,i}=0, \bar{\boldsymbol{l}}_{t,i};\hat{\boldsymbol\alpha}^{refit}(\lambda_n^a)\}w^c_{t}\{\bar{\boldsymbol{a}}_{t-1,i},\bar{\boldsymbol{l}}_{t,i};\hat{\boldsymbol\theta}^{refit}(\lambda_n^c)\}$$ 
where $\hat{\boldsymbol\alpha}(\lambda_n^a)$ and $\hat{\boldsymbol\theta}(\lambda_n^c)$ are parameter estimates under the treatment and censoring model after variable selection by LOAL with the tuning parameters $(\lambda_n^a, \lambda_n^c)$, respectively. Also, 
$$w_{t,i}^{c}=w^c_{t}\{\bar{\boldsymbol{a}}_{t-1,i},\bar{\boldsymbol{l}}_{t,i};\hat{\boldsymbol\theta}^{refit}(\lambda_n^c)\}w^a_{t-1}\{a_{t-1,i},\bar{\boldsymbol{l}}_{t-1,i};\hat{\boldsymbol\alpha}^{refit}(\lambda_n^a)\}$$  is the weight for current censoring for subject $i$ at time $t$ estimated under the proposed LOAL approach of censoring and treatment.
Let $L_{t,k}$ denote the $k^{th}$ component in $\boldsymbol{L}_t$ for $k=(1, \cdots, p_k)$ where $p_k$ represents the number of components of $\boldsymbol{L}_{t}$. Then $\beta_{\tau,t,k}$ represents the coefficients in the structural equations~(equations \ref{outcome model}) for $\tau=(1,\cdots, T-1)$ and $\tau=(2,\cdots, T)$ referring to the treatment model and censoring model, respectively. Then the weighted absolute mean difference of the treatment and of the censoring can be evaluated based on the variable considered at time $t$ respectively weighted by the corresponding structural models coefficient (equations \ref{outcome model}) divided by its standard error. 
\begin{align}
    &\text{wAMD}^a_{\tau,t,k} = \notag\\
    &\frac{\mid\hat{\beta}_{\tau,t,k}\mid }{\sigma_{\hat{\beta}_{\tau,t,k}}}
 \left| \frac{\sum_{i=1}^n a_{\tau,i}l_{t,k,i}w^{a}_{\tau,i}\mathbb{I}(a_{\tau-1,i}=0,c_{\tau,i}=0)}{\sum_{i=1}^n a_{\tau,i}w^{a}_{\tau,i}\mathbb{I}(a_{\tau-1,i}=0,c_{\tau,i}=0)} -
  \frac{\sum_{i=1}^n{(1-a_{\tau,i})l_{t,k,i}}w^{a}_{\tau,i}\mathbb{I}(a_{\tau-1,i}=0,c_{\tau,i}=0)}{\sum_{i=1}^n{(1-a_{\tau,i})}w^{a}_{\tau,i}\mathbb{I}(a_{\tau-1,i}=0,c_{\tau,i}=0)}
 \right|\notag \\
& \text{for~} \tau=(1,\cdots, T-1), t=(1,\cdots, T-1) \text{~and}~t\leq \tau. \notag\\
\notag\\
    &\text{wAMD}^c_{\tau,t,k} = \frac{\mid\hat{\beta}_{\tau,t,k}\mid }{\sigma_{\hat{\beta}_{\tau,t,k}}}
 \Biggl| \frac{\sum_{i=1}^n \mathbb{I}(c_{\tau,i}=0)l_{t,k,i}w^{c}_{\tau,i}}{\sum_{i=1}^n \mathbb{I}(c_{\tau,i}=0)w^{c}_{\tau,i}} -\frac{\sum_{i=1}^n{\{\mathbb{I}(c_{\tau,i}=1)l_{t,k,i}}w^{c}_{\tau,i}\}}{\sum_{i=1}^n{\{\mathbb{I}(c_{\tau,i}=1)}w^{c}_{\tau,i}\}}
 \Biggr| \notag\\
 & \text{for~} \tau=(2,\cdots, T) , t=(1,\cdots, T) \text{~and}~t\leq \tau \notag
\end{align}

We selected the two tuning parameters by minimizing the sum of the balance criterion for the treatment and the balance criterion for the censoring, i.e. the selected $(\lambda_n^a, \lambda_n^c) = \argmin_{\lambda_n^a, \lambda_n^c} (\mathcal M + \mathcal N)$ where 

\begin{align}
\mathcal M = &\sum_{k=1}^{p_1}\text{wAMD}^a_{1,1,k} + \sum_{t=1}^{2}\sum_{k=1}^{p_k}\text{wAMD}^a_{2,t,k} +\cdots +
  \sum_{t=1}^{T-1}\sum_{k=1}^{p_k} \text{wAMD}^a_{T-1,t,k} \notag
\\
=& \sum_{\tau=1}^{T-1}\sum_{t=1}^{\tau}\sum_{k=1}^{p_k} \text{wAMD}^a_{\tau,t,k} \notag\\
\mathcal N = &\sum_{t=1}^{2}\sum_{k=1}^{p_k}\text{wAMD}^c_{2,t,k} + \sum_{t=1}^{3}\sum_{k=1}^{p_k}\text{wAMD}^c_{3,t,k} +\cdots +
  \sum_{t=1}^{T}\sum_{k=1}^{p_k} \text{wAMD}^c_{T,t,k}\notag
\\\notag
=& \sum_{\tau=2}^{T}\sum_{t=1}^{\tau}\sum_{k=1}^{p_k} \text{wAMD}^c_{\tau,t,k}\\\notag
\end{align}

\bmsubsection{Selective fusion}\label{appD.9}
To perform the selective fusion, we initially establish a penalty graph, wherein vertices represent coefficients within the pooled model eligible for fusion. This graph may be structured with cliques connecting elements that share common variable names across different time points. For example, it links the remaining baseline covariates in different treatment model times. As for time-varying covariates, we created cliques to allow coefficient fusion of the  most recent variables of common types across various time points; this allows for the possibility of common effects of historical covariates with common lag time on current exposure initiation. Specifically, the penalty graph connected the same baseline variables across time points, and the same time-varying variables with the same lag across time points (e.g., the $\boldsymbol{L}_{t-1}$ variables are connected when modeling treatment and censoring across times $t$).

For example, consider the propensity score model for $A_4$ and a particular time varying covariate $L_{t,3}$. Suppose that both $L_{3,3}$ (the most recent) and $L_{4,3}$ (the current) were selected into this model. In the model for $A_5$, suppose that $L_{4,3}$ and $L_{5,3}$ were selected.  In the model for $A_6$, suppose that $L_{4,3}$ and $ L_{6,3}$ were selected. The cliques  connect the coefficients of the three current variables, $L_{4,3}, L_{5,3}$, and $L_{6,3}$ across the models for $A_4, A_5$ and $A_6$, respectively, and also connect the two most recent variables, $L_{3,3}$ and $L_{4,3}$ in the models for $A_4$ and $A_5$, respectively. 

Denote $\mathcal{G}$ as the set of all pairwise connected indices of the coefficients in accordance with the fusion graph definition. 
The fused LASSO penalizes the absolute differences between the coefficients of connected variables.

Define $\boldsymbol{\alpha}^{*}$ as the parameter vector of the same length as $\boldsymbol{\alpha}$ that is set to zero at the indices of the zero-elements of $\hat{\boldsymbol{\alpha}}^{refit}(\lambda_n^a)$.Then the generalized Adaptive Fused LASSO for treatments is 
\begin{align*}
    &\argmin_{\boldsymbol{\alpha}^{*}}\sum_{\tau=1}^{T-1}\sum_{i=1}^n \biggl[ a_{\tau,i} \log\{m_{\tau}(\bar{\boldsymbol{l}}_{\tau,i}, \bar{\boldsymbol{a}}_{\tau-1,i};\boldsymbol{\alpha}^{*})\} +(1-a_{\tau,i})\log\{1-m_{\tau}(\bar{\boldsymbol{l}}_{\tau,i}, \bar{\boldsymbol{a}}_{\tau-1,i};\boldsymbol{\alpha}^{*})\}\biggr]\notag\\
& + \lambda_{1,n}^a\sum_{(\mathcal{J}^a_{\tau, t}, \mathcal{J}^a_{\tau', t'})\in \mathcal{G}^a}\frac{\lvert \boldsymbol{\alpha}^{*}_{\mathcal{J}^a_{\tau, t}}-\boldsymbol{\alpha}^{*}_{\mathcal{J}^a_{\tau', t'}} \rvert 
}{{\lvert \hat{\boldsymbol{\alpha}}^{refit}_{\mathcal{J}^a_{\tau, t}}(\lambda_n^a)-\hat{\boldsymbol{\alpha}}^{refit}_{\mathcal{J}^a_{\tau', t'}}(\lambda_n^a) \rvert}^{\tau}}.
 \end{align*}
 where $ (\mathcal{J}^a_{\tau, t}, \mathcal{J}^a_{\tau', t'})$ is a pair of indices in the graph for treatment $\mathcal{G}^a$. Note that $\tau, \tau', t, t'$ all in $ (1,\cdots,T-1)$, and $t\leq \tau, t'\leq\tau', \tau \neq \tau', t\neq t'$, and  $\tau-\tau'=t-t'$ based on the penalty graph we defined.

The penalty graph for censoring $\mathcal{G}^c$ was created in the same way as the graph for treatment model. Note that all coefficients corresponding to the treatment terms were not allowed to fuse. Denote $\boldsymbol{\theta}^{*}$ as the vector of the same length as $\boldsymbol{\theta}$ that is set to zero at the indices of the zero-elements of $\hat{\boldsymbol{\theta}}(\lambda_n^c)$. The generalized Adaptive Fused LASSO for the censoring is 
\begin{align*}
    &\argmin_{\boldsymbol{\theta}^{*}}\sum_{\tau=2}^T\sum_{i=1}^n  \biggl[ (1-c_{{\tau},i})\log\{h_{\tau}(\bar{\boldsymbol{l}}_{{\tau},i},\bar{\boldsymbol{a}}_{{\tau}-1,i};\boldsymbol{\theta}^{*})\} +c_{{\tau},i}\log\{1-h_{\tau}( \bar{\boldsymbol{l}}_{{\tau},i},\bar{\boldsymbol{a}}_{{\tau}-1,i};\boldsymbol{\theta}^{*})\}\biggr]\notag\\
& + \lambda_{1,n}^c\sum_{(\mathcal{J}^c_{\tau, t}, \mathcal{J}^c_{\tau', t'})\in\mathcal{G}^c}\frac{\lvert \boldsymbol{\theta}^{*}_{\mathcal{J}^c_{\tau, t}}-\boldsymbol{\theta}^{*}_{\mathcal{J}^c_{\tau', t'}} \rvert 
}{{\lvert \hat{\boldsymbol{\theta}}^{refit}_{\mathcal{J}^c_{\tau, t}}(\lambda_n^c)-\hat{\boldsymbol{\theta}}^{refit}_{\mathcal{J}^c_{\tau', t'}}(\lambda_n^c) \rvert}^{\tau}},
 \end{align*}
where  $(\mathcal{J}^c_{\tau, t}, \mathcal{J}^c_{\tau', t'})$ is a pair of connected indices  in the graph for censoring $\mathcal{G}^c$ and  $\tau, \tau', t, t'$ all in $ (2,\cdots,T)$, and $t\leq \tau, t'\leq\tau', \tau \neq \tau', t\neq t'$, and  $\tau-\tau'=t-t'$ . We utilized the archived \texttt{FusedLasso} package to implement the fusion step. The selection of optimal $\lambda_{1,n}^a$ and $\lambda_{1,n}^c$ values for the treatment and censoring models was determined based on the summation of the Bayesian Information Criterion (BIC) of the treatment and censoring models.

\bmsubsection{NDIT results}\label{appD.10}

The intention-to-treat analysis of the NDIT data included eight baseline covariates and six time-varying covariates. The pooled treatment model included 130 variables, while the pooled censoring model conditioned on 175 variables. To regularize our models, we  set the tuning parameter $\gamma$ to 2.5 and set 20 possible values for the tuning parameters $\lambda^a$ and $\lambda^c$ (refer to Table \ref{lambda table}). The selected tuning parameters,  $[\lambda_n^a, \lambda_n^c]$, were found to be $[3.728, 67.392]$. These values corresponded to a strong penalty for the treatment model and a relatively light penalty for the censoring model. For the fusion step, we considered 20 possible values for $\lambda_{1,n}^a$ within the range $[e^{-10}, e^{-1}]$ and for $\lambda_{1,n}^c$ within the range $[e^{-5}, 1]$. Ultimately, the minimum summation of BICs corresponding to the treatment and censoring was achieved with $\lambda_{1,n}^a$ in $[0.002, 0.368]$ and $\lambda_{1,n}^c$ in $[0.206, 1]$.

 The initial treatment model, with 135 parameters (including five intercepts), was  reduced to 37 parameters. Consequently, the fusion step further reduced the number of parameters to 23 (see Table in the main manuscript). The variables sex, country of birth, current depressive symptoms, and worry about weight  were selected for inclusion in each time period, and their corresponding coefficients were then fused.  The variables mother education, ever smoked, family stress, other stress, and team sports were selected into the models for some time-points but were not fused. 
 For the censoring models, initially, there were 180 parameters (including five intercepts and 15 past treatments), which were later reduced to 112 due to selection and further fused to produce a total of 55 parameters. The selected and fused variables in this case included sex, country of birth, current depressive symptoms, ever smoked, family-related stress, other stress, participation in team sports and worry about weight (Table with results is in the main manuscript).

We also applied LTMLE to estimate the target parameters of the MSM. Based on the same outcome models which involved all covariate main terms and the interaction terms between sex and treatments, we implemented: 1) LTMLE using propensity scores on the full set of covariates; 2) LTMLE using propensity scores on the selected set of covariates by LOAL; 3) LTMLE using propensity scores after LOAL selection and fusion. \textcolor{black}{In addition, we also applied LTMLE with machine learning (\texttt{superlearner R} package) for stratified treatment models, censoring models and outcome models in which we included the algorithms: ``SL.mean'', ``SL.glm'', ``SL.gam'',  ``SL.gam, screen.randomForest'', ``SL.glm.interaction'', ``SL.glm.interaction, screen.randomForest'', ``SL.earth'', ``SL.earth, screen.randomForest''. }Estimated coefficients and standard errors are presented in the Table in the main manuscript. We used the \texttt{sandwich R} package to estimate the robust sandwich variance of IPTW and the variance of LTMLE was estimated based on influence function. 

All methods consistently demonstrated that being female was associated with more severe depressive symptoms when compared to males, who served as the reference group. IPTW full\textcolor{black}{, LTMLE full, and LTMLE SL} had point-estimates that suggested that early alcohol initiation was linked to detrimental effects on depressive symptoms in males. All IPTW point-estimates suggested that earlier alcohol initiation was beneficial for females, while the LTMLE results indicated harmful or null effects for females.  Furthermore, propensity scores derived from covariates selected by LOAL  and then fused led to apparent reduced estimation variance in both IPTW and LTMLE analyses. \textcolor{black}{In addition, to assess the extent of practical positivity violations, we examined the summary of the cumulative product probabilities of treatment and censoring under the full model and the model selected by LOAL, as used in the IPTW analyses. The results are presented in Table \ref{ps_ndit}, showing that using full models and models selected by LOAL yield high medians near 0.90, with LOAL exhibiting slightly less extreme minimum and maximum values.}

\begin{table}[ht]
\centering \color{black}
\caption{\textcolor{black}{NDIT analysis: Summary of the cumulative product probabilities of treatment and censoring using full models or with models selected by LOAL}}\label{ps_ndit}
\begin{tabular}{lrrrrrrr}
  \toprule
 Method & Min. & 10\% & 25\% & 50\% & 75\% & 90\% &  Max.\\ 
  \midrule
Full & 0.0008 & 0.1858 & 0.4595 & 0.9031 & 0.9720 & 0.9898 & 0.9991 \\ 
  LOAL & 0.0012 & 0.1749 & 0.4409 & 0.9038 & 0.9590 & 0.9775 & 0.9893 \\ 
   \bottomrule
\end{tabular}
\end{table}

\begin{sidewaystable} 
\caption{{NDIT analysis: Grid of balance criteria values based on 20 values of $\lambda^a$ and $\lambda^c$, used in the selection step for the treatment and censoring outcome-adaptive LASSOs. The first column represents 20 values of $\lambda^a$ (treatment model) and the first row represents 20 values of $\lambda^c$ (censoring model). In each cell, the left number represents the balance criterion value related to treatment  and the right number represents the balance criterion value related to censoring  for the corresponding $\lambda^a$ and $\lambda^c$ pair. The pair of numbers highlighted in red indicates minimum sum of the balance criteria.}} \label{lambda table}
\resizebox{\textwidth}{!}{
\begin{tabular}{rr|ccccccccccc}
 \toprule
 & \multirow{2}{*}{\textbf{$\lambda_n^a/\lambda_n^c$} } & 1 & 2 & 3 & 4 & 5 & 6 & 7 & 8 & 9 & 10 \\ 
 &  & 22026.466 & 10001.86 & 4541.682 & 2062.304 & 936.459 & 425.231 & 193.09 & 87.679 & 39.814 & 18.079 \\ 
  \midrule
 1 & 2980.958 & (16.369, 46.532) & (16.369, 46.532) & (16.31, 46.578) & (16.377, 47.207) & (16.306, 48.167) & (16.292, 48.351) & (16.256, 48.497) & (16.218, 47.631) & (16.242, 47.788) & (16.181, 48.484) \\ 
  2 & 1585.129 & (14.432, 49.587) & (14.432, 49.587) & (14.389, 49.472) & (14.455, 48.651) & (14.334, 48.255) & (14.313, 48.281) & (14.272, 48.263) & (14.202, 47.601) & (14.196, 47.703) & (14.172, 47.805) \\ 
  3 & 842.895 & (13.629, 51.104) & (13.629, 51.104) & (13.593, 51.027) & (13.632, 50.253) & (13.504, 50.003) & (13.484, 50.035) & (13.45, 50.08) & (13.389, 49.672) & (13.398, 49.559) & (13.408, 49.453) \\ 
  4 & 448.211 & (13.517, 51.445) & (13.517, 51.445) & (13.482, 51.374) & (13.521, 50.695) & (13.394, 50.409) & (13.366, 50.433) & (13.331, 50.577) & (13.283, 50.157) & (13.294, 49.956) & (13.325, 49.929) \\ 
  5 & 238.337 & (12.925, 49.895) & (12.925, 49.895) & (12.886, 49.772) & (12.928, 49.003) & (12.882, 48.973) & (12.882, 49.058) & (12.85, 49.057) & (12.78, 48.514) & (12.734, 48.773) & (12.691, 48.791) \\ 
  6 & 126.736 & (13.011, 49.078) & (13.011, 49.078) & (12.974, 49.043) & (13.024, 49.144) & (12.987, 52.875) & (12.988, 53.166) & (12.977, 53.071) & (12.916, 52.259) & (12.863, 52.665) & (12.793, 53.189) \\ 
  7 & 67.392 & (13.233, 48.36) & (13.233, 48.36) & (13.196, 48.367) & (13.25, 47.889) & (13.138, 48.241) & (13.142, 48.33) & (13.128, 48.433) & (13.102, 47.836) & (13.078, 47.709) & (13.001, 47.548) \\ 
  8 & 35.836 & (13.769, 48.901) & (13.769, 48.901) & (13.729, 49.341) & (13.782, 48.709) & (13.8, 53.452) & (13.889, 53.535) & (13.923, 53.453) & (13.809, 52.908) & (13.757, 52.433) & (13.534, 53.065) \\ 
 9 & 19.056 & (12.23, 63.313) & (12.23, 63.313) & (12.182, 63.526) & (12.242, 65.978) & (12.221, 70.768) & (12.281, 70.792) & (12.3, 70.589) & (12.196, 69.671) & (12.147, 69.004) & (12, 69.429) \\ 
  10 & 10.133 & (14.396, 79.943) & (14.396, 79.943) & (14.352, 80.08) & (14.425, 81.565) & (14.339, 81.133) & (14.36, 81.155) & (14.372, 81.368) & (14.318, 82.405) & (14.272, 81.373) & (14.184, 82.627) \\ 
  11 & 5.388 & (13.033, 50.673) & (13.033, 50.673) & (12.976, 50.8) & (13.07, 51.984) & (12.953, 56.357) & (12.953, 56.32) & (12.931, 55.807) & (12.924, 54.88) & (12.896, 55.083) & (12.845, 55.337) \\ 
  12 & 2.865 & (11.737, 78.484) & (11.737, 78.484) & (11.681, 78.818) & (11.779, 78.971) & (11.66, 79.397) & (11.653, 79.338) & (11.63, 79.895) & (11.645, 79.97) & (11.624, 78.424) & (11.582, 79.129) \\ 
  13 & 1.524 & (10.21, 73.849) & (10.21, 73.849) & (10.159, 74.314) & (10.256, 74.081) & (10.101, 75.355) & (10.085, 75.27) & (10.057, 75.701) & (10.091, 76.048) & (10.08, 74.789) & (10.075, 75.687) \\ 
  14 & 0.81 & (10.125, 78.234) & (10.125, 78.234) & (10.069, 78.646) & (10.166, 77.191) & (10.069, 78.387) & (10.075, 78.158) & (10.062, 78.571) & (10.056, 79.251) & (10.035, 78.483) & (9.979, 78.849) \\ 
  15 & 0.431 & (9.926, 77.562) & (9.926, 77.562) & (9.872, 77.969) & (9.964, 76.484) & (9.85, 77.671) & (9.848, 77.439) & (9.835, 77.861) & (9.84, 78.573) & (9.815, 77.751) & (9.784, 78.117) \\ 
  16 & 0.229 & (10.198, 82.721) & (10.198, 82.721) & (10.137, 82.982) & (10.232, 82.354) & (10.147, 82.951) & (10.145, 82.762) & (10.127, 82.964) & (10.128, 83.23) & (10.105, 83.2) & (10.051, 83.456) \\ 
  17 & 0.122 & (9.681, 85.639) & (9.681, 85.639) & (9.626, 85.92) & (9.724, 85.554) & (9.629, 85.726) & (9.614, 85.573) & (9.587, 85.704) & (9.586, 85.678) & (9.571, 86.078) & (9.547, 86.678) \\ 
  18 & 0.065 & (10.333, 85.462) & (10.333, 85.462) & (10.278, 85.721) & (10.376, 85.195) & (10.271, 85.544) & (10.264, 85.396) & (10.238, 85.512) & (10.25, 85.467) & (10.238, 85.73) & (10.223, 86.233) \\ 
  19 & 0.034 & (10.376, 84.707) & (10.376, 84.707) & (10.32, 85.021) & (10.414, 84.404) & (10.314, 84.796) & (10.307, 84.648) & (10.285, 84.789) & (10.296, 84.762) & (10.288, 85.052) & (10.265, 85.627) \\ 
  20 & 0.018 & (9.842, 82.699) & (9.842, 82.699) & (9.786, 83.157) & (9.878, 82.589) & (9.775, 82.927) & (9.772, 82.786) & (9.763, 82.662) & (9.754, 82.957) & (9.747, 83.58) & (9.702, 84.764) \\  
  \midrule
    & \multirow{2}{*}{\textbf{$\lambda_n^a/\lambda_n^c$} }  & 11 & 12 & 13 & 14 & 15 & 16 & 17 & 18 & 19 & 20 \\ 
   &  & 8.209 & 3.728 & 1.693 & 0.769 & 0.349 & 0.158 & 0.072 & 0.033 & 0.015 & 0.007 \\ \midrule
   1 & 2980.958 & (16.241, 47.392) & (16.268, 47.381) & (16.304, 47.619) & (16.196, 48.132) & (16.202, 48.08) & (16.211, 48.124) & (16.206, 48.47) & (16.214, 48.083) & (16.251, 48.359) & (16.251, 48.362) \\ 
 2 & 1585.129 & (14.201, 47.442) & (14.22, 47.336) & (14.259, 47.534) & (14.135, 48.18) & (14.129, 47.388) & (14.139, 47.278) & (14.133, 47.424) & (14.139, 47.375) & (14.176, 47.953) & (14.176, 47.936) \\ 
  3 & 842.895 & (13.379, 49.057) & (13.4, 49.067) & (13.464, 49.183) & (13.317, 49.981) & (13.305, 49.313) & (13.313, 49.209) & (13.31, 49.466) & (13.312, 49.396) & (13.348, 49.893) & (13.349, 49.881) \\ 
  4 & 448.211 & (13.283, 49.489) & (13.306, 49.489) & (13.371, 49.691) & (13.218, 50.605) & (13.192, 49.802) & (13.2, 49.708) & (13.195, 49.853) & (13.189, 49.82) & (13.223, 50.437) & (13.224, 50.426) \\ 
  5 & 238.337 & (12.675, 48.203) & (12.691, 48.246) & (12.705, 48.394) & (12.634, 49.274) & (12.617, 48.925) & (12.625, 48.857) & (12.622, 48.992) & (12.618, 48.9) & (12.648, 49.573) & (12.649, 49.561) \\ 
  6 & 126.736 & (12.791, 51.678) & (12.804, 51.85) & (12.821, 51.639) & (12.757, 52.187) & (12.756, 53.047) & (12.765, 53.082) & (12.759, 53.512) & (12.753, 53.23) & (12.782, 53.336) & (12.782, 53.344) \\ 
  7 & 67.392 & (12.99, 47.300) & \color{red}(13.009, 47.213) & (13.047, 47.37) & (12.941, 47.901) & (12.984, 47.415) & (12.991, 47.361) & (12.986, 47.613) & (12.994, 47.469) & (13.021, 47.791) & (13.02, 47.794) \\ 
  8 & 35.836 & (13.561, 52.193) & (13.559, 51.975) & (13.515, 51.59) & (13.477, 50.41) & (13.534, 51.194) & (13.541, 51.249) & (13.541, 51.16) & (13.558, 51.292) & (13.592, 50.791) & (13.588, 50.83) \\ 
  9 & 19.056 & (12.029, 68.203) & (12.035, 68.258) & (12.035, 67.988) & (11.94, 66.67) & (12.021, 67.544) & (12.028, 67.56) & (12.015, 68.187) & (12.028, 68.354) & (12.06, 68.077) & (12.058, 68.107) \\ 
  10 & 10.133 & (14.224, 81.347) & (14.235, 81.301) & (14.266, 81.324) & (14.184, 80.782) & (14.259, 80.037) & (14.267, 80.021) & (14.262, 80.026) & (14.272, 80.354) & (14.305, 80.168) & (14.302, 80.212) \\ 
  11 & 5.388 & (12.872, 54.095) & (12.895, 54.104) & (12.926, 53.702) & (12.87, 52.07) & (12.914, 53.444) & (12.919, 53.514) & (12.913, 53.945) & (12.92, 54.03) & (12.95, 53.677) & (12.949, 53.71) \\ 
  12 & 2.865 & (11.608, 79.496) & (11.64, 79.382) & (11.684, 79.465) & (11.583, 78.679) & (11.629, 78.595) & (11.634, 78.551) & (11.614, 78.531) & (11.618, 78.872) & (11.65, 78.364) & (11.647, 78.405) \\ 
  13 & 1.524 & (10.086, 75.889) & (10.114, 75.651) & (10.184, 75.592) & (10.051, 74.761) & (10.09, 74.363) & (10.095, 74.311) & (10.073, 74.413) & (10.075, 74.792) & (10.106, 74.1) & (10.103, 74.144) \\ 
  14 & 0.81 & (10.032, 79.517) & (10.051, 79.263) & (10.094, 79.325) & (9.983, 78.107) & (10.028, 78.435) & (10.033, 78.4) & (10.014, 78.353) & (10.019, 78.587) & (10.053, 78.176) & (10.049, 78.218) \\ 
  15 & 0.431 & (9.811, 78.784) & (9.837, 78.52) & (9.891, 78.6) & (9.79, 77.35) & (9.837, 77.661) & (9.843, 77.63) & (9.825, 77.613) & (9.818, 77.864) & (9.846, 77.454) & (9.843, 77.496) \\ 
  16 & 0.229 & (10.106, 83.874) & (10.115, 83.644) & (10.157, 83.668) & (10.042, 82.409) & (10.084, 82.672) & (10.09, 82.666) & (10.078, 82.637) & (10.073, 82.794) & (10.105, 82.672) & (10.102, 82.716) \\ 
  17 & 0.122 & (9.547, 86.89) & (9.548, 86.636) & (9.6, 86.675) & (9.535, 85.576) & (9.58, 85.272) & (9.589, 85.272) & (9.573, 85.225) & (9.549, 85.513) & (9.578, 85.992) & (9.576, 86.036) \\ 
  18 & 0.065 & (10.219, 86.494) & (10.223, 86.24) & (10.287, 86.264) & (10.184, 85.157) & (10.23, 84.971) & (10.239, 84.966) & (10.223, 84.91) & (10.207, 85.175) & (10.235, 85.479) & (10.233, 85.523) \\ 
  19 & 0.034 & (10.267, 85.835) & (10.268, 85.56) & (10.334, 85.58) & (10.242, 84.484) & (10.285, 84.148) & (10.294, 84.143) & (10.278, 84.112) & (10.262, 84.407) & (10.293, 84.797) & (10.291, 84.842) \\ 
  20 & 0.018 & (9.741, 84.284) & (9.741, 83.937) & (9.775, 83.957) & (9.733, 82.972) & (9.773, 82.141) & (9.781, 82.143) & (9.769, 82.121) & (9.756, 82.356) & (9.772, 83.187) & (9.77, 83.231) \\ 
   \bottomrule
\end{tabular}}
\end{sidewaystable}

\end{document}